\documentclass{aa} 

\usepackage{graphicx}
\usepackage{color}
\usepackage{float}
\usepackage{amssymb, amsmath}
\begin{document}

   \title{A 1.3 mm SMA Survey of 29 Variable Young Stellar Objects}

   \subtitle{}

   \author{Hauyu Baobab Liu\inst{1}
          \and
          Michael M. Dunham\inst{2,3}
          \and
          Ilaria Pascucci\inst{4}
          \and
          Tyler L. Bourke\inst{5}
          \and
          Naomi Hirano\inst{6}
          \and
          Steven Longmore\inst{7}
          \and
          Sean Andrews\inst{3}
          \and
          Carlos Carrasco-Gonz\'{a}lez\inst{8}
          \and
          Jan Forbrich\inst{9}
          \and
          Roberto Galv\'{a}n-Madrid\inst{8}
          \and
          Josep M. Girart\inst{10}
          \and
          Joel D. Green\inst{11}
          \and
          Carmen Ju\'{a}rez\inst{10}
          \and
          \'{A}gnes K\'{o}sp\'{a}l\inst{12,13}
          \and
          Carlo F. Manara\inst{1}
          \and
          Aina Palau\inst{8}
          \and
          Michihiro Takami\inst{6}
          \and
          Leonardo Testi\inst{1}
          \and
          Eduard I. Vorobyov\inst{14,15,16}                        
          }

   \institute{European Southern Observatory (ESO), Karl-Schwarzschild-Str. 2, D-85748 Garching, Germany \\
                 \email{baobabyoo@gmail.com}
           \and
           Department of Physics, State University of New York at Fredonia, 280 Central Ave, Fredonia, NY 14063
           \and
           Harvard-Smithsonian Center for Astrophysics, 60 Garden Street, Cambridge, MA 02138, USA
           \and
           Lunar and Planetary Laboratory, University of Arizona, Tucson, AZ 85721, USA
           \and
           SKA Organisation, Jodrell Bank Observatory, Lower Withington, Macclesfield SK11 9DL, UK
           \and
           Institute of Astronomy and Astrophysics, Academia Sinica, 11F of Astronomy-Mathematics Building, National Taiwan University, No 1, Sec., Roosevelt Road, Taipei, 10617 Taiwan
           \and
           Astrophysics Research Institute, Liverpool John Moores University, IC2, Liverpool Science Park, 146 Brownlow Hill, Liverpool L3 5RF, UK
           \and 
           Instituto de Radioastronom\'ia y Astrof\'isica, UNAM, A.P. 3-72, Xangari, Morelia, 58089, Mexico
           \and
           Centre for Astrophysics Research, School of Physics, Astronomy and Mathematics, University of Hertfordshire, College Lane, Hatfield AL10 9AB, UK
           \and
           Institut de Ci\`{e}ncies de l'Espai (IEEC-CSIC), Can Magrans, S/N, E-08193 Cerdanyola del Vall\'{e}s, Catalonia, Spain
           \and
           Space Telescope Science Institute, Baltimore, MD 21218, USA
           \and 
           Konkoly Observatory, Research Centre for Astronomy and Earth Sciences, Hungarian Academy of Sciences, Hungarian Academy of Science, Konkoly-Thege Mikl\'os\'ut 15-17, 1121 Budapest, Hungary
           \and 
           Max-Planck-Institut f\"or Astronomie, K\"onigstuhl 17, D-69117 Heidelberg, Germany
           \and
           Institute of Fluid Mechanics and Heat Transfer, TU Wien, 1060 Vienna, Austria
           \and
           Department of Astrophysics, University of Vienna, Vienna 1180, Austria
           \and
           Research Institute of Physics, Southern Federal University, Rostov-on-Don 344090, Russia
             }

   \date{Received 14 September, 2017; accepted 24 October, 2017}

  \abstract
    {Young stellar objects (YSOs) may undergo periods of active accretion (outbursts), during which the protostellar accretion rate is temporarily enhanced by a few orders of magnitude. Whether or not these accretion outburst YSOs possess similar dust/gas reservoirs to each other, and whether or not their dust/gas reservoirs are similar as quiescent YSOs, are issues not yet clarified.
    } 
   {The aim of this work is to characterize the millimeter thermal dust emission properties of a statistically significant sample of long and short duration accretion outburst YSOs (i.e., FUors and EXors) and the spectroscopically identified candidates of accretion outbursting YSOs (i.e., FUor-like objects).}
   {We have carried out extensive Submillimeter Array (SMA) observations mostly at $\sim$225 GHz (1.33 mm) and $\sim$272 GHz (1.10 mm), from 2008 to 2017. We covered accretion outburst YSOs located at $<$1 kpc distances from the solar system.
   }
   { We analyze all the existing SMA data of such objects, both published and unpublished, in a coherent way to present a millimeter interferometric database of 29 objects. We obtained 21 detections at $>$3-$\sigma$ significance. 
   Detected sources except for the two cases of V883\,Ori and NGC\,2071\,MM3 were observed with $\sim$1$''$ angular resolution. 
   Overall our observed targets show a systematically higher millimeter luminosity distribution than those of the $M_{*}>$0.3 $M_{\odot}$ Class\,II YSOs in the nearby ($\lesssim$400 pc) low-mass star-forming molecular clouds (e.g., Taurus, Lupus, Upp Scorpio, and Chameleon\,I). In addition, at 1\,mm our observed confirmed binaries or triple-system sources are systematically fainter than the rest of the sources even though their 1 mm fluxes are broadly distributed.
We may have detected $\sim$30-60\% millimeter flux variability from V2494\,Cyg and V2495\,Cyg, from the observations separated by $\sim$1 year. 
   } 
   {
}


\titlerunning{SMA survey towards FU Orionis objects and EXors}

   \maketitle


\section{Introduction}

\begin{figure*}
\hspace{-0.7cm}
\begin{tabular}{ p{4.3cm} p{4.3cm} p{4.3cm} p{4.3cm} }
 &
 \includegraphics[width=4.8cm]{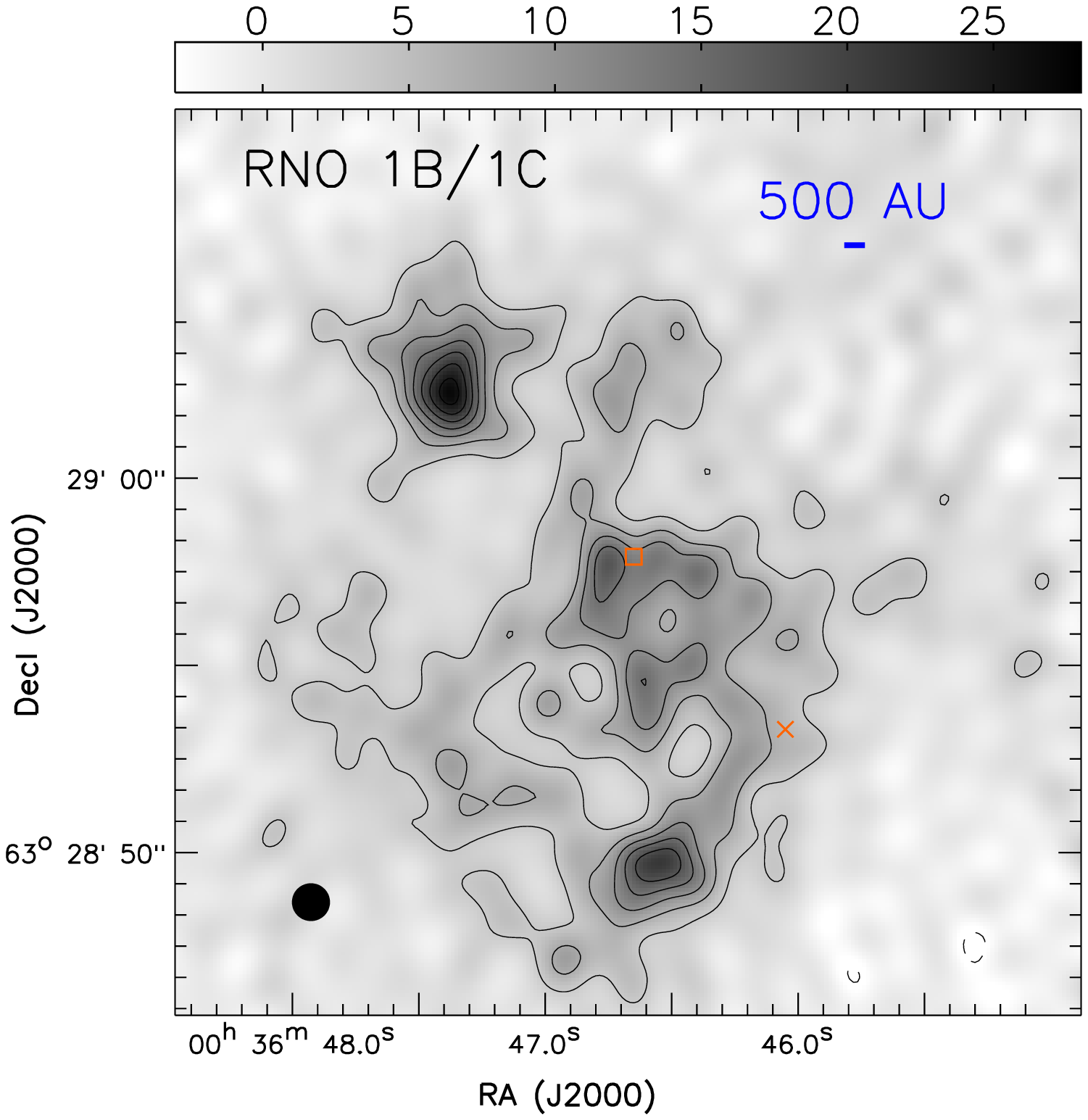} &
 \includegraphics[width=4.8cm]{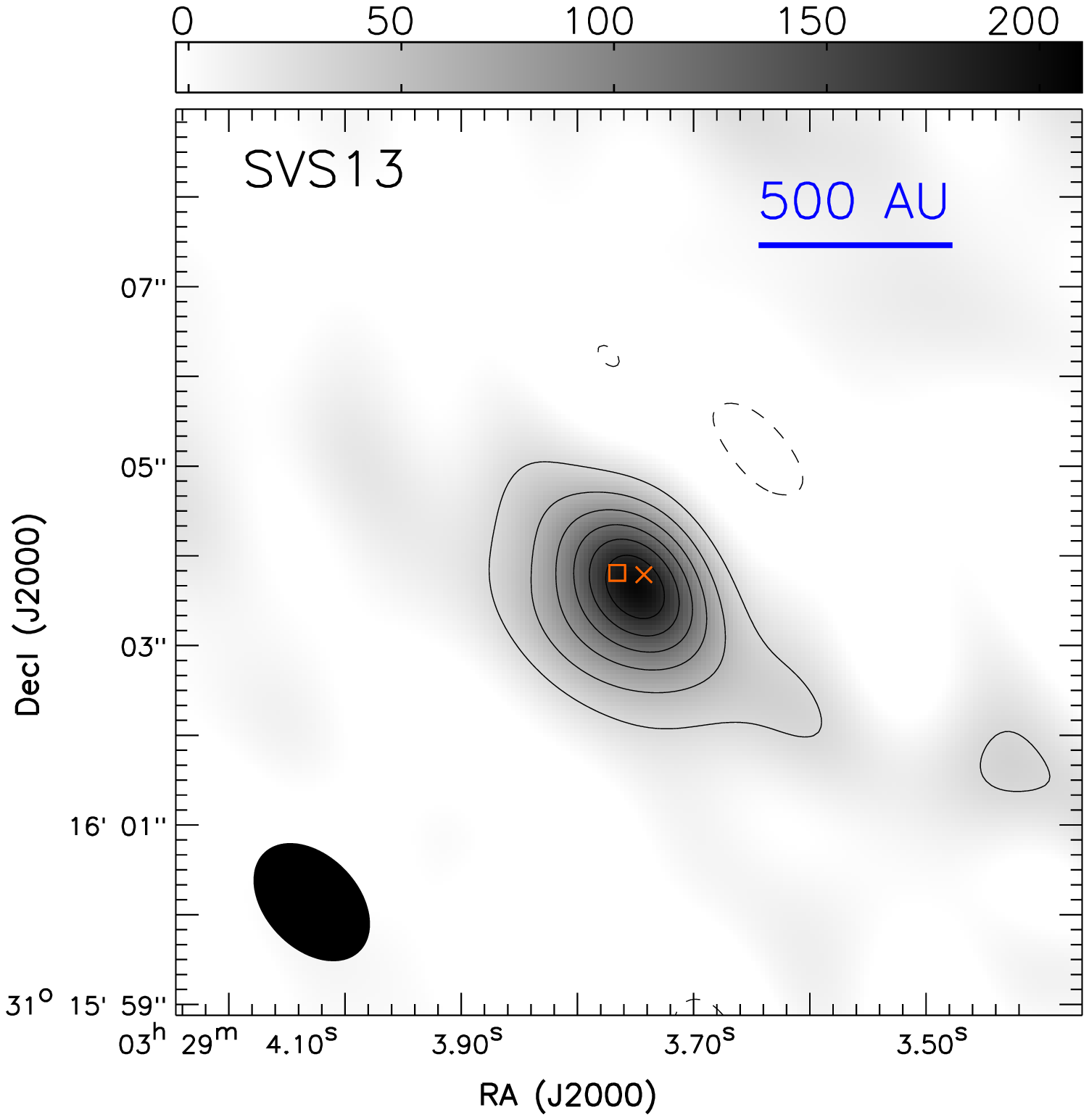} &
 \includegraphics[width=4.8cm]{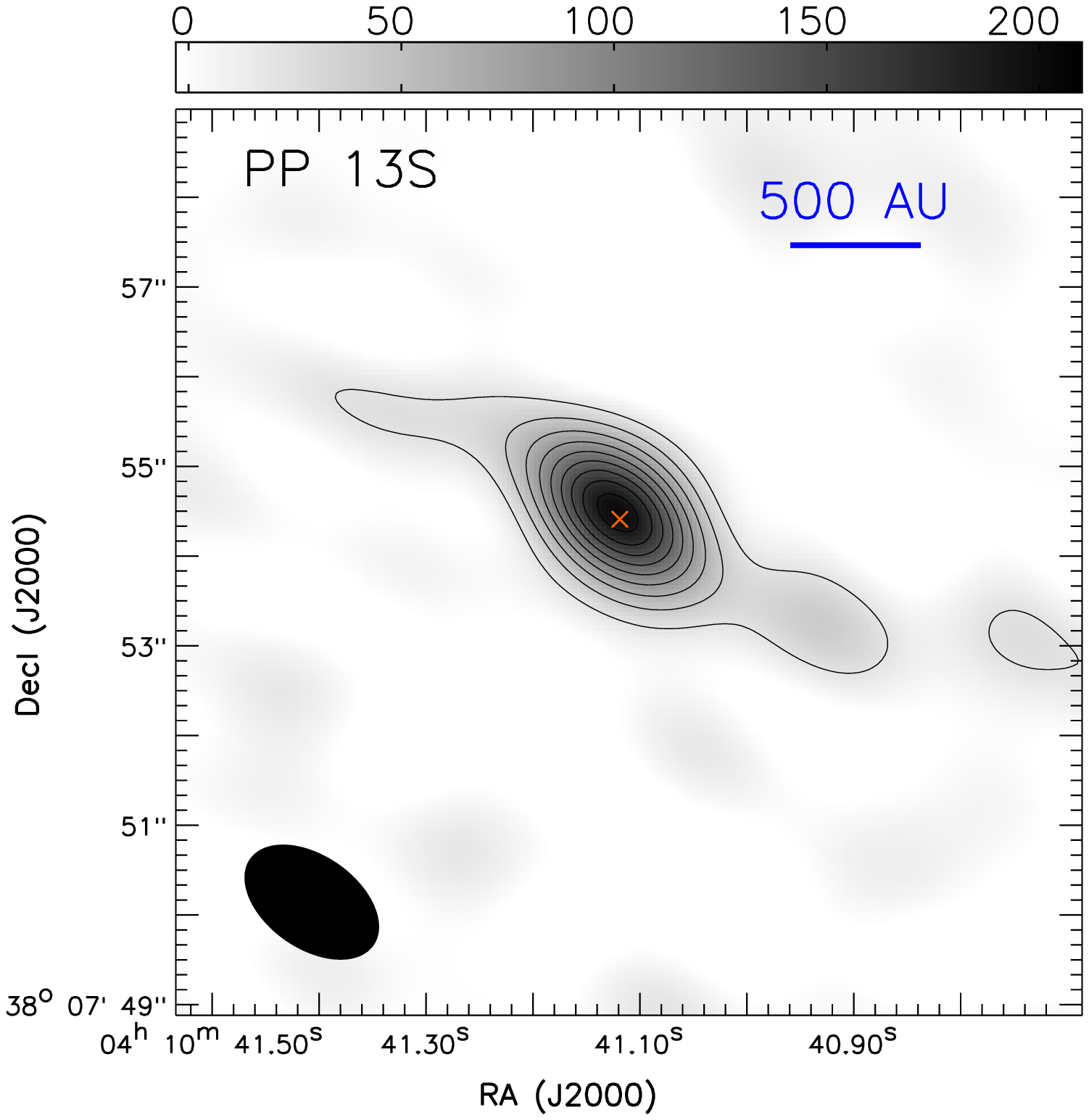} 
 \\
\end{tabular}

\hspace{-0.7cm}
\begin{tabular}{ p{4.3cm} p{4.3cm} p{4.3cm} p{4.3cm} }
\includegraphics[width=4.8cm]{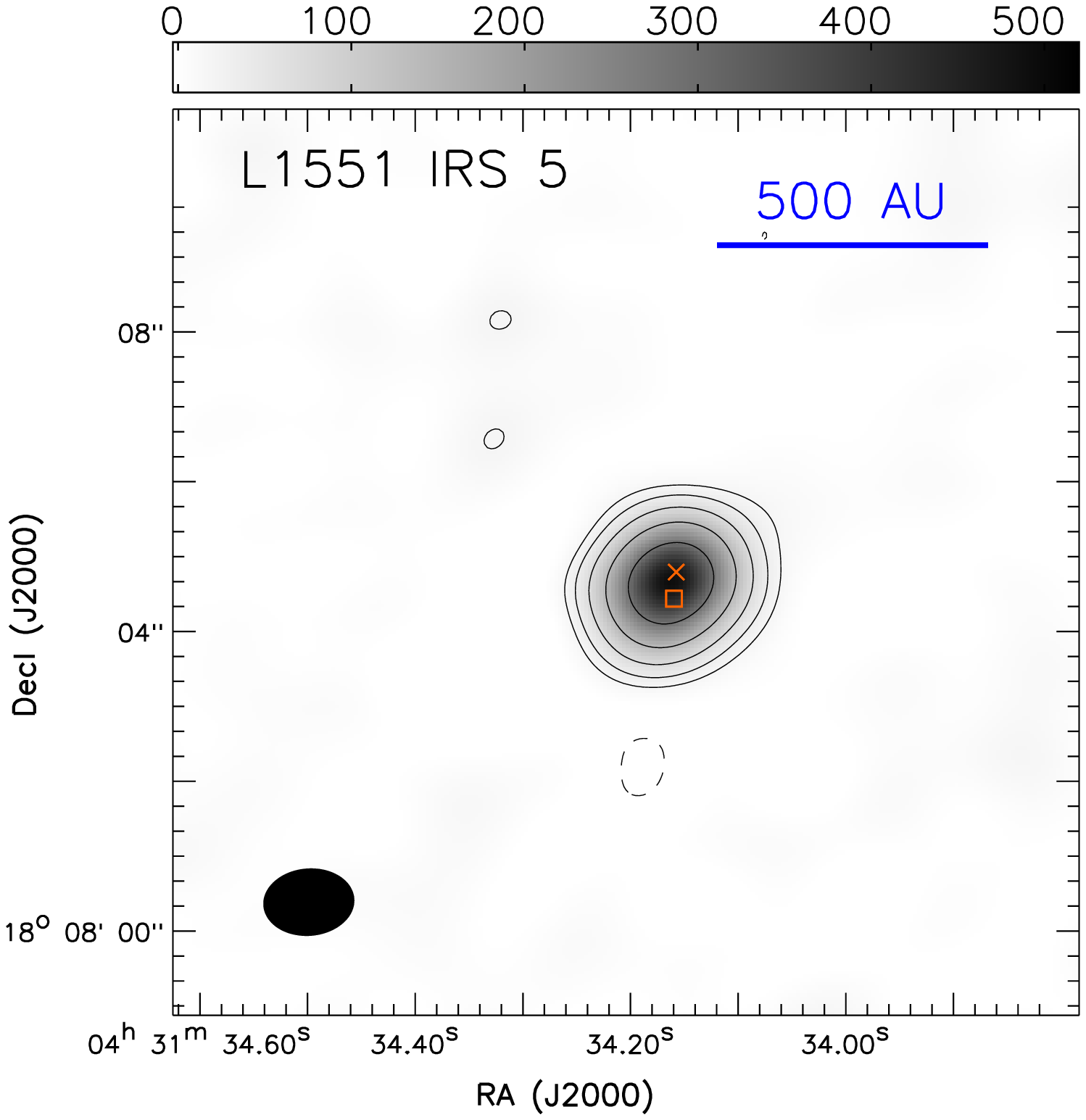} &
\includegraphics[width=4.8cm]{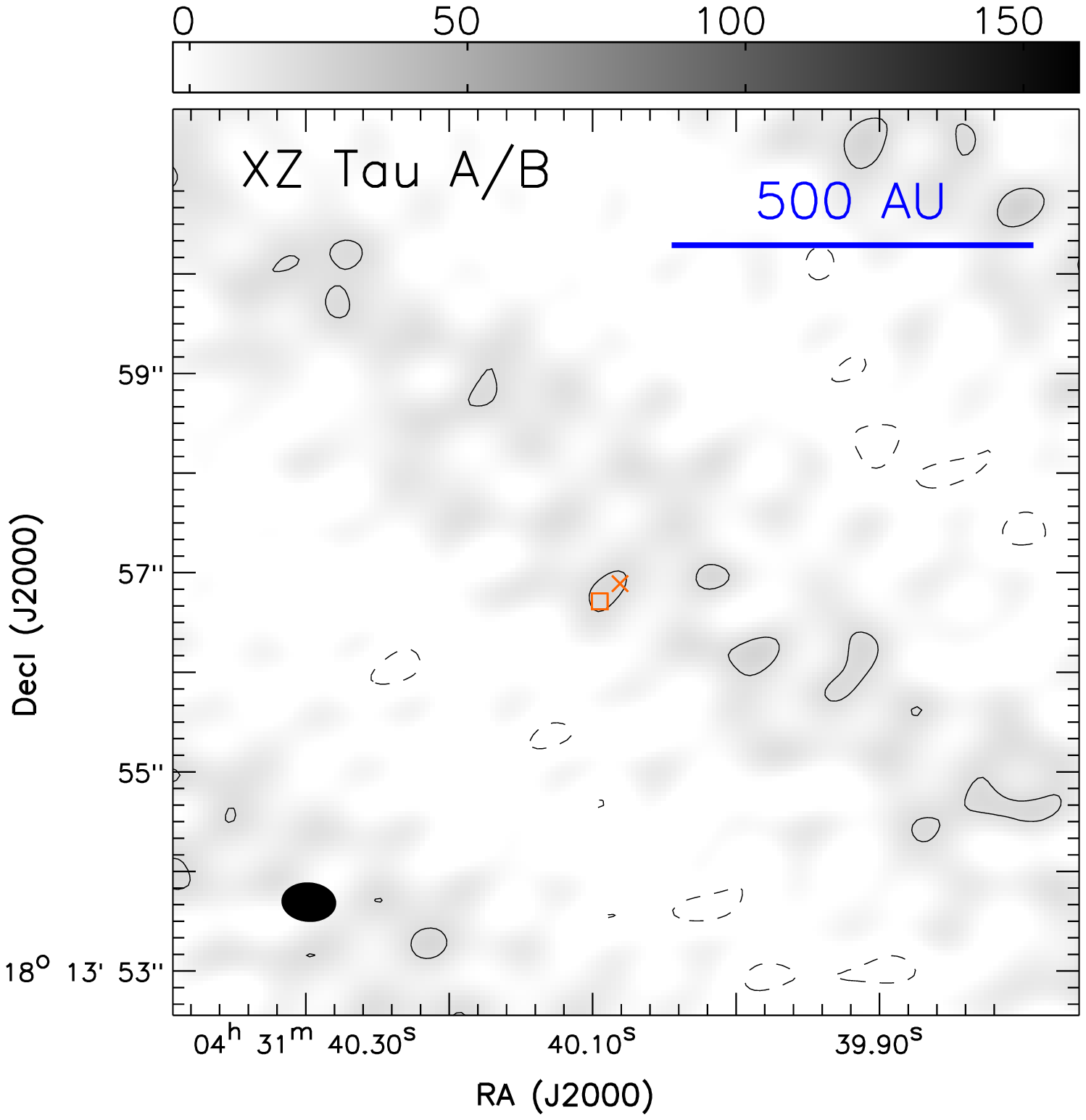} &
\includegraphics[width=4.8cm]{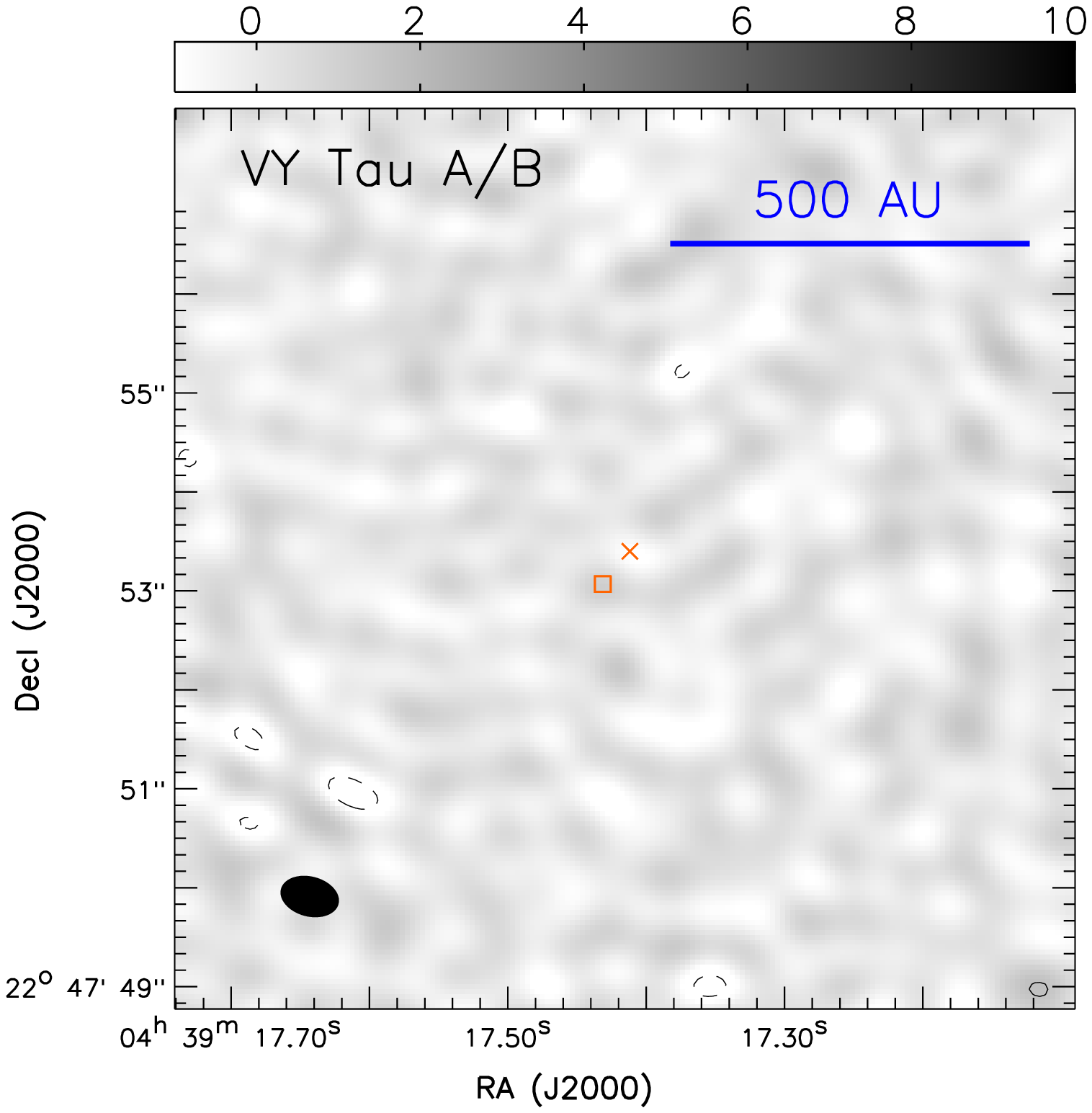} &
\includegraphics[width=4.8cm]{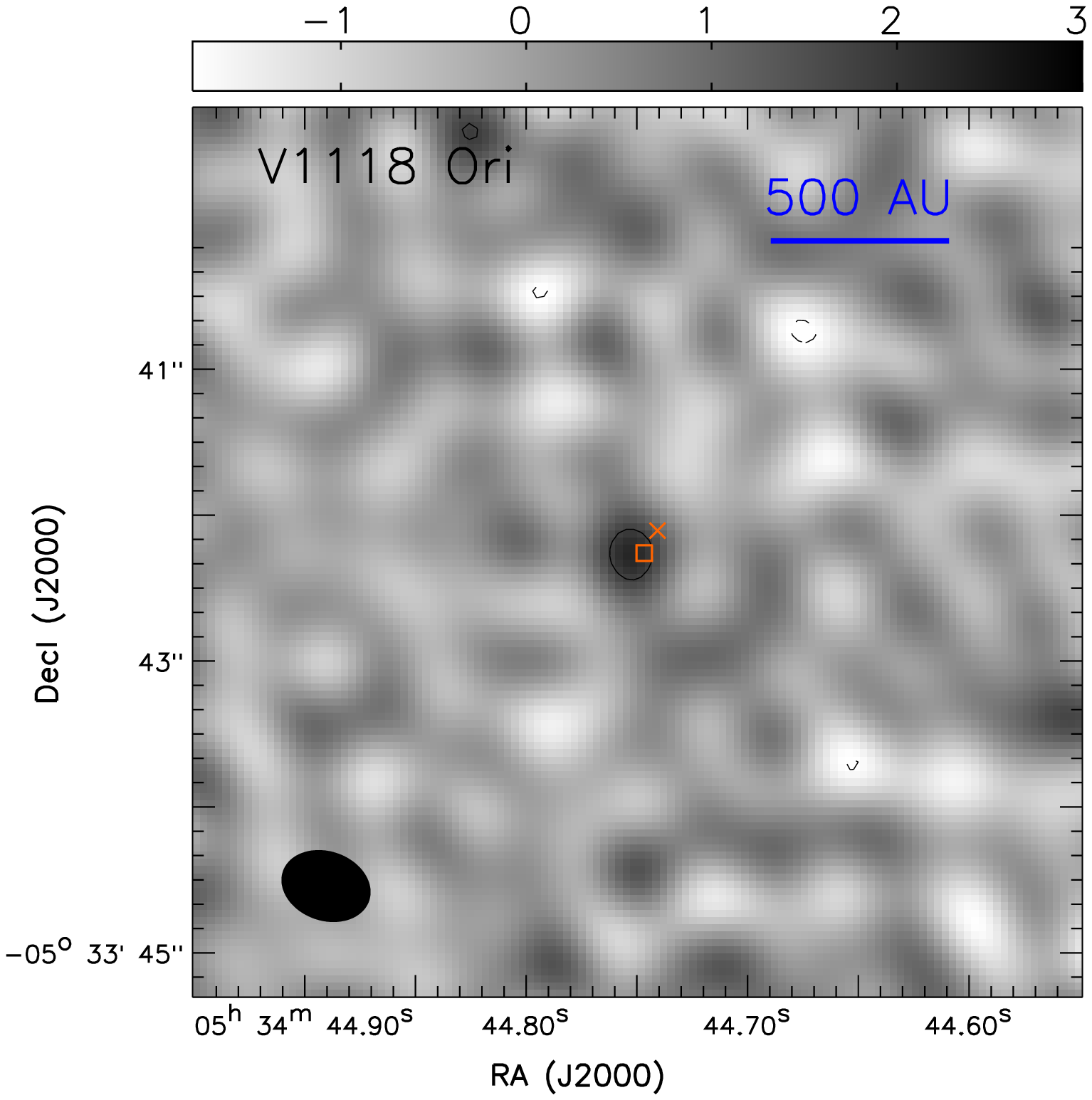}
 \\
\end{tabular}

\hspace{-0.7cm}
\begin{tabular}{ p{4.3cm} p{4.3cm} p{4.3cm} p{4.3cm} }
\includegraphics[width=4.8cm]{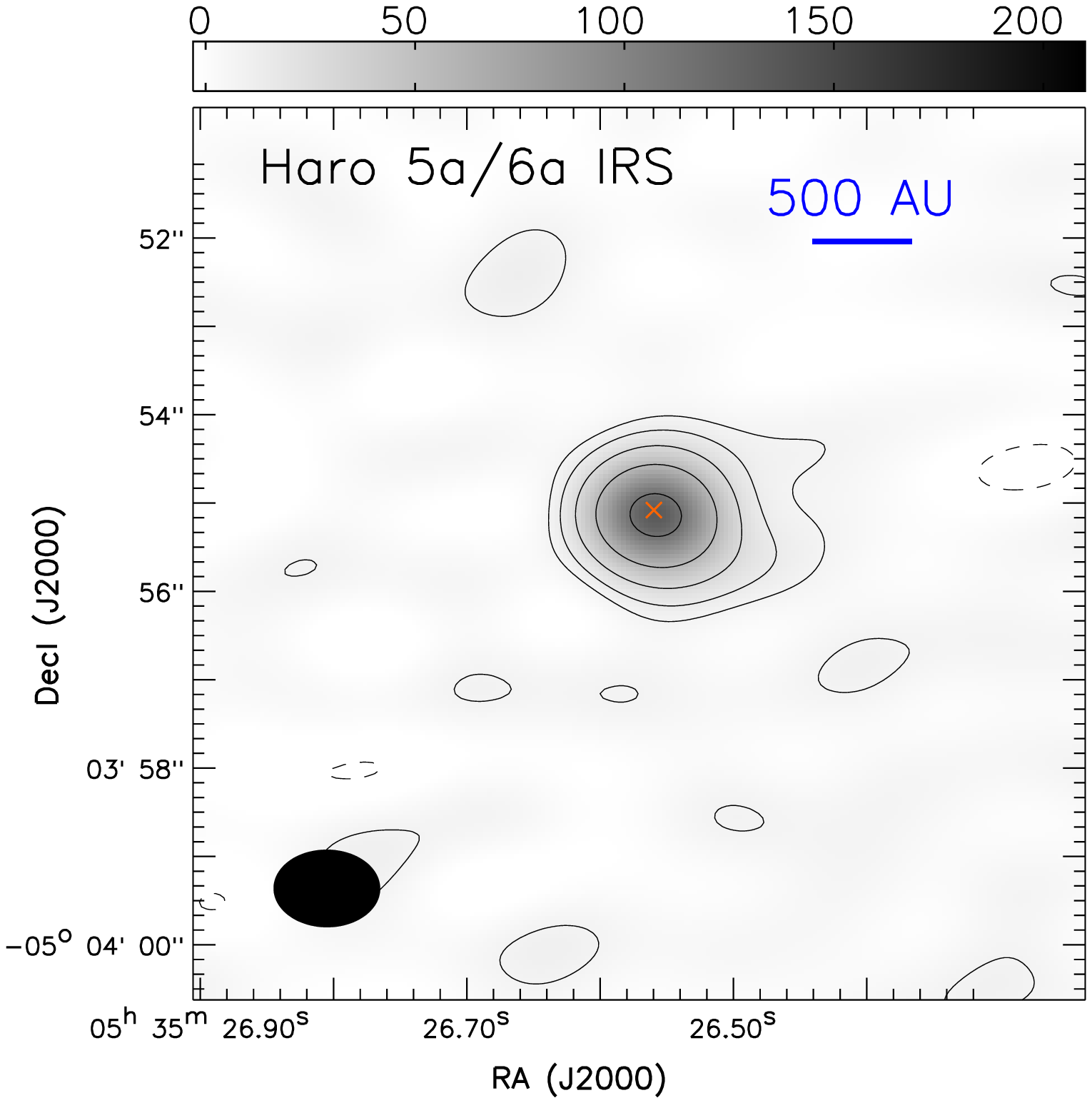} &
\includegraphics[width=4.8cm]{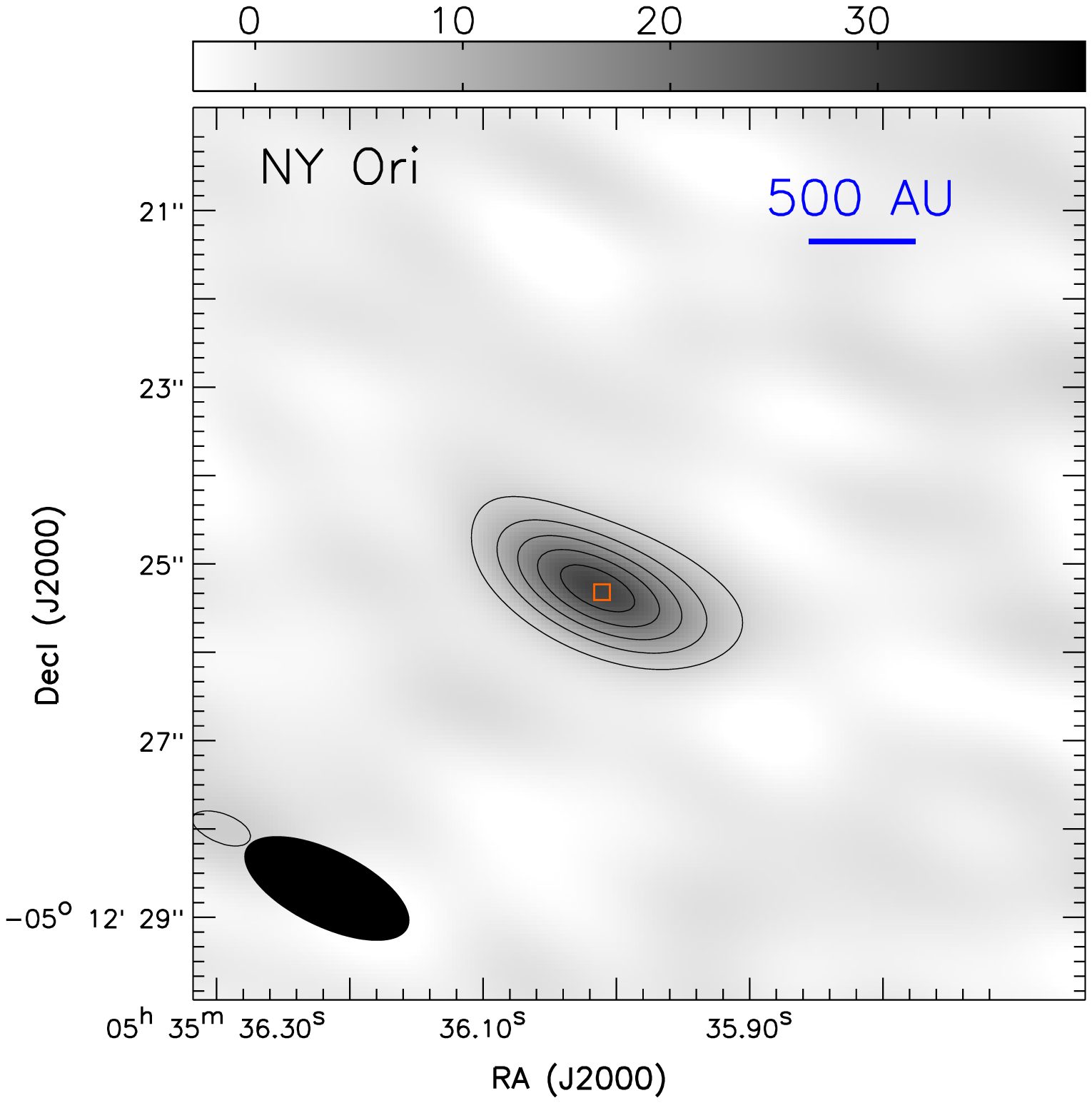} &
\includegraphics[width=4.8cm]{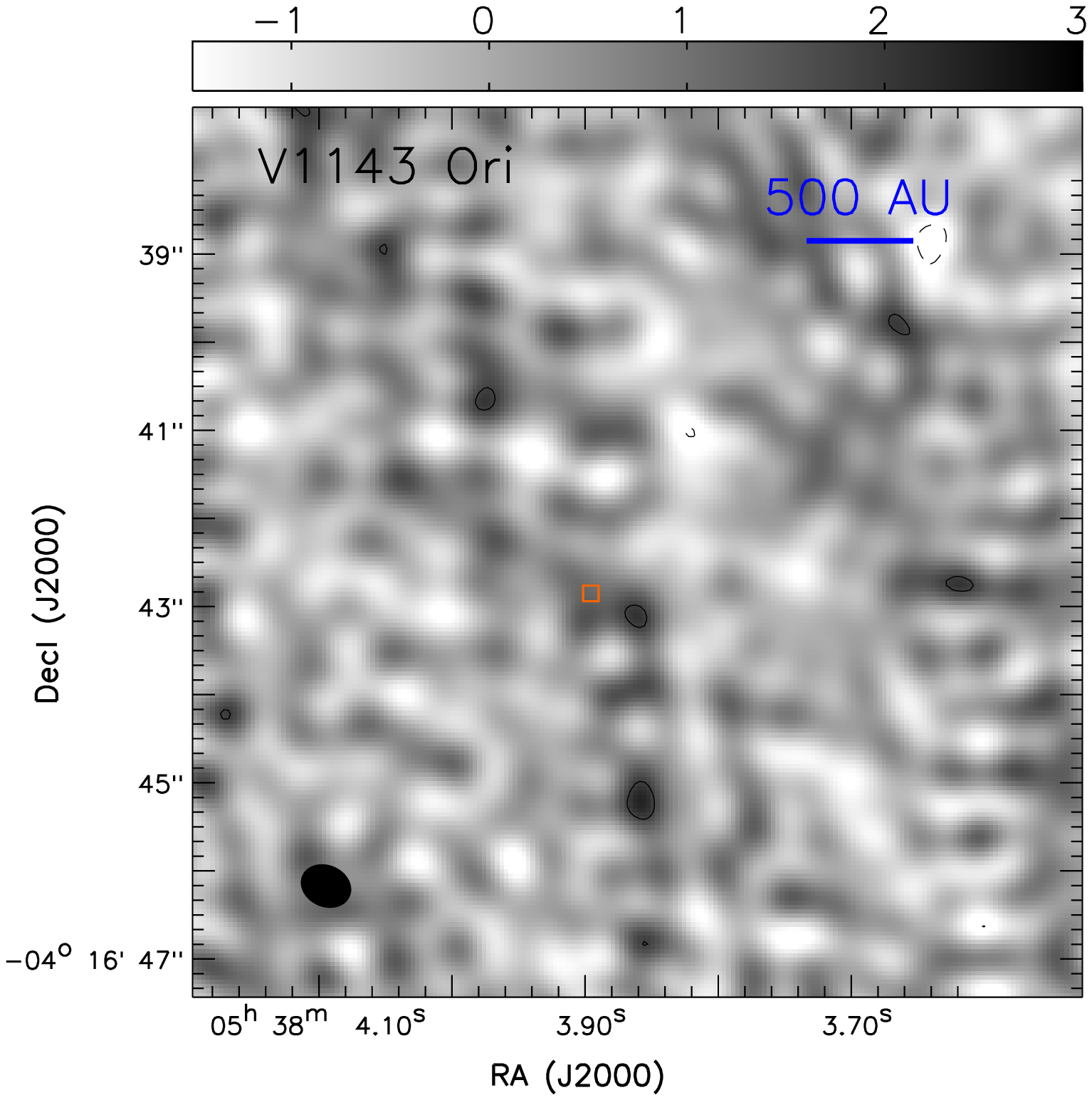} &
\includegraphics[width=4.8cm]{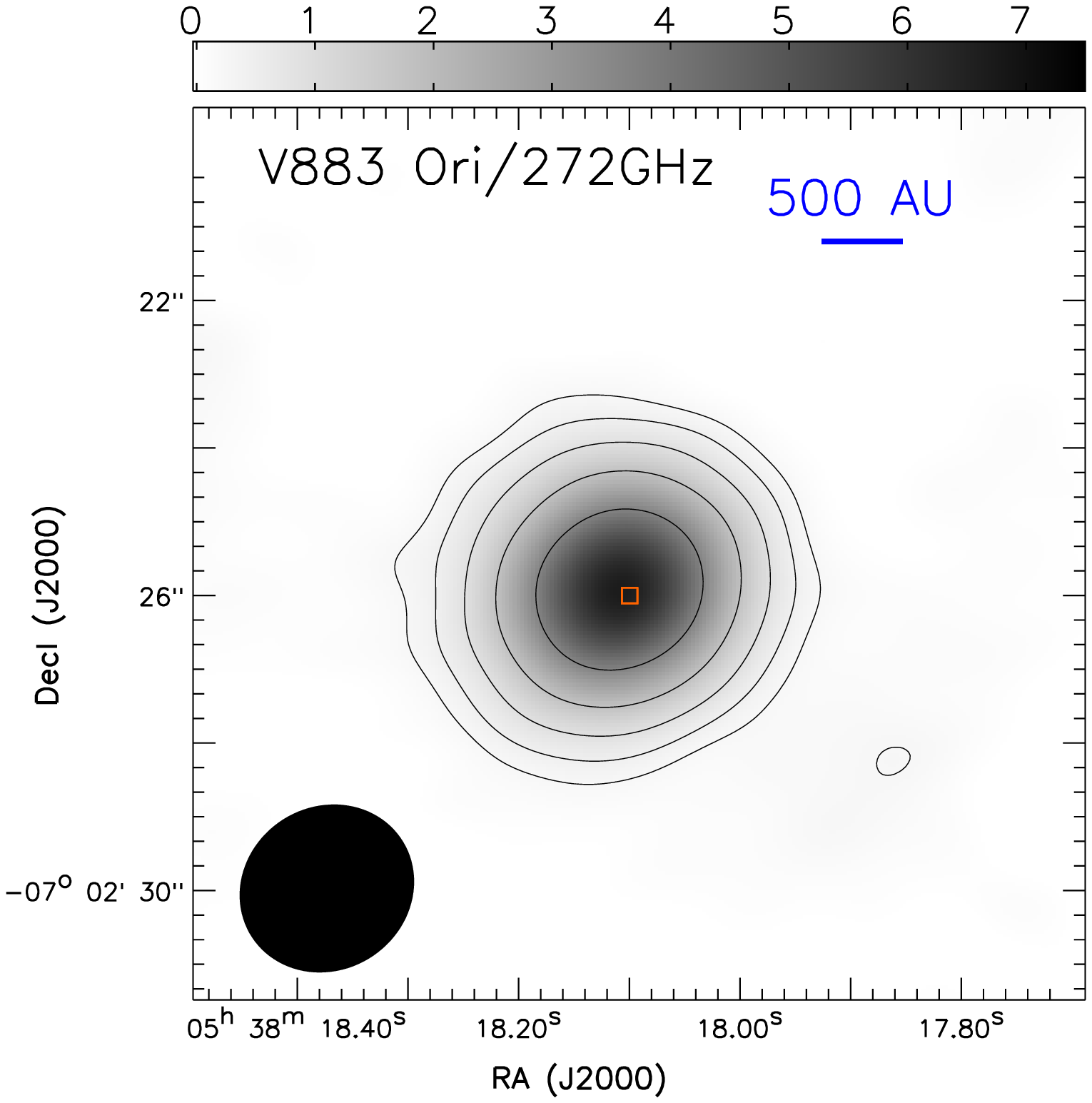}
 \\
\end{tabular}

\hspace{-0.7cm}
\begin{tabular}{ p{4.3cm} p{4.3cm} p{4.3cm} p{4.3cm} }
\includegraphics[width=4.8cm]{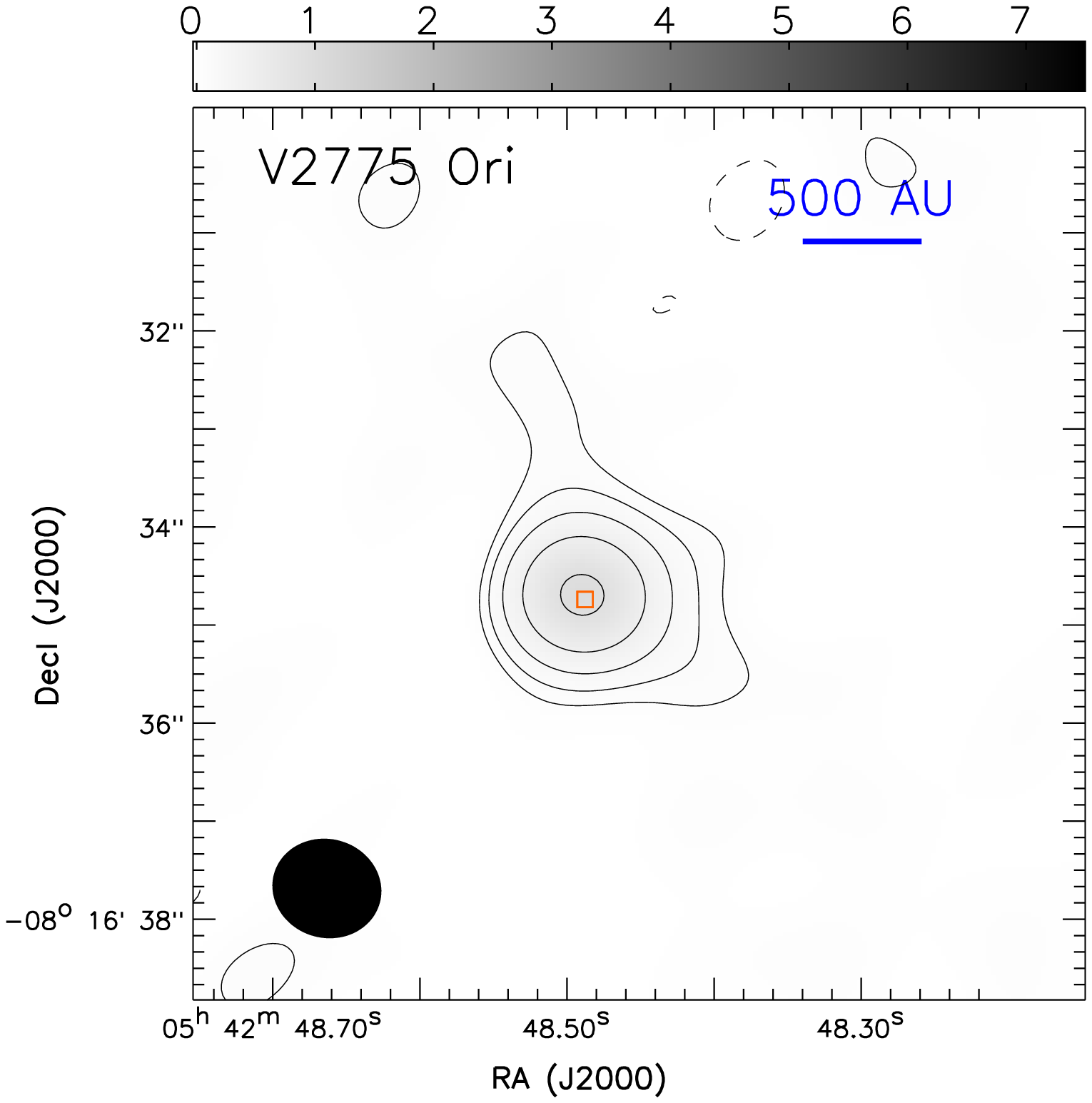} &
\includegraphics[width=4.8cm]{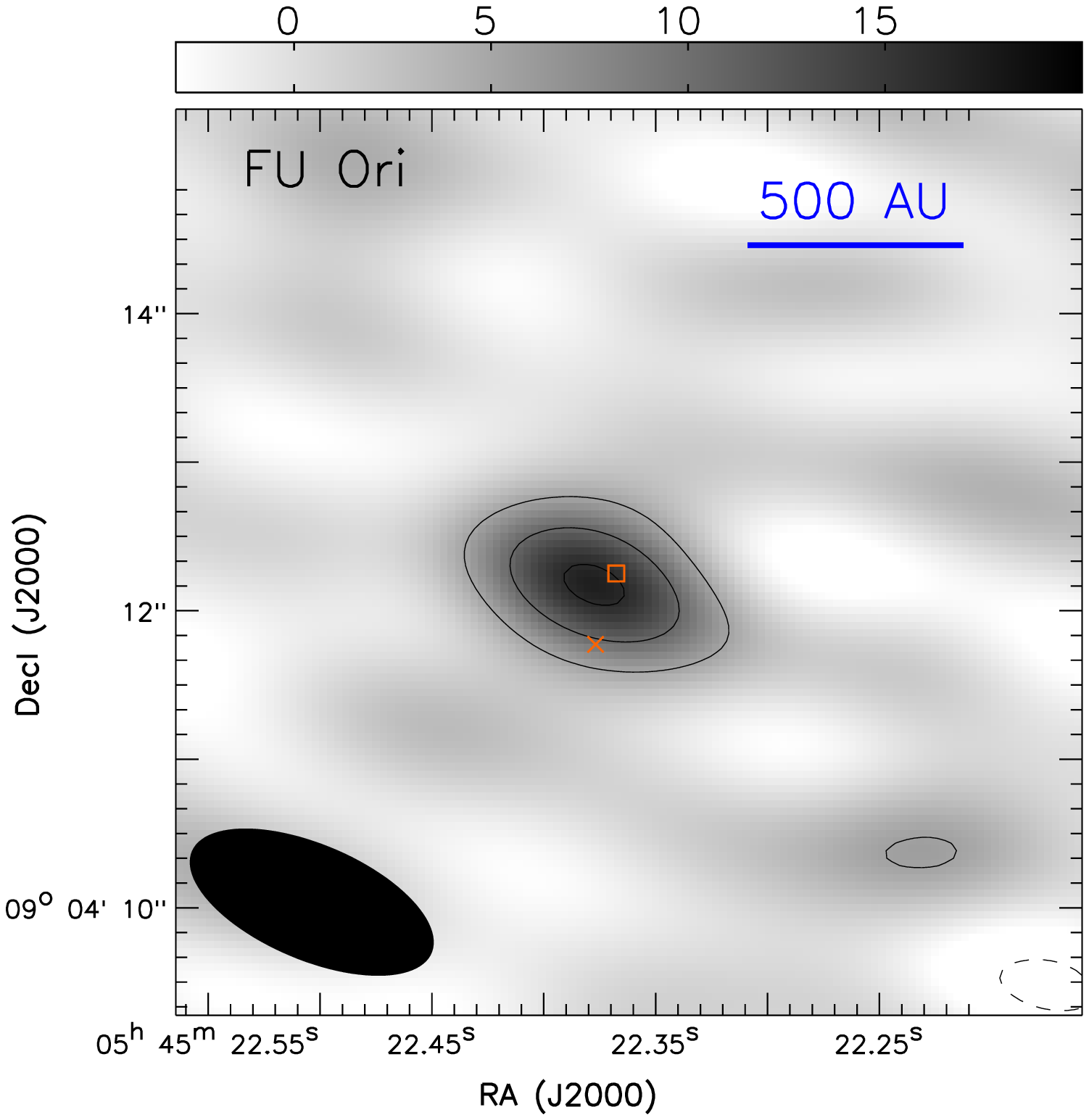} &
\includegraphics[width=4.8cm]{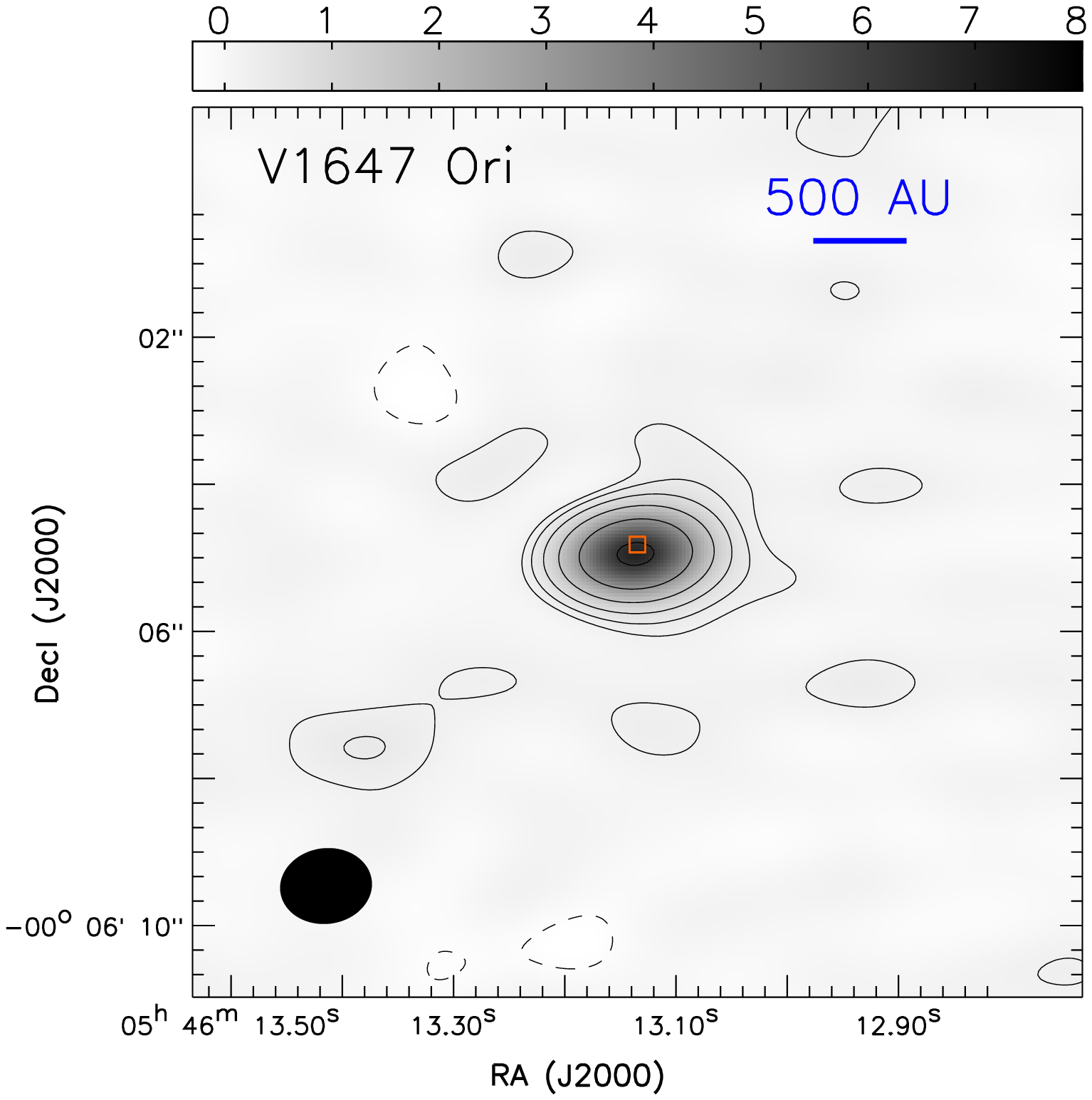} &
\includegraphics[width=4.8cm]{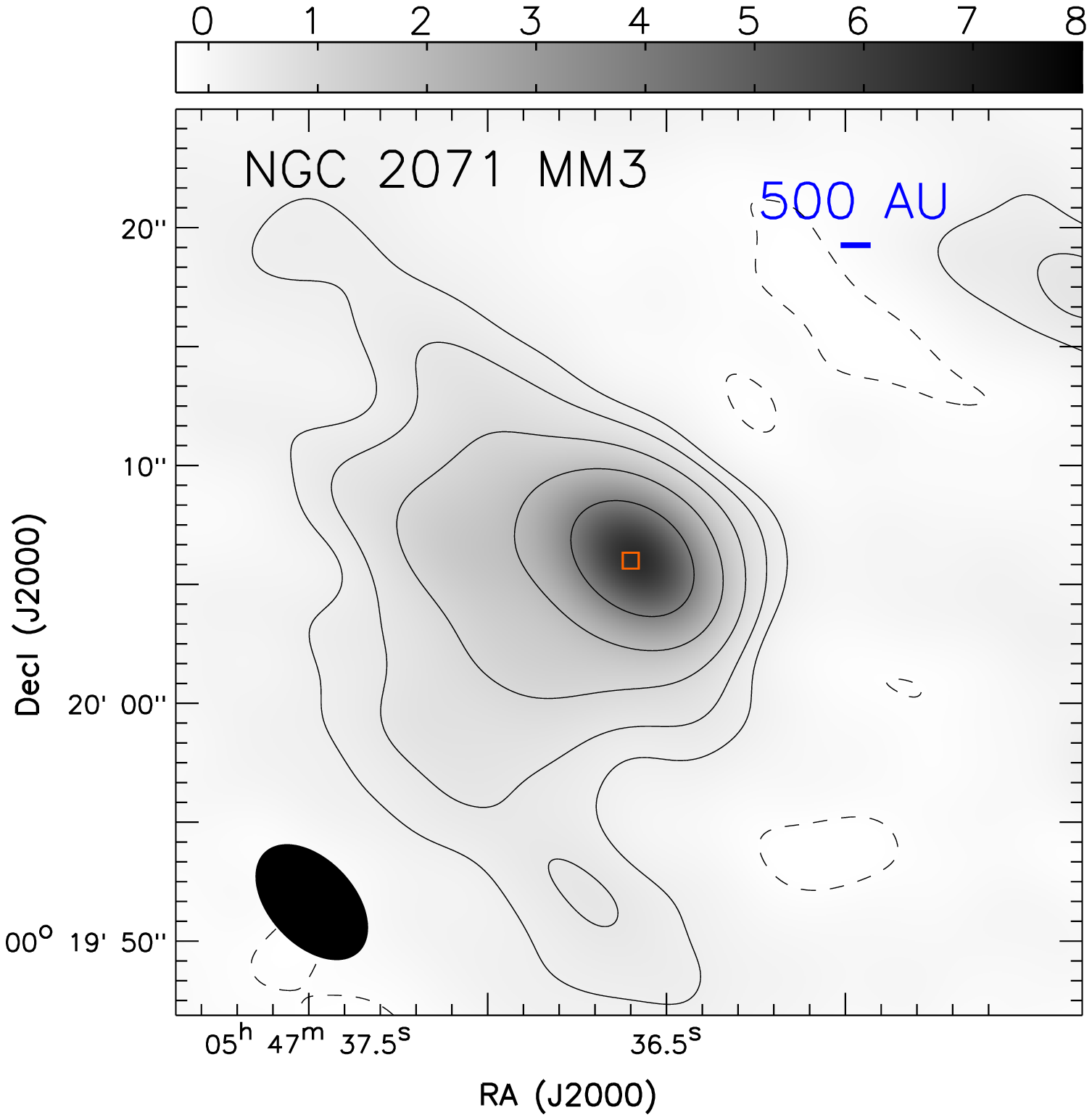}
 \\
\end{tabular}
\caption{\footnotesize{
SMA images of the observed FUors, EXors, and FUor-like objects. 
Images are taken at the mean frequency of 224-225 GHz (1.33 mm) if not specifically annotated.
Synthesized beam is shown in the bottom left corner of each panel.
Color bars are in units of mJy\,beam$^{-1}$.
Contours are in steps of 3-$\sigma$ (c.f. Table \ref{tab:summary}) if not specifically mentioned.
Contours of L1551\,IRS\,5 are 6 mJy\,beam$^{-1}$ (2$\sigma$) $\times$ [-3, 3, 6, 12, 24, 48].
Contours of Haro\,5a/6a\,IRS are 2.4 mJy\,beam$^{-1}$ (1$\sigma$) $\times$ [-3, 3, 6, 12, 24, 48].
Contours of V883\,Ori are 7.6 mJy\,beam$^{-1}$ (1$\sigma$) $\times$ [-3, 3, 6, 12, 24, 48].
Contours of V2775\,Ori are 1.9 mJy\,beam$^{-1}$ (1$\sigma$) $\times$ [-3, 3, 6, 12, 24].
Contours of V1647\,Ori are 0.66 mJy\,beam$^{-1}$ (1$\sigma$) $\times$ [-3, 3, 6, 12, 24, 48, 96].
Image of NGC\,2071\,MM3 is presented with a $\theta_{\mbox{\scriptsize{maj}}}$ $\times$ $\theta_{\mbox{\scriptsize{min}}}$=5\farcs8$\times$3\farcs6 (P.A.=43$^{\circ}$) synthesized beam to better present its extended envelope; contours are 0.66 mJy\,beam$^{-1}$ (1$\sigma$) $\times$ [-3, 3, 6, 12, 24, 48].
Rectangle and cross in the panel of RNO\,1B/1C mark the locations of RNO 1B and 1C quoted from Quanz et al. (2007a); in the panel of SVS\,13 they mark Per-emb-44 A and B (Anglada et al. 2004); in the panel of L1551\,IRS\,5 they mark the southern and northern binary components quoted from Lim et al. (2016); in the panel of XZ\,Tau\,A/B they mark the locations of XZ\,Tau\,A and B quoted from Forgan et al. (2004); in the panel of VY\,Tau\,A/B they mark the location of VY\,Tau A and B (e.g., Dodin et al. 2015); in the panel of V1118\,Ori they mark V1118\,Ori and its companion (Reipurth et al. 2007); in the panel of FU\,Ori they mark the locations of FU\,Ori and FU\,Ori\,S (Liu et al. 2017).
}}
\label{fig:poststamp1}
\end{figure*}

\begin{figure*}
\hspace{-0.7cm}
\begin{tabular}{ p{4.3cm} p{4.3cm} p{4.3cm} p{4.3cm} }
\includegraphics[width=4.8cm]{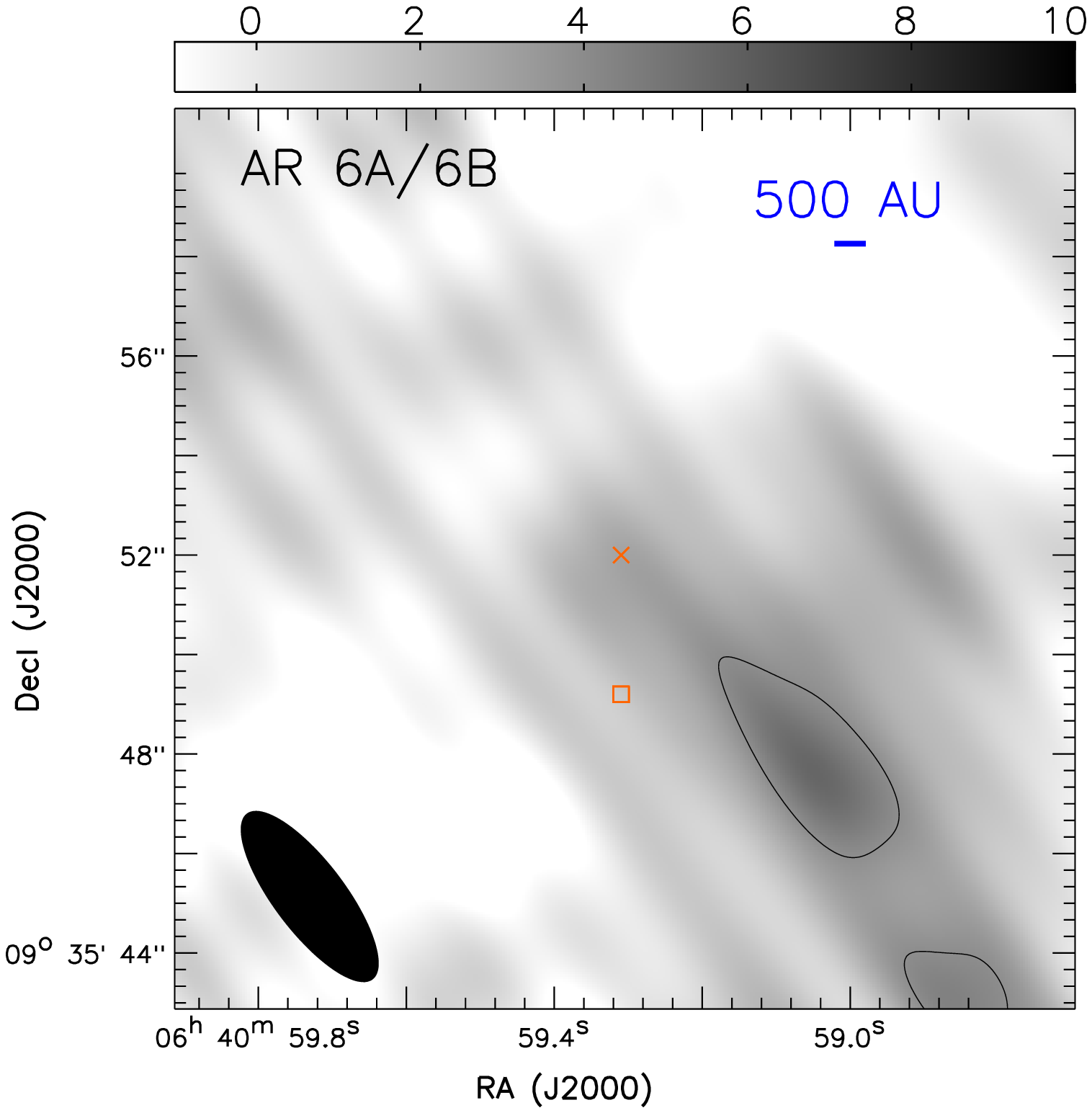} &
\includegraphics[width=4.8cm]{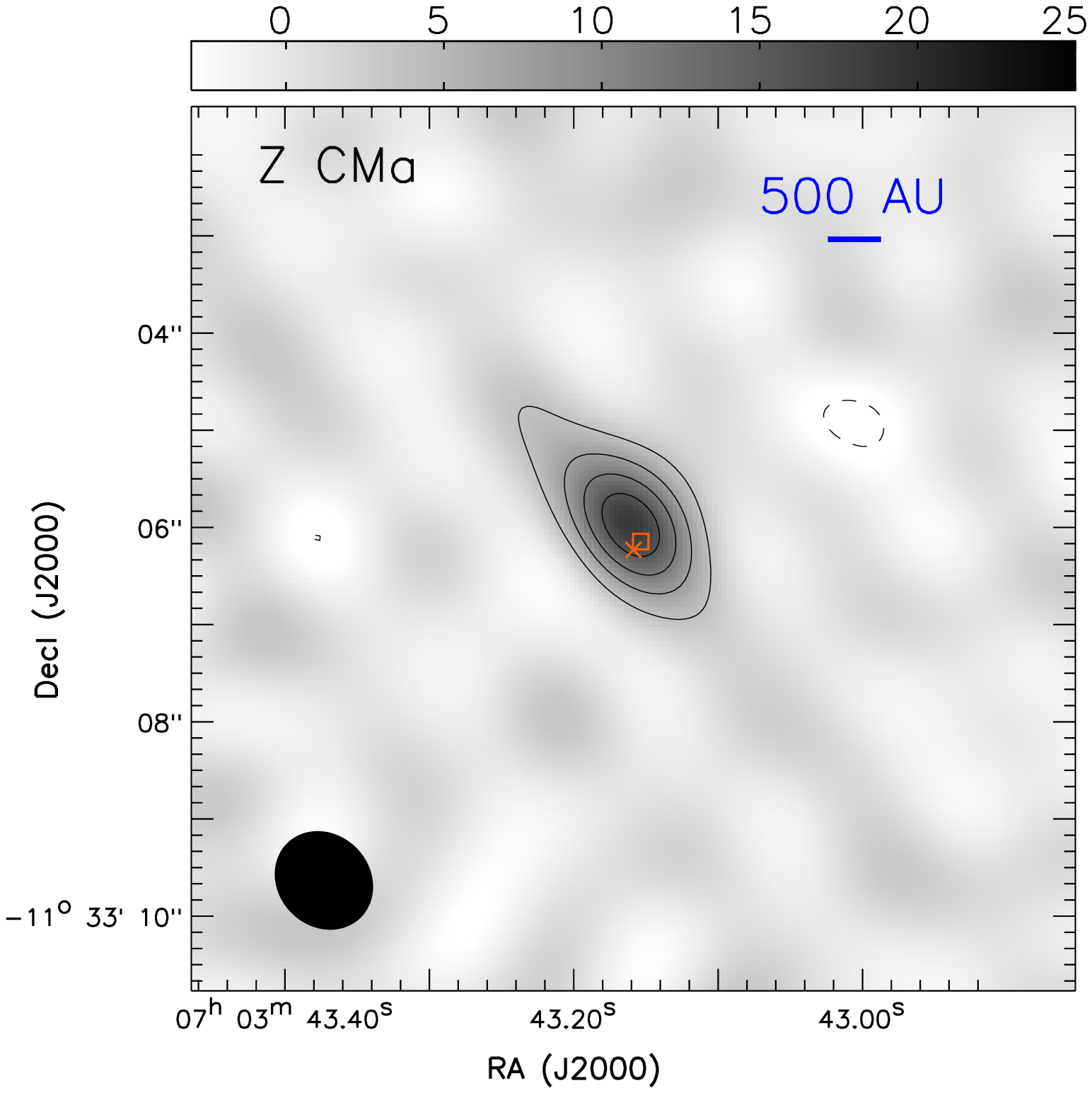} &
\includegraphics[width=4.8cm]{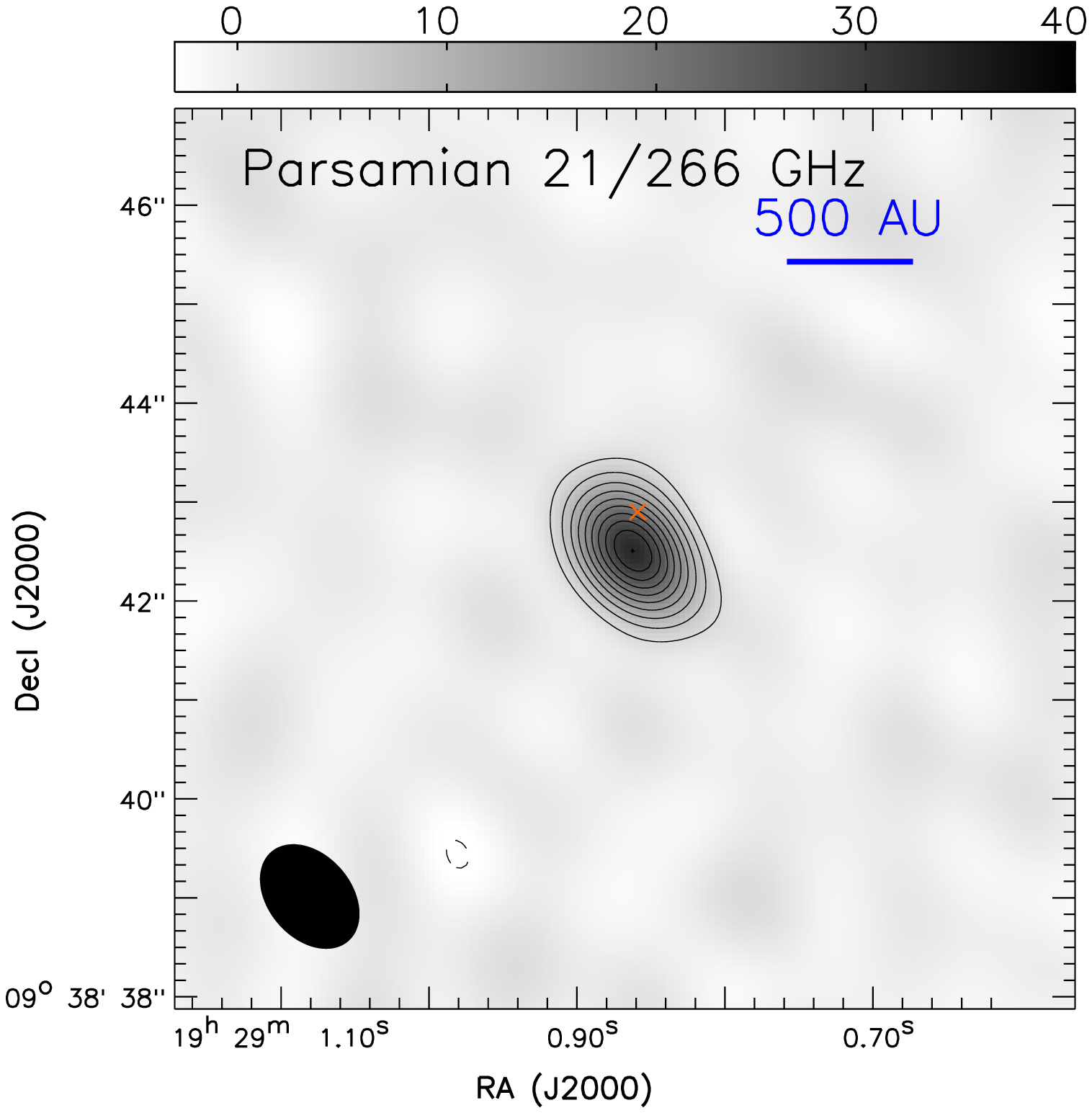} &
 \includegraphics[width=4.8cm]{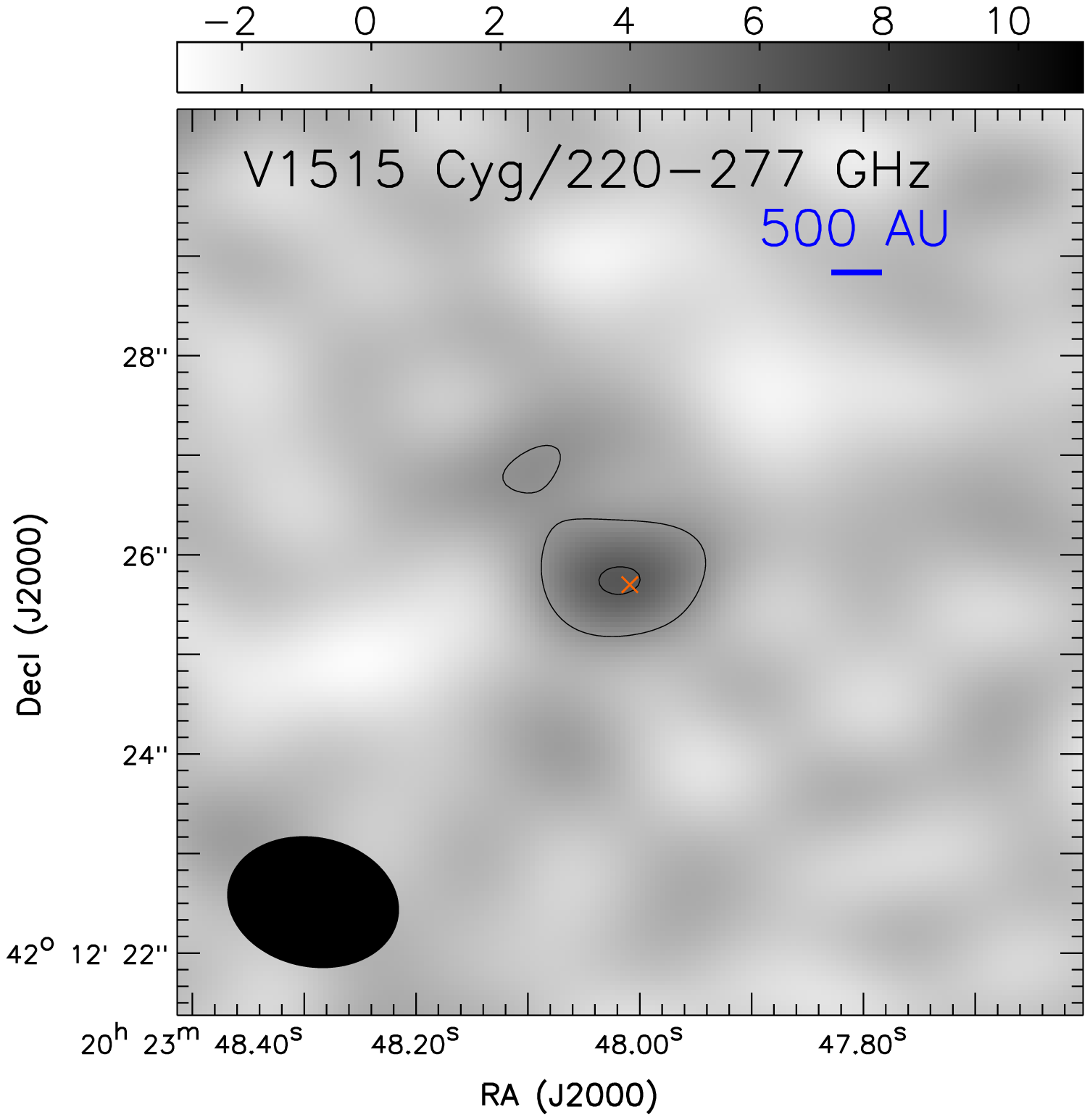} 
 \\
\end{tabular}

\hspace{-0.7cm}
\begin{tabular}{ p{4.3cm} p{4.3cm} p{4.3cm} p{4.3cm} }
\includegraphics[width=4.8cm]{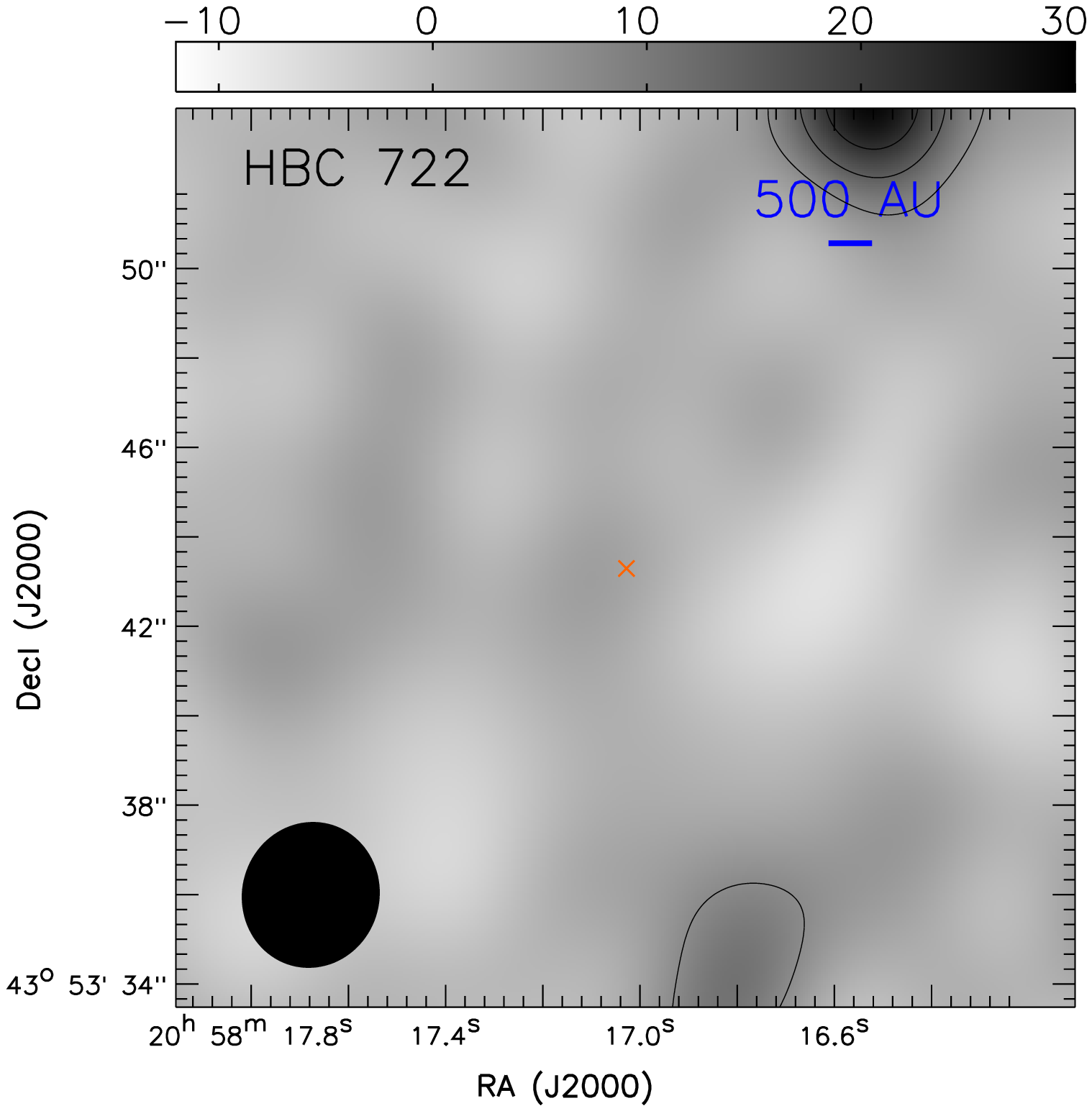} &
\includegraphics[width=4.8cm]{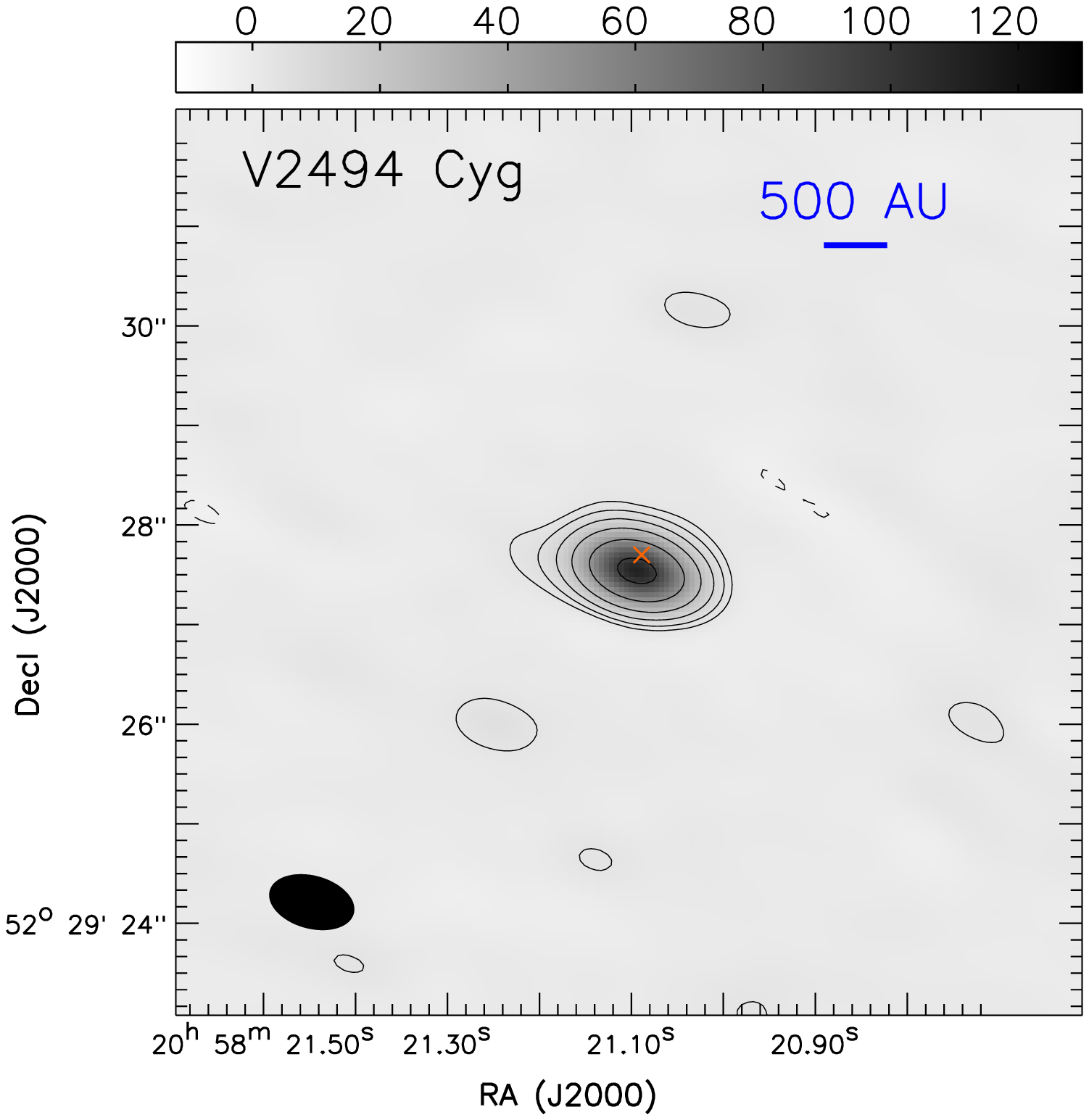} &
\includegraphics[width=4.8cm]{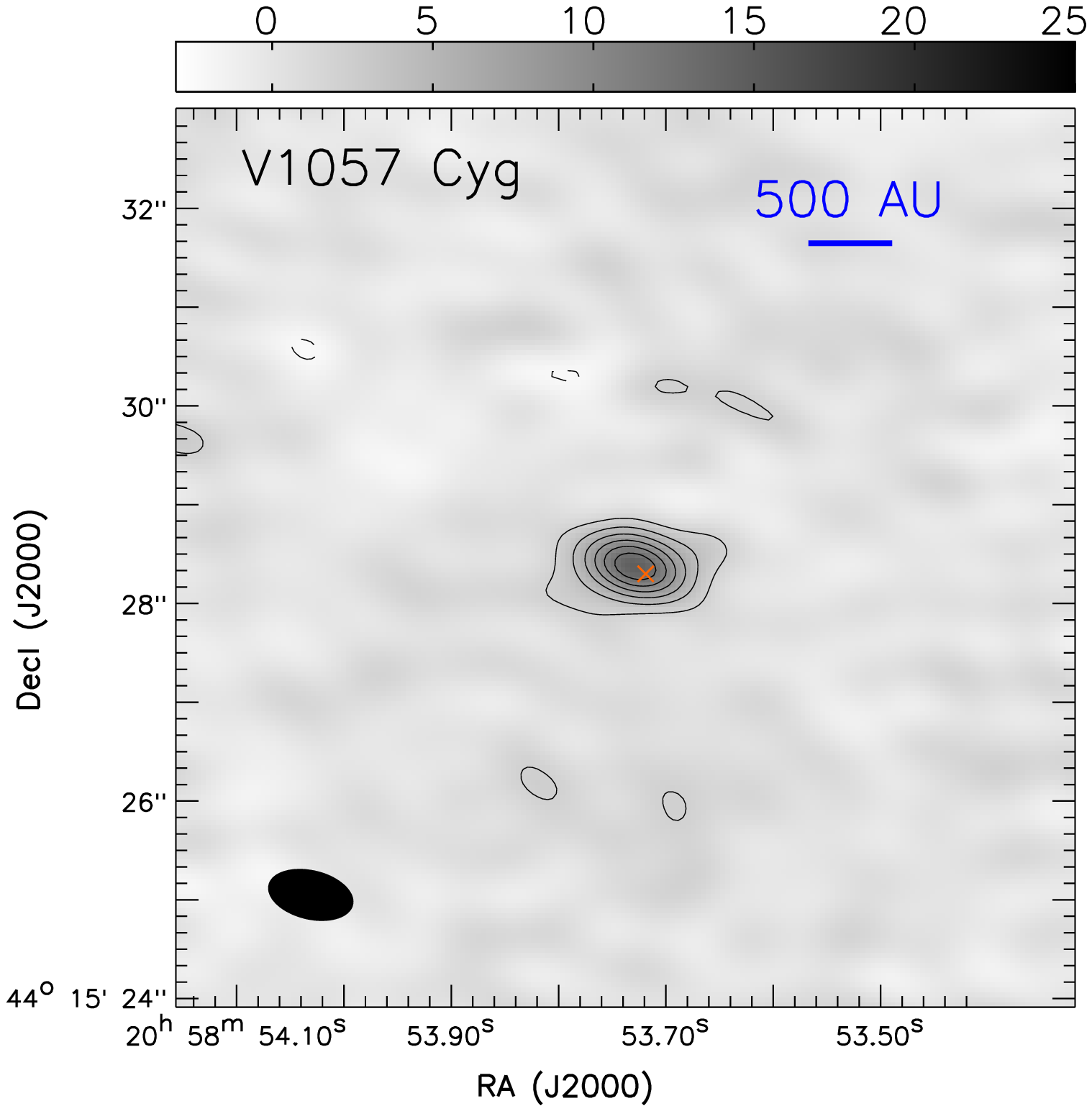} &
\includegraphics[width=4.8cm]{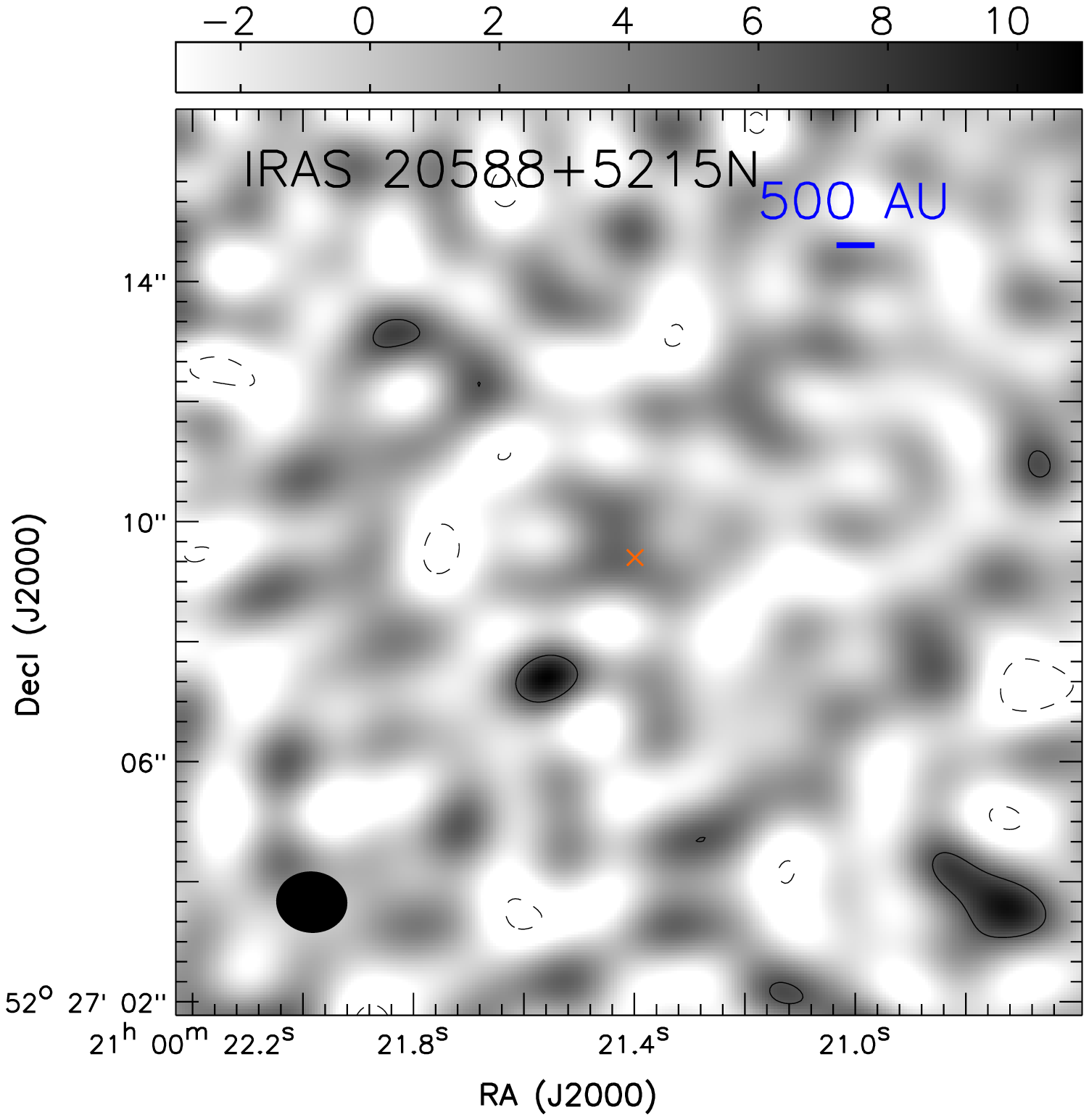}
 \\
\end{tabular}

\hspace{-0.7cm}
\begin{tabular}{ p{4.3cm} p{4.3cm} p{4.3cm} p{4.3cm} }
\includegraphics[width=4.8cm]{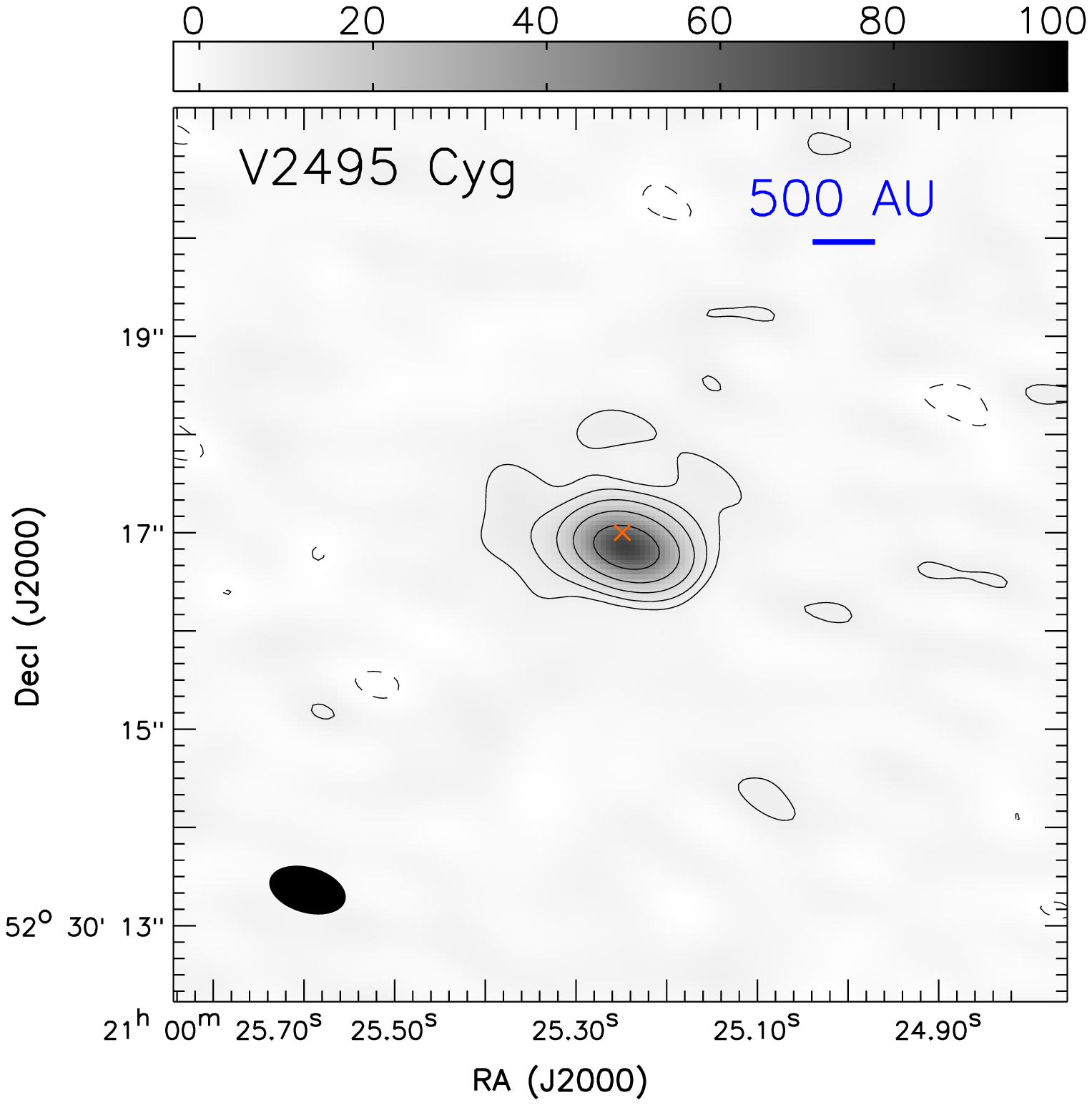} &
\includegraphics[width=4.8cm]{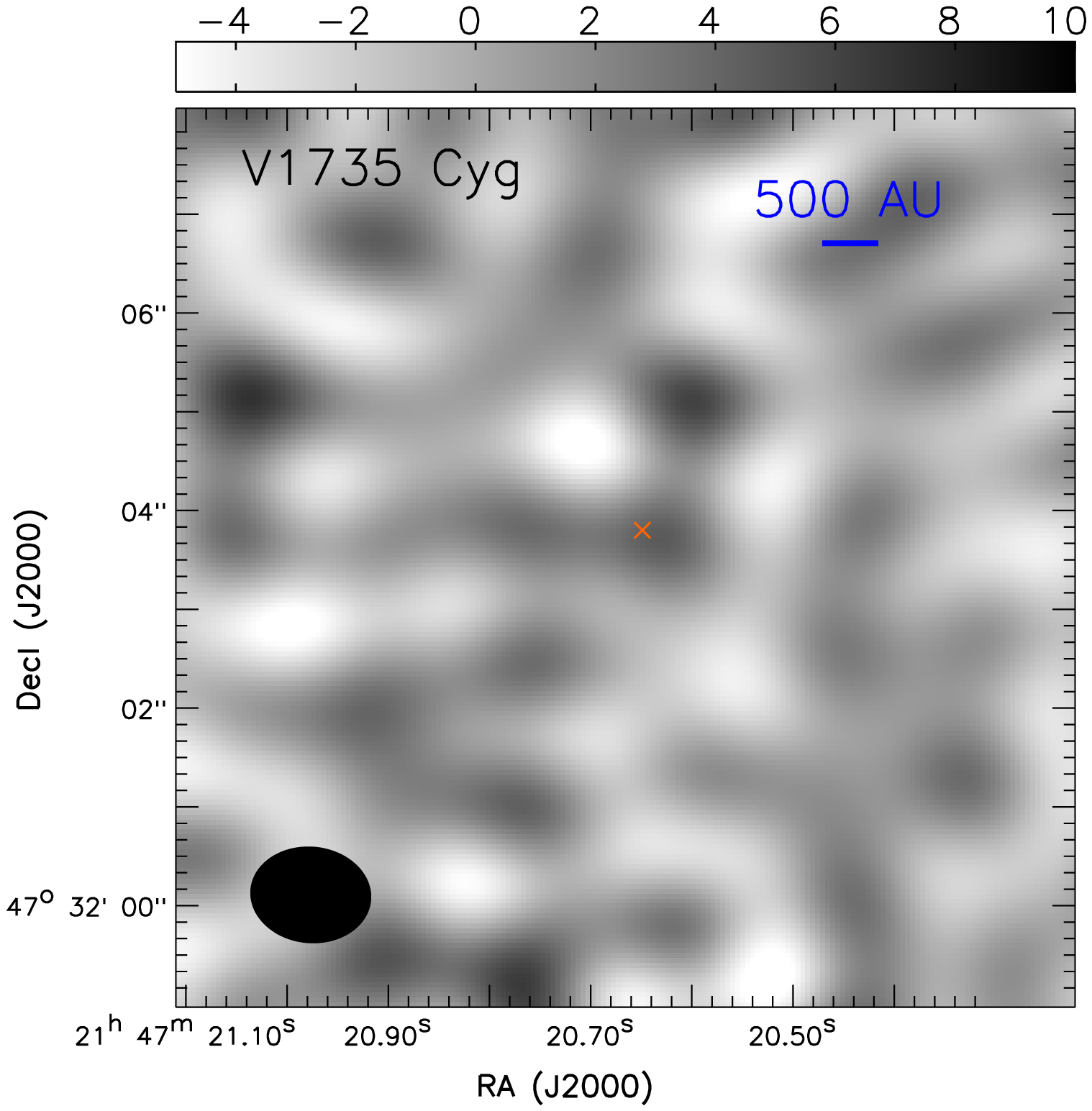} &
\includegraphics[width=4.8cm]{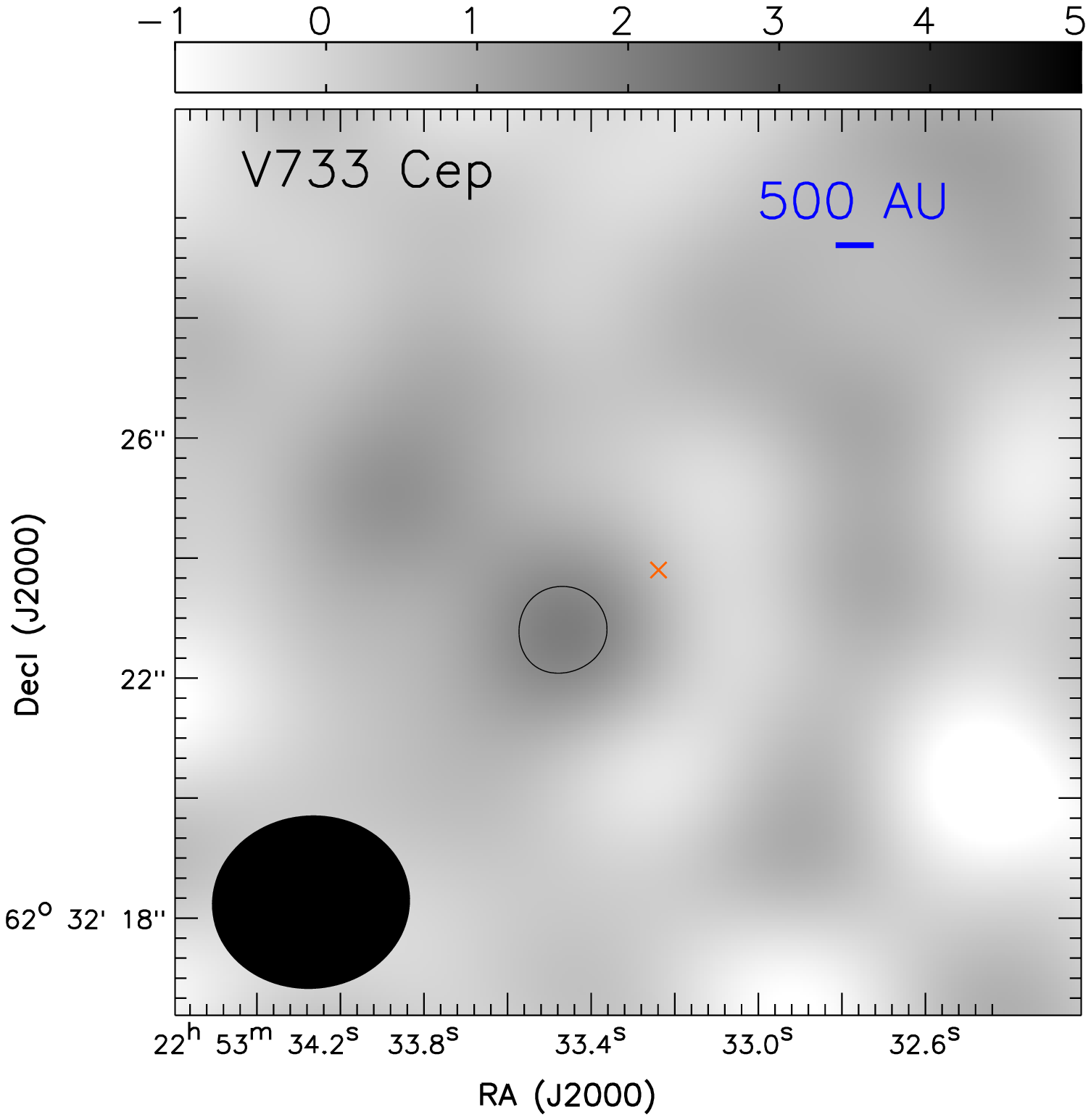}  &
 \\
\end{tabular}

\caption{\footnotesize{
Continuation of Figure \ref{fig:poststamp1}.
Contours of V2494\,Cyg are 1.0 mJy\,beam$^{-1}$ (1$\sigma$) $\times$ [-3, 3, 6, 12, 24, 48, 96].
Rectangle and cross in the panel of AR\,6A/6B marks the locations of AR\,6A and 6B quoted from Aspin \& Reipurth (2003); those in the panel of Z\,CMa mark the locations of Z\,CMa\,NW and SE, respectively (Szeifert et al. 2010).
}}
\label{fig:poststamp2}
\end{figure*}

\begin{figure*}
\hspace{-0.7cm}
\begin{tabular}{ p{5.8cm} p{5.8cm} p{5.8cm} }
\includegraphics[width=6.5cm]{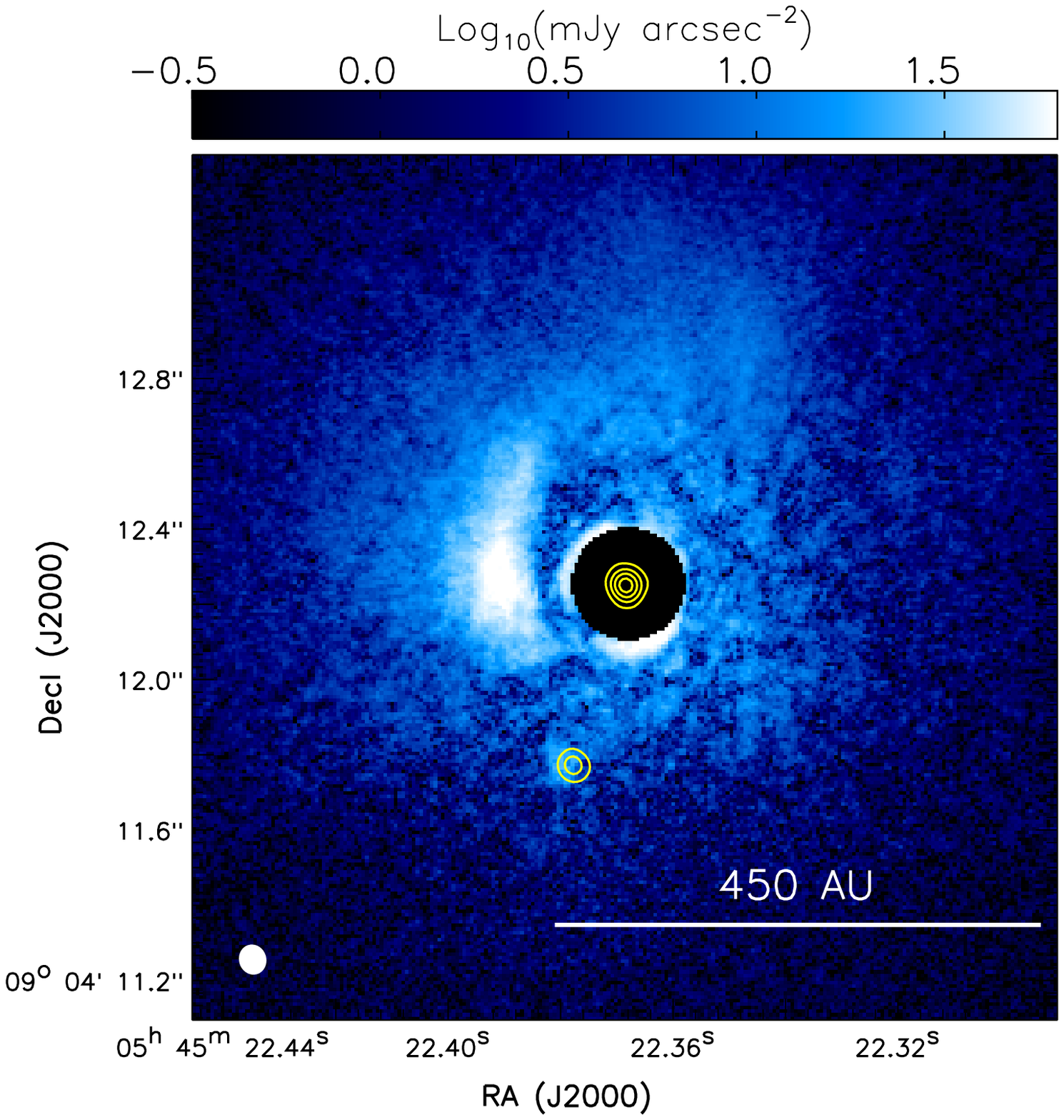} &
\includegraphics[width=6.5cm]{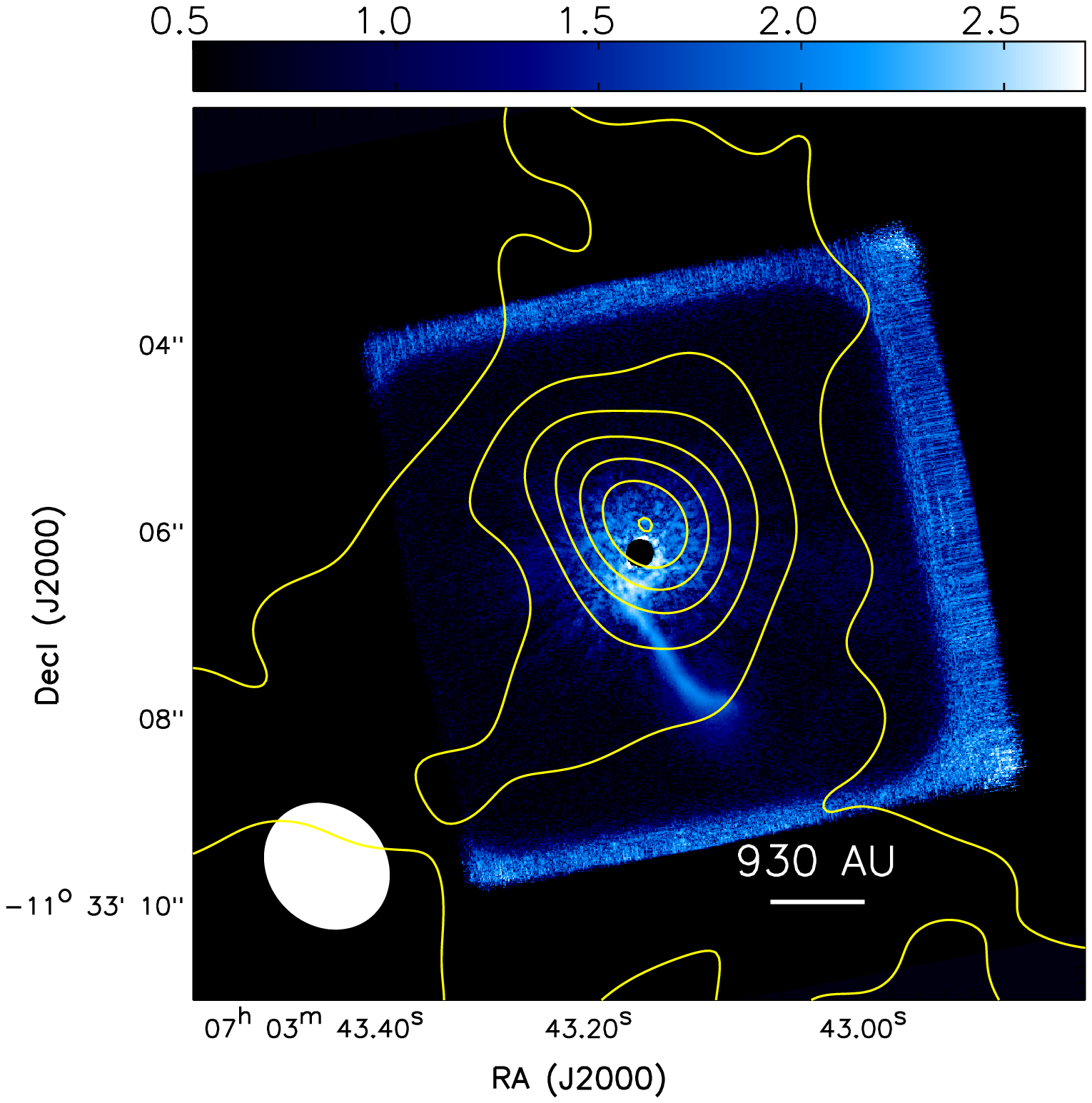} &
\includegraphics[width=6.5cm]{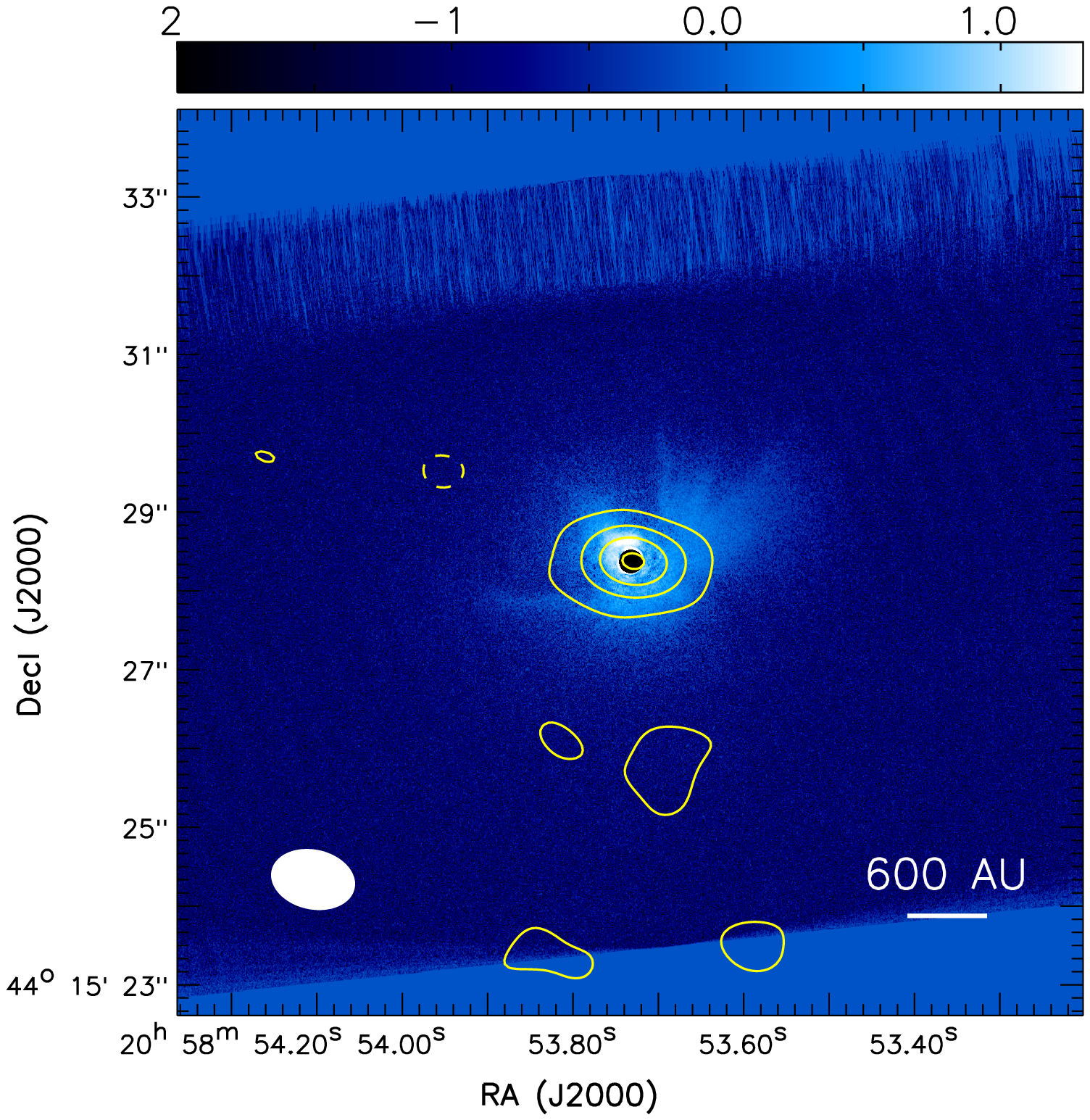} \\
\end{tabular}
\caption{\footnotesize{
Comparisons of the millimeter images with the Subaru-HiCIAO near infrared polarization intensity images, for FU\,Ori (left), Z\,CMa (middle), and V1057\,Cyg (right).
In each panel, the angular resolution of the millimeter images are presented in bottom left.
Left:-- The Subaru-HiCIAO H-band image (color), overplotted with the JVLA 33 GHz continuum image. Contours are 36 $\mu$Jy\,beam$^{-1}$ (10-$\sigma$) $\times$ [1, 2, 3, 4, 5] (reproduced from Liu et al. 2017).
Middle:-- The Subaru-HiCIAO K-band image (color), overplotted with the SMA 225 GHz continuum image. Contours are 5 mJy\,beam$^{-1}$ (5-$\sigma$) $\times$ [1, 2, 3, 4, 5, 6, 7]. The spatial offset between the millimeter emission peak and the coronagraphic mask can be casued by a combination of the astrometric uncertainty of SMA due to phase self-calibration, and the astrometric uncertainty of Subaru-HiCIAO.
Right:-- The Subaru-HiCIAO H-band image (color), overplotted with the SMA 225 GHz continuum image. Contours are 0.72 mJy\,beam$^{-1}$ (3-$\sigma$) $\times$ [-1, 1, 3, 5, 7].
}}
\label{fig:hiciao}
\end{figure*}

\begin{figure}
\hspace{-0.3cm}
\begin{tabular}{ p{8cm}  }
\includegraphics[width=8.5cm]{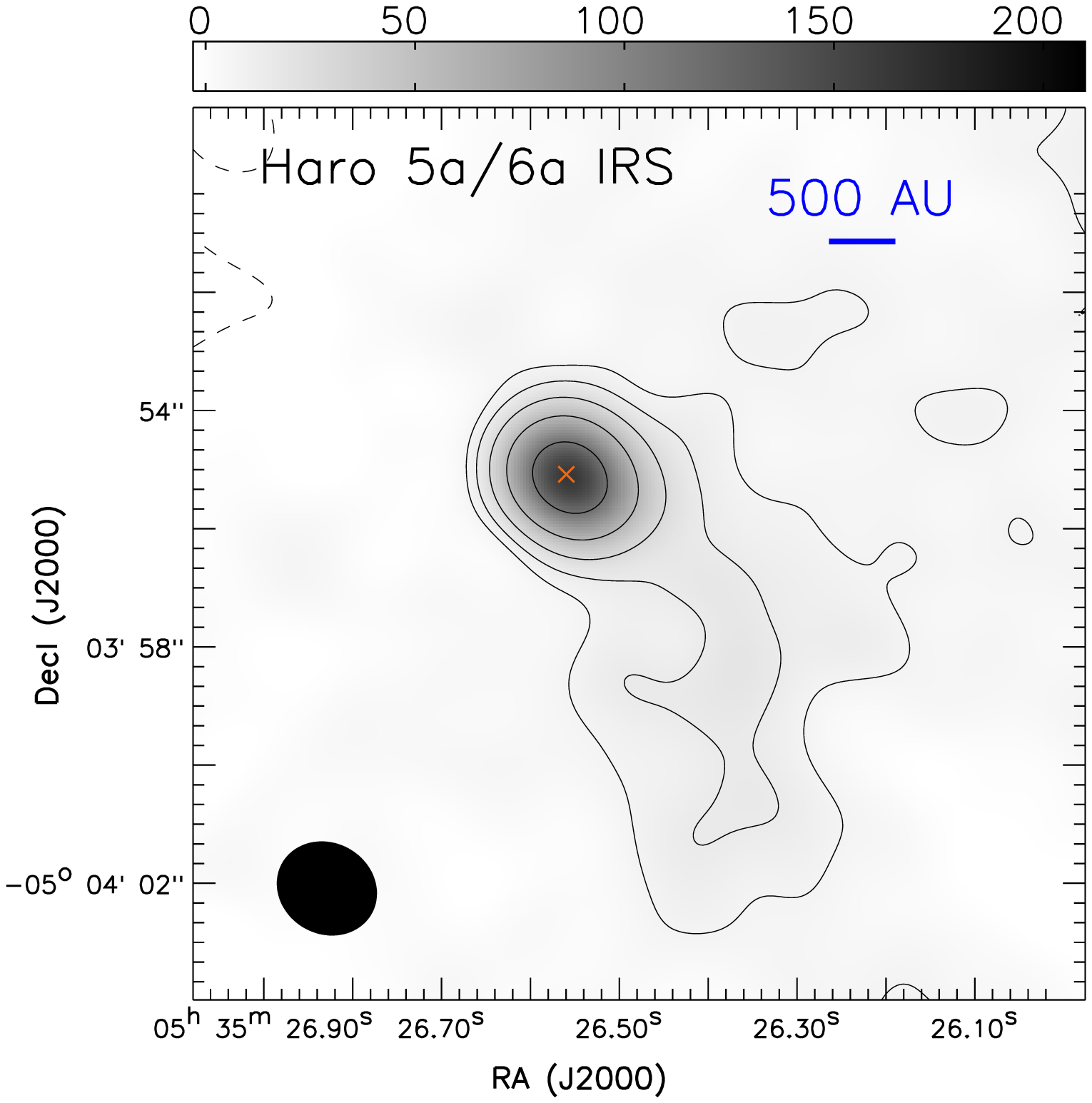}
\end{tabular}
\caption{\footnotesize{
A lower angular resolution 225 GHz continuum image of Haro\,5a/6a\,IRS.
The synthesized beam ($\theta_{\mbox{\scriptsize{maj}}}$ $\times$ $\theta_{\mbox{\scriptsize{min}}}$; P.A.) is 1$\farcs$7$\times$1$\farcs$2; P.A.=59$^{\circ}$.
Contours are 2.4 mJy\,beam$^{-1}$ (1$\sigma$) $\times$ [-3, 3, 6, 12, 24, 48].
}}
\label{fig:haro5a6a}
\end{figure}

Previous optical and infrared (OIR) monitoring observations have detected luminous outbursts from some young stellar objects (YSOs; for a recent review see Audard et al. 2014).
These outbursts are commonly interpreted as due to a temporarily increased accretion rate onto the host YSO, resulting in an enhanced accretion shock luminosity.
The YSOs which have longer outburst duration (a few tens of years or longer) are referred to as FU Orionis objects (FUors, hereafter) after the archetype source FU\,Ori (see Hartmann \& Kenyon 1996 for a review).
Those which have shorter outburst duration (a few hundred days to a few years), and sometimes present repetitive outburst events, are referred to as EXors after the archetype source EX\,Lupi (see Herbig 1977 for a review).
We note that whether or not there is a well defined boundary to separate FUors and EXors is yet uncertain (e.g., K{\'o}sp{\'a}l et al. 2011b).
There are also objects which present OIR spectral features similar to FUors although no OIR outbursts events were detected for them in monitoring observations.
They are referred to as FUor-like objects, of which the accretion outburst may have been onset before humans started to quantitatively record the OIR brightness (see discussion in Hartmann \& Kenyon 1996).
There are indirect evidences suggesting accretion bursts are common throughout the embedded stage, and that FUors are the optical manifestation of that process (e.g., Dunham \& Vorobyov 2012 and references therein).

The OIR outburst events are rare, and a majority of them are located at several hundreds parsec distances from the solar system.
In the early 1980s there were only a handful of YSOs with confirmed OIR outbursts.
Thanks to the persistent OIR monitoring surveys to discover more of them over the last few decades, now it becomes possible to study a large sample of accretion outburst YSOs in systematic surveys.
We refer to Audard et al. (2014) for a thorough summary about the emission properties of these accretion outburst YSOs at $<$100 $\mu$m wavelengths.

\begin{figure}
\hspace{-1.2cm}
\begin{tabular}{p{8.5cm} }
\includegraphics[width=10.25cm]{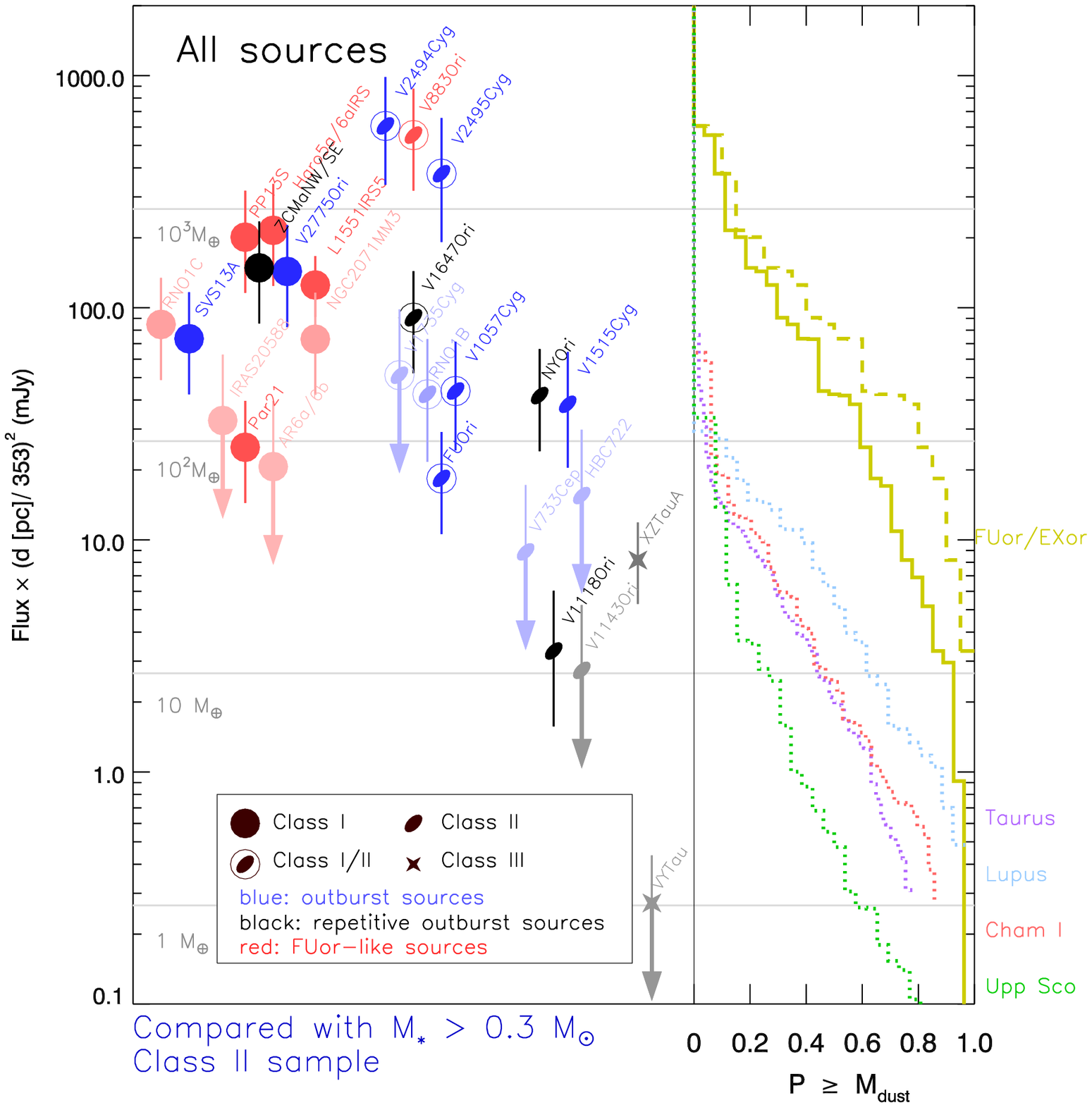} 
\end{tabular}
\caption{\footnotesize{
Millimeter fluxes of the observed target sources. 
The millimeter fluxes are from Table \ref{tab:summary}. 
All presented fluxes from our observations have been scaled to the measurements at 225 GHz by assuming an identical spectral index of $\alpha$ $=$ $3.8$.
In addition, the plotted flux of each source has been scaled by ($d$ [pc] / 353 [pc])$^{2}$, where $d$ is the source distance and 353 pc is the parallax distance of FU\,Ori.
Blue, black, and red symbols are the FUors, EXors, and FUor-like objects, respectively.
Class I, I/II, II, and III sources are distinguished by different symbol shapes, which are introduced in the figure legend. 
Symbols for target sources which are not detected (presented 3-$\sigma$ upper limit) or are largely confused by the parent cloud structures, are presented in lighter colors.
Symbols for non-detected sources are connected with downward arrows.
The horizontal displacements of symbols do not have physical meaning although in general the more embedded sources are more to the left.
The four thin horizontal gray lines show the corresponding dust mass evaluated by $M_{\mbox{\tiny dust}}$ [$M_{\odot}$] $=$ 1.13 $\times$ 10$^{-5}$ $F_{\mbox{\tiny 1.33 mm}}^{\mbox{\tiny 353 pc}}$[mJy], which is a formulation introduced by Ansdell et al. (2017) but was rescaled to a 353 pc distance.
Solid and dashed yellow lines show the cumulative distribution function of the present sample: the former includes both the detected and non-detected sources by assigning their fluxes as the 1-$\sigma$ noise, and the later only includes the detected sources.
Purple, light blue, red, and green dotted lines present the $M_{*}>$0.3 $M_{\odot}$ Class II objects in the Taurus, Lupus, Chameleon I, and Upper Sco regions, respectively (quoted from Pascucci et al. 2016; the original observations can be found in Andrews et al. 2013, Carpenter et al. 2014, Ansdell et al. 2016, Barenfeld et al. 2016, and Pascucci et al. 2016).
}}
\label{fig:fluxsummary}
\end{figure}

\begin{figure}[!]
\begin{tabular}{ c }
\includegraphics[width=9.5cm]{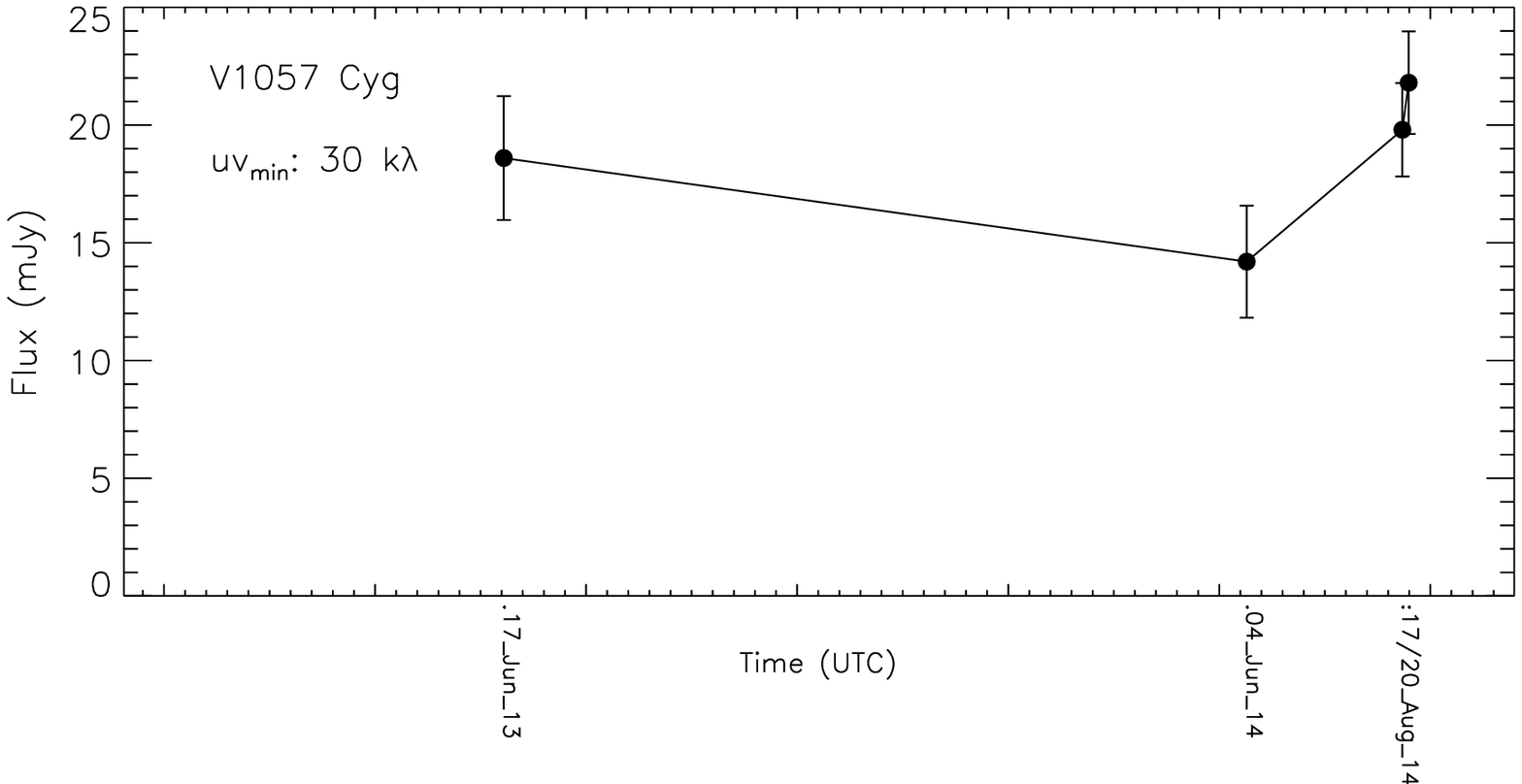} \\
\includegraphics[width=9.5cm]{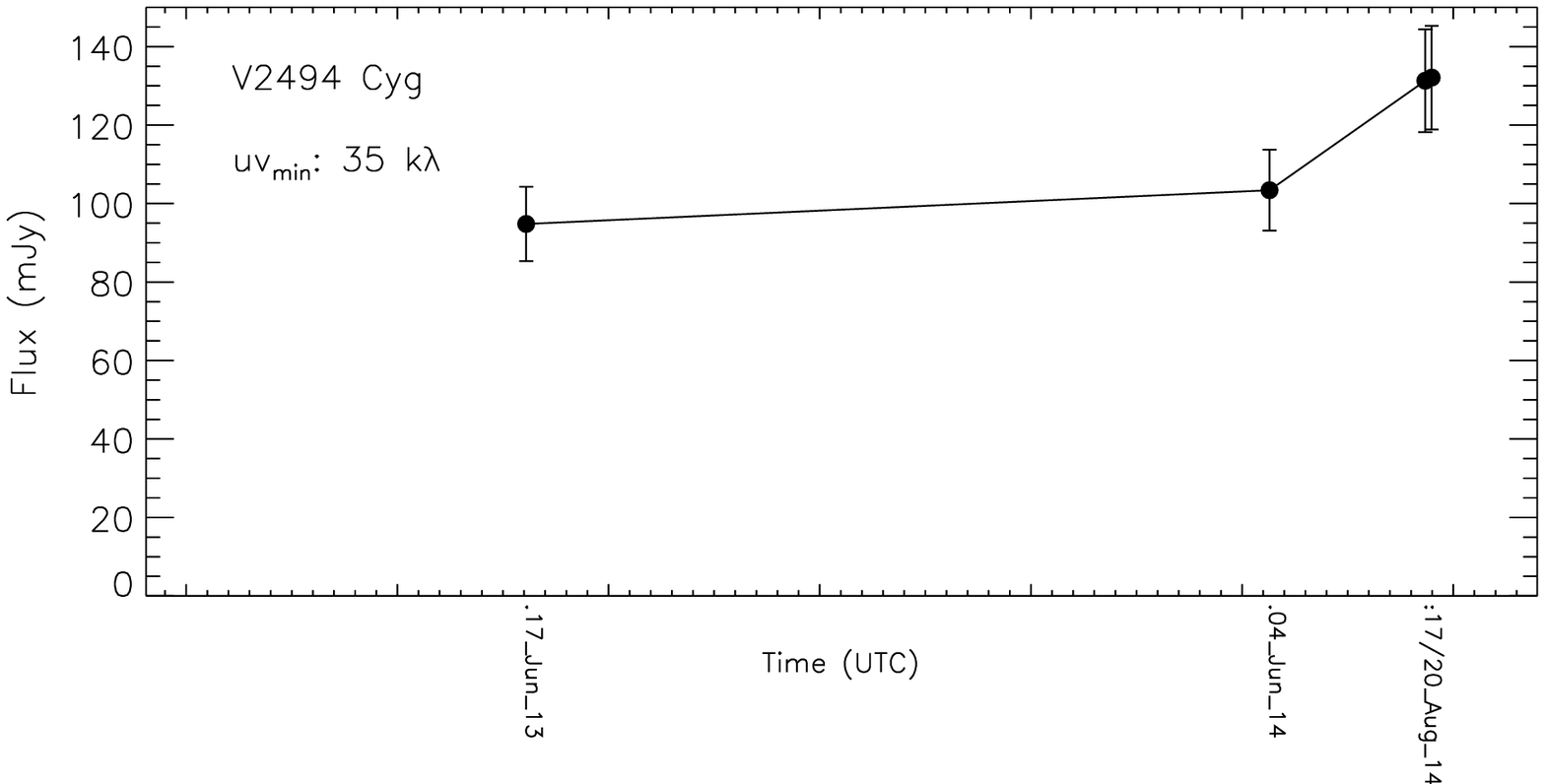} \\
\includegraphics[width=9.5cm]{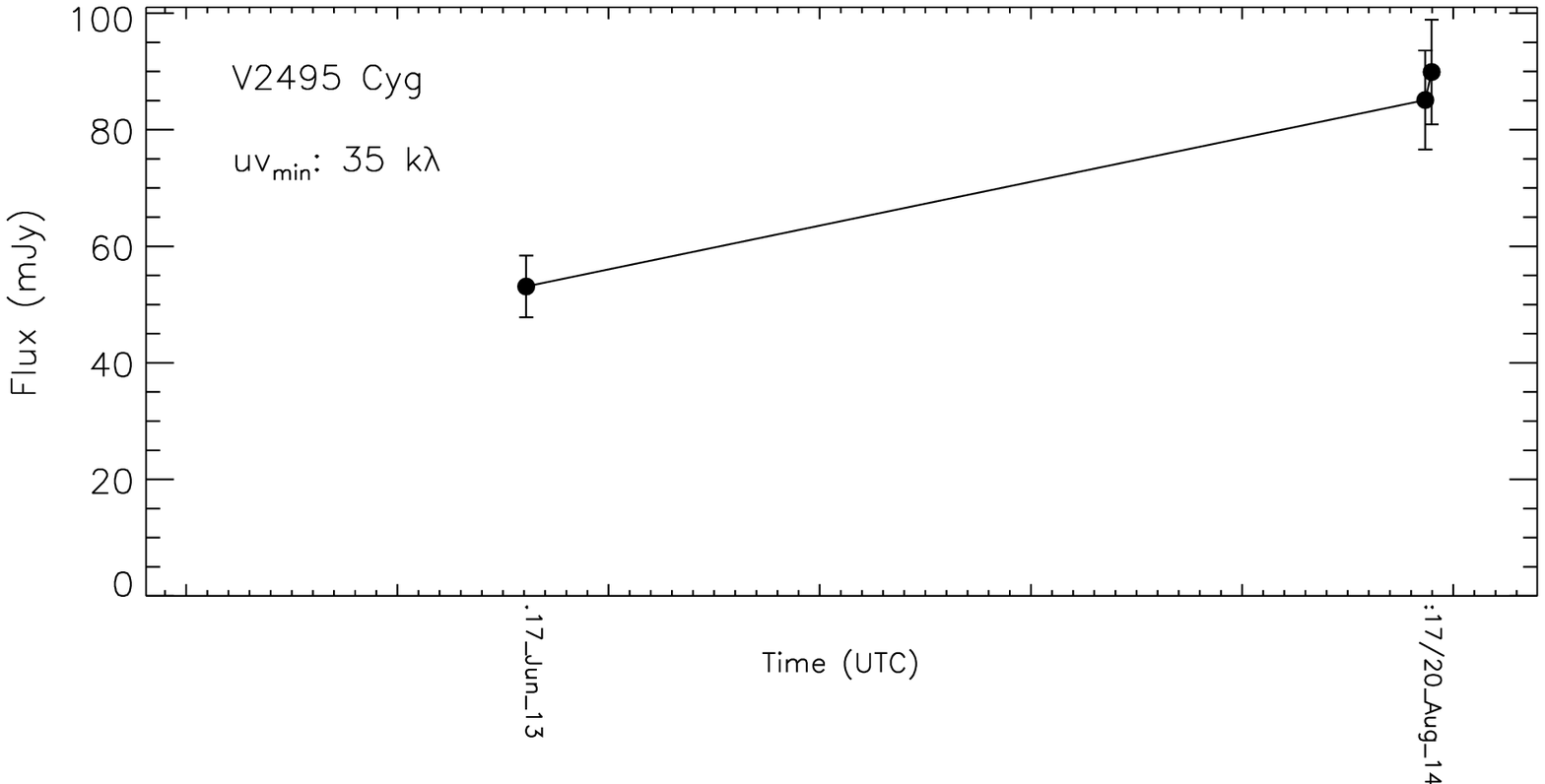} \\
\includegraphics[width=9.5cm]{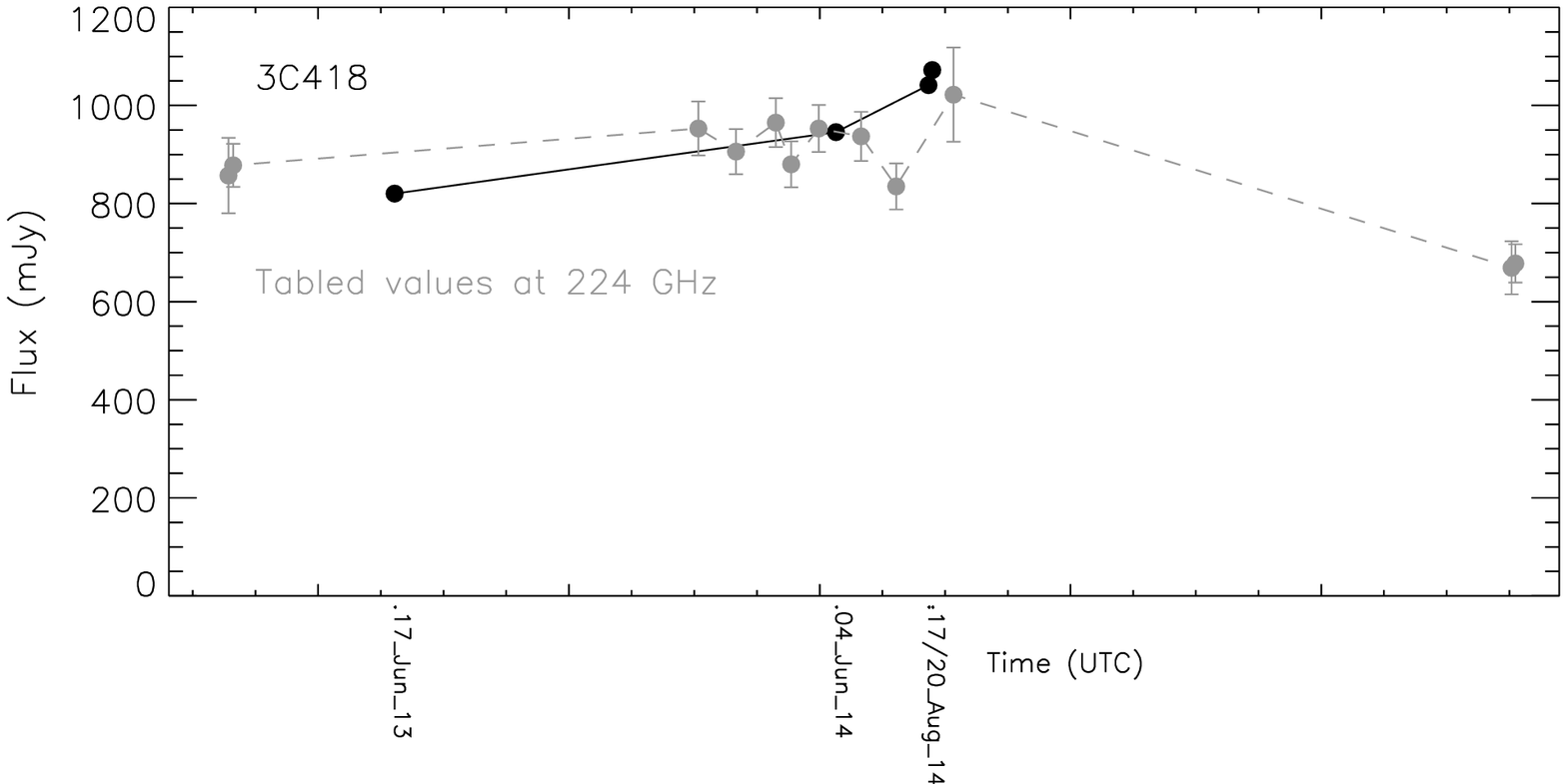} \\
\end{tabular}
\caption{\footnotesize{
Observed fluxes of V1057\,Cyg, V2494\,Cyg, and V2495\,Cyg.
Measurements were made at 224-225 GHz if not specifically annotated.
Black line in the bottom panel shows the measured and applied flux values of the gain phase and amplitude calibrator for these observations, 3C418.
The absolute flux reference sources of these observations are listed in Table \ref{tab:summary}.
In the panel of 3C418, gray symbols and dotted line quote the tablized 224-225 GHz flux values and errors from the SMA Calibrator List, which is maintain by Mark Gurwell.
For the target source measurements, the chosen minimum {\it uv} distances are annotated in individual panels.
For each source, we chose an identical minimum {\it uv} distance for each (time) epoch of measurement.
}}
\label{fig:fluxvarCygnus}
\end{figure}

\begin{figure}[!]
\begin{tabular}{ c }
\includegraphics[width=9.5cm]{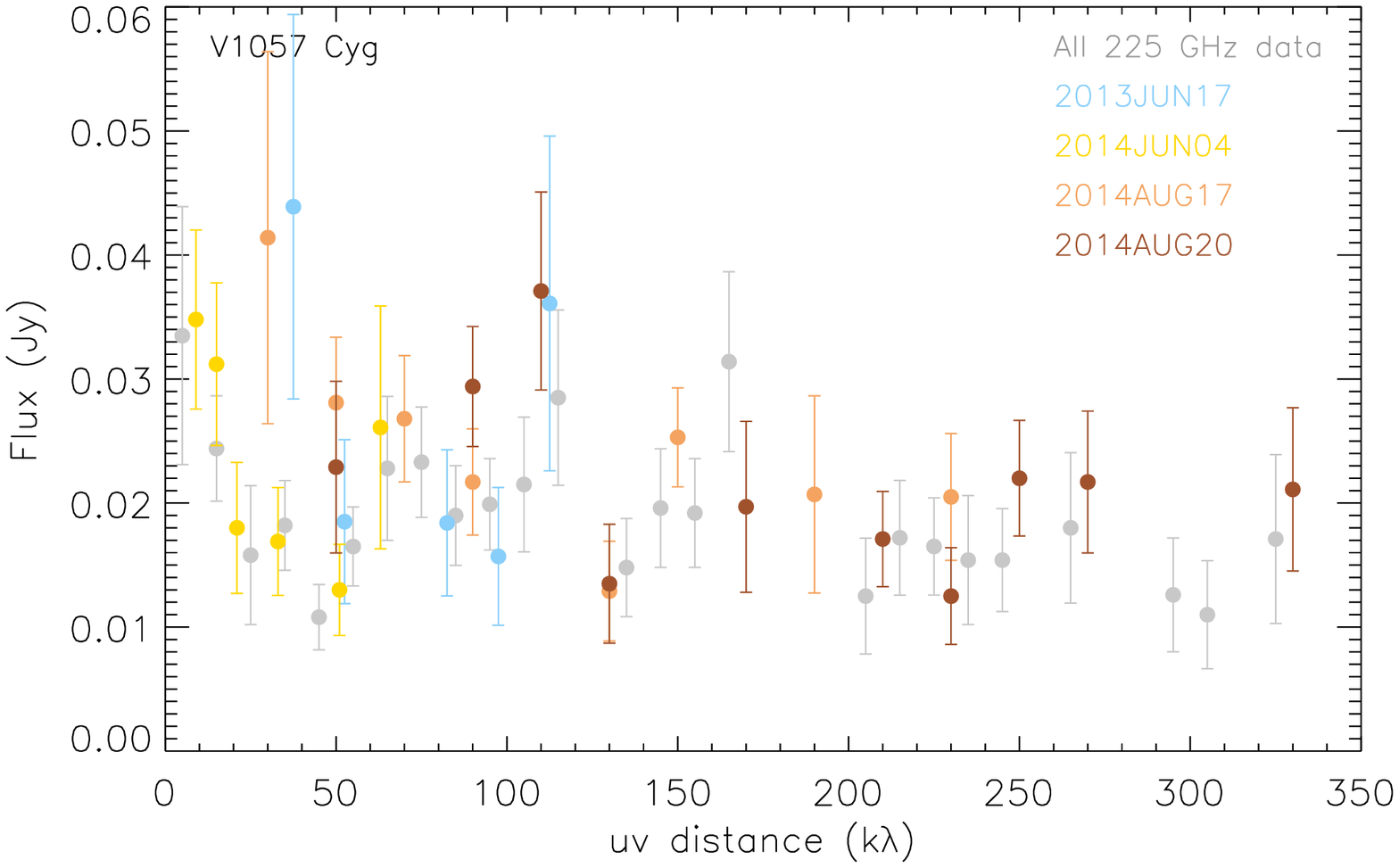} \\
\includegraphics[width=9.5cm]{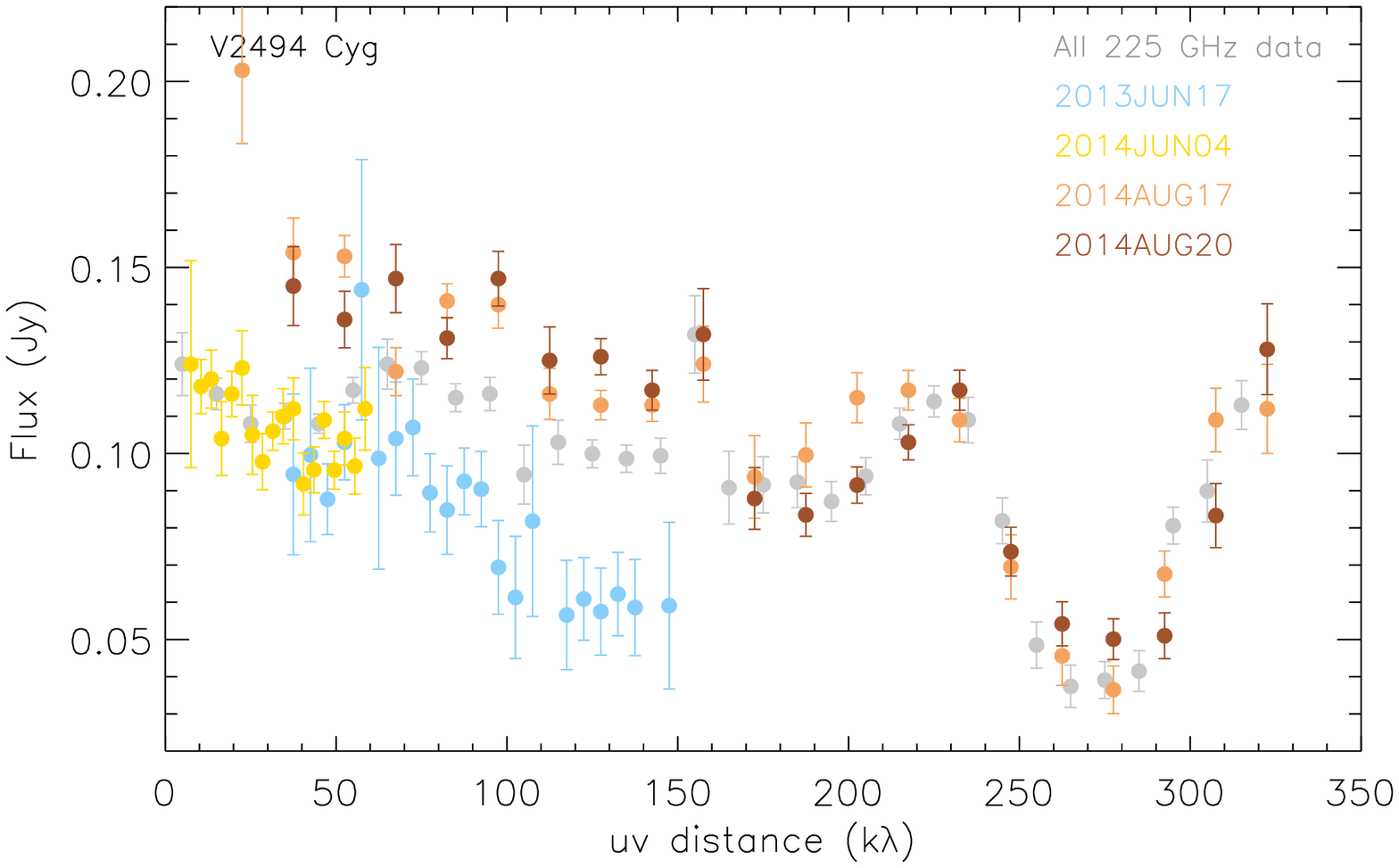} \\
\includegraphics[width=9.5cm]{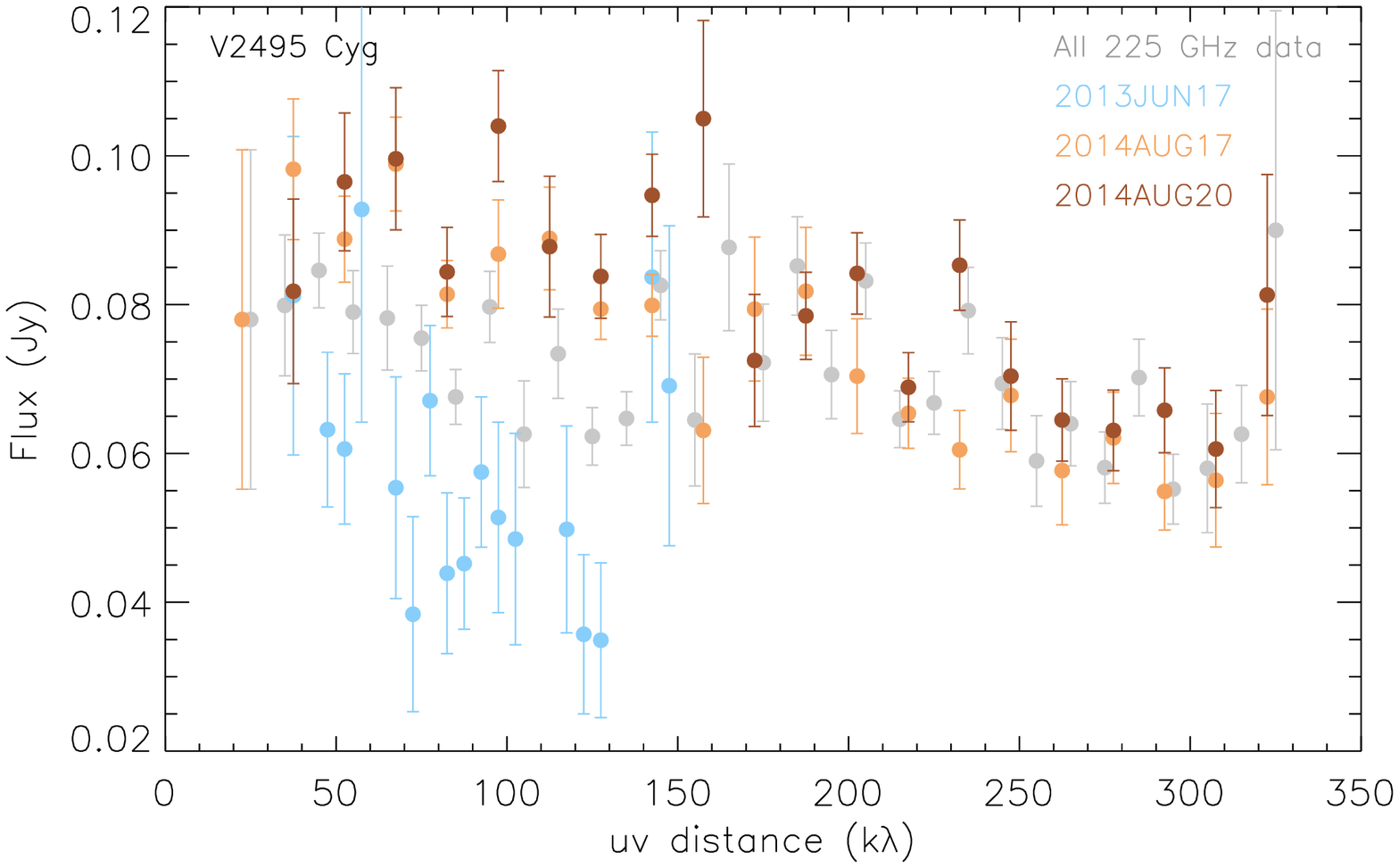} \\
\end{tabular}
\vspace{-0.6cm}
\caption{\footnotesize{
Visibility amplitudes of V1057\,Cyg, V2494\,Cyg, V2495\,Cyg.
In each panel, we plot the azimuthally vector-averaged visibility amplitudes for individual epochs of observations in different colors.
Gray symbols and dashed lines show the averaged visibility amplitudes from all available observations.
We only present the measured amplitudes which are more than two times higher than the expected amplitude assuming pure noise, to suppress the confusion from some poorly sampled {\it uv} distance ranges in our observations.
We still cannot fully avoid the issue related to poor {\it uv} sampling, which in some cases artificially biases the amplitudes higher at the shortest and/or the longest {\it uv} distance.
}}
\label{fig:uvampCygnus}
\end{figure}

To understand the bulk properties of the cool gas/dust reservoirs which are feeding accretion outbursts (e.g., circumstellar disks, inner envelopes), observations in the dust optically thin regime (e.g., mm wavelengths) are necessary.
Previous far infrared and submillimeter surveys of this type of YSOs were done with single-dish observations which lacked the angular resolution to properly resolve the dust emission and were affected by cloud contamination and dust optical depth effects (e.g., Sanders \& Weintraub 2001; Green et al. 2013).
There are numerous higher angular resolution interferometric observations at submillimeter and millimeter bands towards individual target sources (e.g., Lim \& Takakuwa 2006; Alonso-Albi et al. 2009; P\'erez et al. 2010; K{\'o}sp{\'a}l 2011b; Dunham et al. 2012; Hales et al. 2015; Cieza et al. 2016; K{\'o}sp{\'a}l et al. 2017; Lim et al. 2016; Liu et al. 2016a; Liu et al. 2017; Zurlo et al. 2017; Ru{\'{\i}}z-Rodr{\'{\i}uez et al. 2017).
However, a systematic comparison of the (sub)millimeter emission property from high angular resolution observations is still lacking.

From our extensive observations using the Submillimeter Array (SMA)\footnote{
The Submillimeter Array is a joint project between the Smithsonian Astrophysical Observatory and the Academia Sinica Institute of Astronomy and Astrophysics, and is funded by the Smithsonian Institution and the Academia Sinica (Ho et al. 2004).} we compile a sample of 29 accretion outburst YSOs (FUors, EXor, and FUor-like objects) except for the active accretion YSO IRAS\,20588+5215N which has a less well clarified nature (Aspin et al. 2009).
We have also processed the archival SMA observations of these types of objects.
The observed target sources are summarized in Table \ref{tab:summary}.
The selected target AR\,6A/6B is a visual binary object, of which both of the host YSOs are FUor-like objects.
For the visual binary Z\,CMa, one host protostar is a repetitive short-duration outburst YSO while the other is a FUor-like object.
The sources RNO\,1B and 1C were previously considered as binary outburst sources.
However, they are $\sim$5000 AU apart and are in fact located in a condensed low-mass cluster-forming regions (e.g., Staude \& Neckel 1991; Anglada et al. 1994; Quanz et al. 2007).
They should be regarded as independent sources, of which the binarity is not yet resolved.
We refer to Audard et al. (2014), Gramajo et al. (2014), and references therein for the properties of the selected target sources, and omit duplicating those descriptions.

Most of these observations were carried out at the central frequencies of $\sim$225 GHz (1.33 mm) and $\sim$271 GHz (1.1 mm), which can be compared with the recent Atacama Large Millimeter Array (ALMA) 1.33 mm surveys of Class\,II YSOs in the nearby molecular clouds, and can be compared with the 0.88 mm surveys after adopting certain assumed (sub)millimeter spectral indices (e.g., Carpenter et al. 2014; Ansdell et al. 2016; Pascucci et al. 2016; Ansdell et al. 2017; and see also Andrews et al. 2013).
The majority of our target sources were observed with $\sim$1$''$ synthesized beams, which is sufficient for tracing dust emission from their circumstellar disks and inner envelopes.
The presented observations and the details of data reduction are introduced in Section \ref{sec:observations}.
We summarize the spatially resolved structures, the continuum emission, and the time variability of millimeter fluxes of a subset of sources in Section \ref{sub:structure}, \ref{sub:mmflux}, and \ref{sub:mmvariability}, respectively.
The present paper intends to focus on presenting measurements. 
Nevertheless, we briefly provide our preliminary hypotheses to explain some identified features from our observations in Section \ref{sec:discussion}.
We defer the detailed analysis of (sub)millimeter spectral energy distributions (SEDs) to forthcoming papers. 
Our conclusion is given in Section \ref{sec:conclusion}. 
Appendix \ref{appendix:obs} presents a summary for our SMA observations.
Appendix \ref{appendix:variability} introduces the millimeter flux variability of some sources which were observed in multiple epochs but have limited data quality.
Appendix \ref{appendix:uvamp} archives the visibility amplitude distributions of some sources that are not addressed in the main text.

\section{Observations}\label{sec:observations}
Extensive SMA observations at $\sim$1 mm band towards FU Orionis objects and EXors were carried out from 2008 to 2014, under the management of Michael Dunham (Project codes: 2011A-S030, 2013A-S085 , 2013B-S078, 2013B-S092, 2014A-S011, 2015A-S065, 2015B-S078), Hauyu Baobab Liu (Project codes: 2013B-A004, 2014B-A001), Tyler Bourke (Project code: 2008B-S002, 2013A-S057), Naomi Hirano (Project code: 2013A-A018), and Steve Longmore (2008A-S101, 2009A-S066).
We also retrieved archival data of the two sources V883\,Ori and V1647\,Ori taken by Dave Principe (unknown project code) and un-identified PIs from 2004 to 2012.
The observations of HBC\,722 taken from project 2011A-S030 have been reported in Dunham et al. (2012).
The observations of FU\,Ori taken from projects 2008B-S002 and 2013B-A004 have been reported in Liu et al. (2017).
The observations of EXors taken from project 2014B-A001 have been reported in Liu et al. (2016a).
A brief summary including all the observational setups is provided in Appendix \ref{appendix:obs} for the sake of the integrity of information.
We have re-calibrated some of the already reported data or have re-imaged them for our present purposes.

Some of these observations were carried out with a track-sharing observational strategy. 
All of these observations utilized the Application-Specific Integrated Circuit (ASIC) correlator, which provided  4–8 GHz intermediate frequency (IF) coverages in the upper and the lower sidebands, with 48 spectral windows in each sideband. 
Some observations taken after January of 2015 further include the  SMA Wideband Astronomical ROACH2 Machine (SWARM) correlator, which provided two additional spectral windows covering the 8-9.5 GHz and 10.5-12 GHz IFs, respectively. 
However, we omit using data in the 10.5-12 GHz IF because of poor response, which was also pointed out in Liu et al. (2016a).
Not all of the presented observations were originally carried out for the same scientific purposes.
In addition, some of these observations were taken utilizing filler observing time, during when the spectral tunings had to follow those of the regular time projects.
Therefore, there are deviations of the central frequencies.
Nevertheless, the local oscillator frequencies of most of these observations are either in the range of  224-225 GHz ($\sim$1.33 mm; e.g., for simultaneously observing the $^{12}$CO 2-1 isotopologues), or are in the range of 271-272 GHz ($\sim$1.10 mm; e.g., for simultaneously observing HCN/HCO$^{+}$/N$_{2}$H$^{+}$ lines).
We will conveniently refer to the former as 225 GHz observations, and the later as 272 GHz observations, since we do not have enough of accuracy to distinguish flux measurements at $\sim$1 GHz frequency offsets.
The differences in the observing wavelength is not particularly significant in terms of our major science purpose of constraining the dust emission around $\sim$1 mm wavelength (more discussion in Section \ref{sec:results} and \ref{sec:discussion}).
For similar reasons, the projected baseline lengths covered by the observations of individual sources can be very different, which do have an impact on our interpretation.
We will provide fair and relevant cautions at the related scientific analysis and discussion.

\begin{figure}[h!]
\begin{tabular}{ c }
\includegraphics[width=9.5cm]{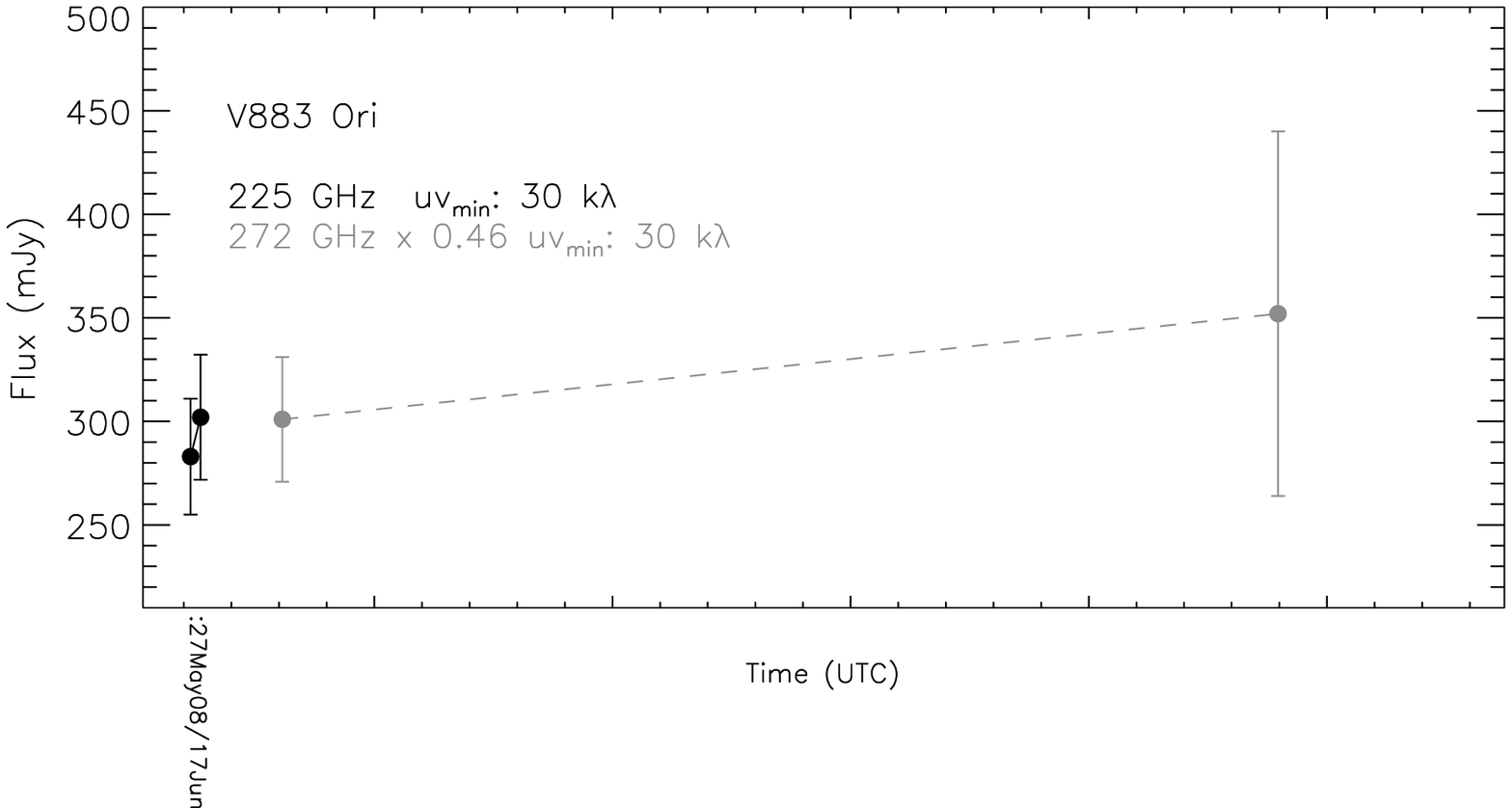} \\
\includegraphics[width=9.5cm]{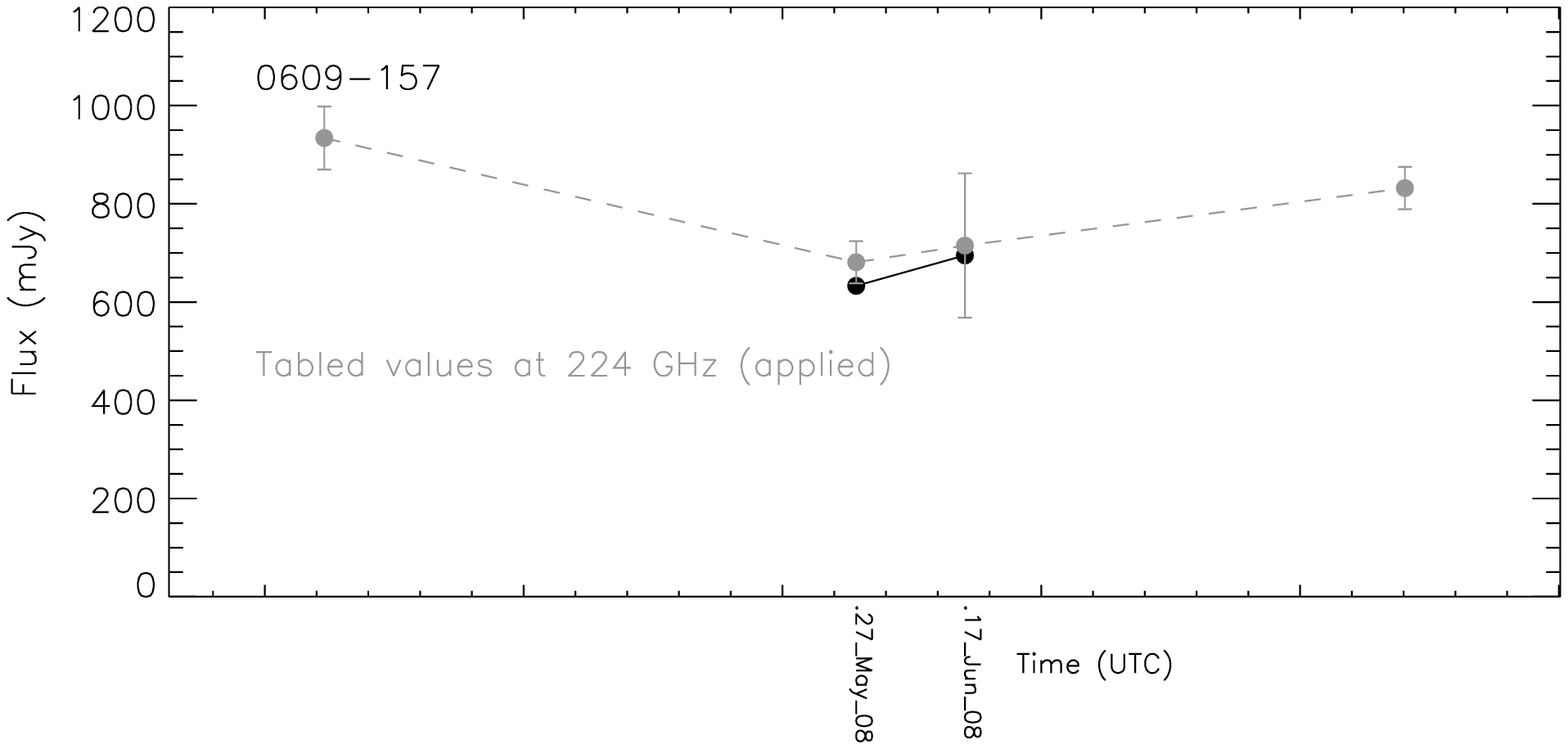} \\
\includegraphics[width=9.5cm]{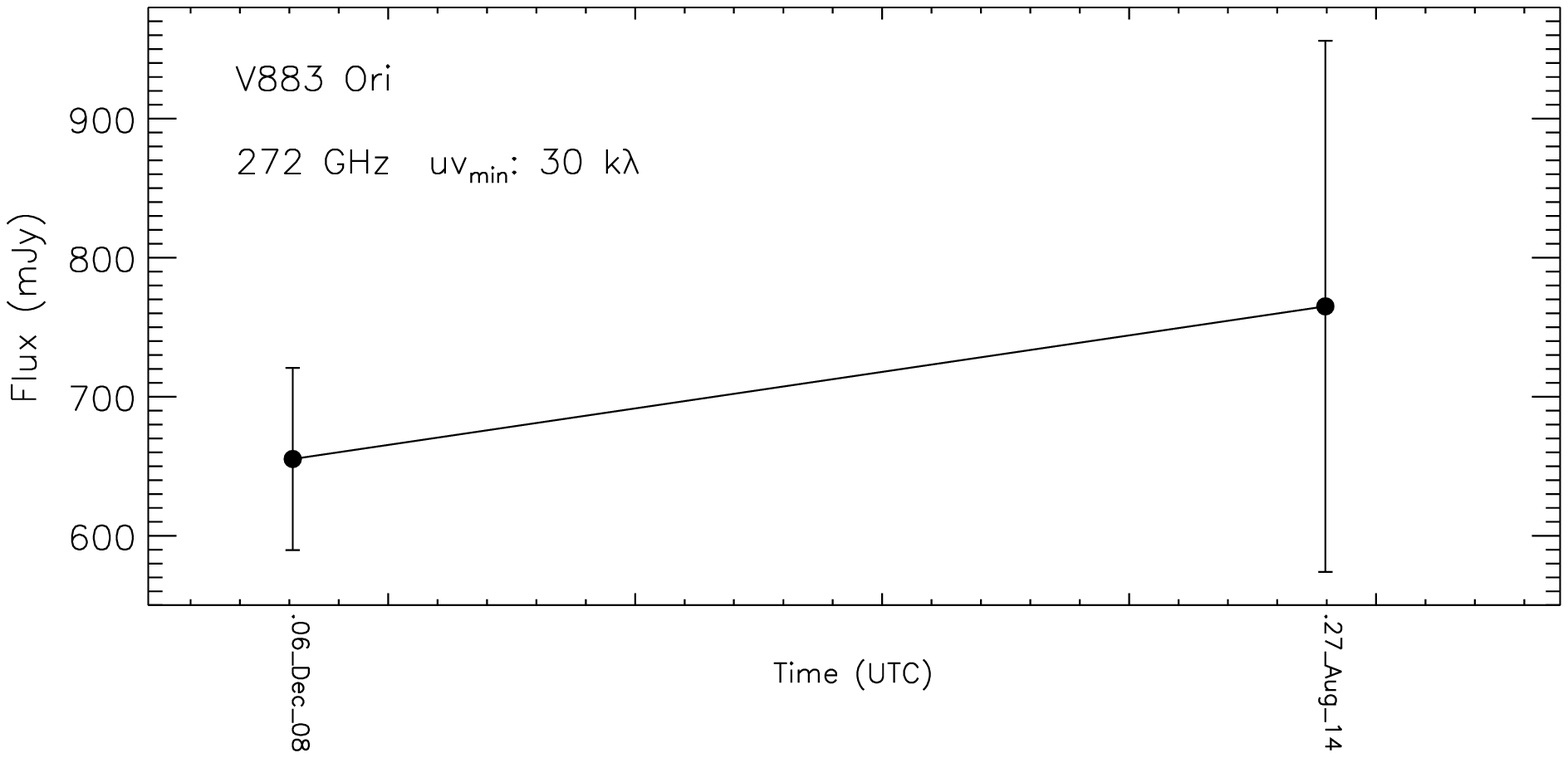} \\
\includegraphics[width=9.5cm]{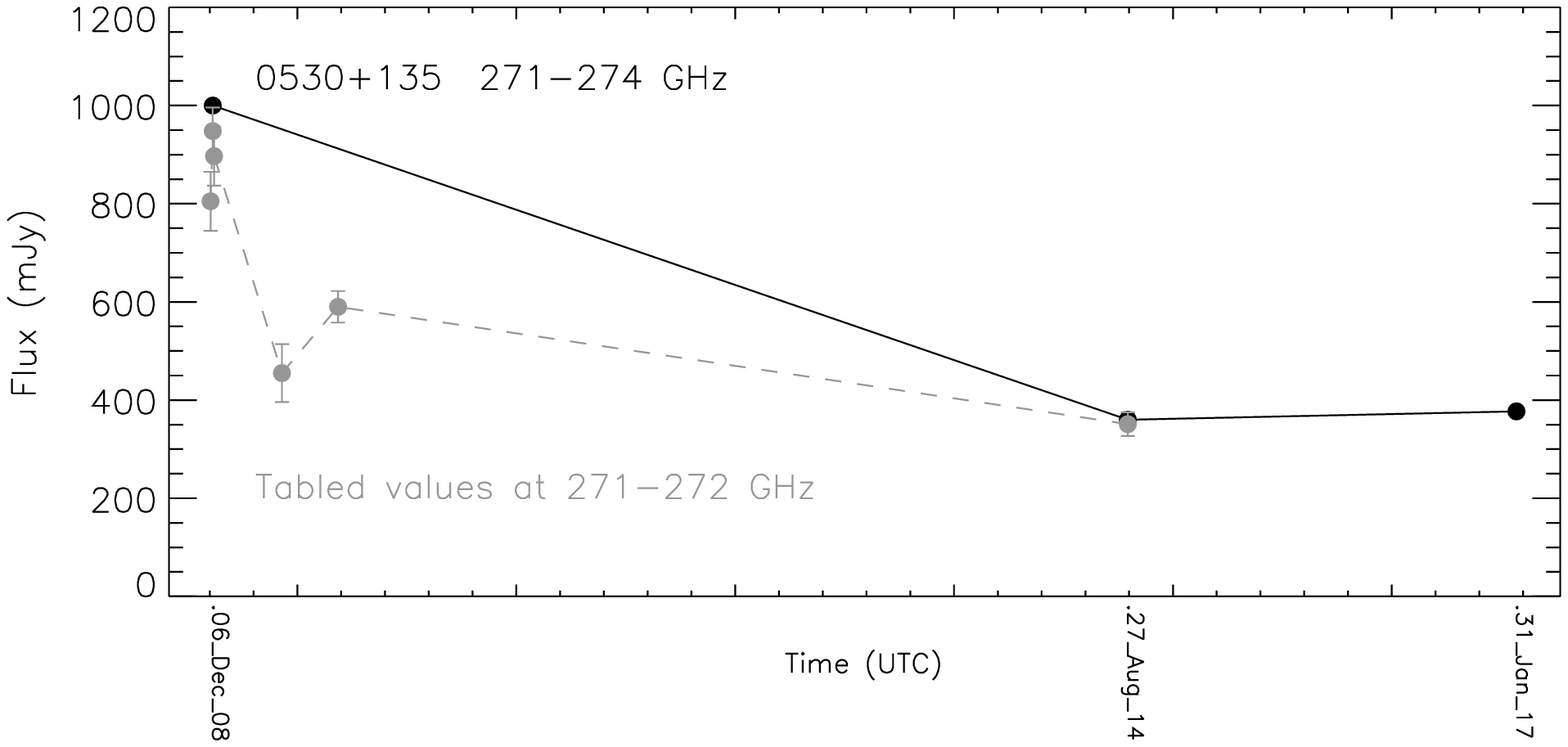} \\
\end{tabular}
\caption{\footnotesize{
Similar to Figure \ref{fig:fluxvarCygnus}, but for V883\,Ori.
Top two panels present the observations at 225 GHz, which were gain calibrated by the observations of quasar 0609-157.
Bottom two panels present the observations at 272 GHz, which were gain calibrated by the observations of quasar 0530+135.
Quasar 0609-157 happened to be flux monitored by SMA at 224-225 GHz on the dates of our 225 GHz observations, May 27 and June 17 of 2008.
We applied the tabulated flux values of 0609-157 on these two dates instead of applying our own measurements, since the measurements made by Mark Gurwell are well vetted.
We scale the 272 GHz measurements to 225 GHz by multiplying a factor of 0.46 (i.e., assuming spectral index $\alpha$ $\sim$ 4.0, and overplot to the top panel.
We adopted a relatively large (25\%) potential absolute flux error for the measurements on August 27, 2014, due to the large phase dispersion during the observations.
}}
\label{fig:fluxvarv883ori}
\end{figure}

\begin{figure}
\begin{tabular}{ c }
\includegraphics[width=9.5cm]{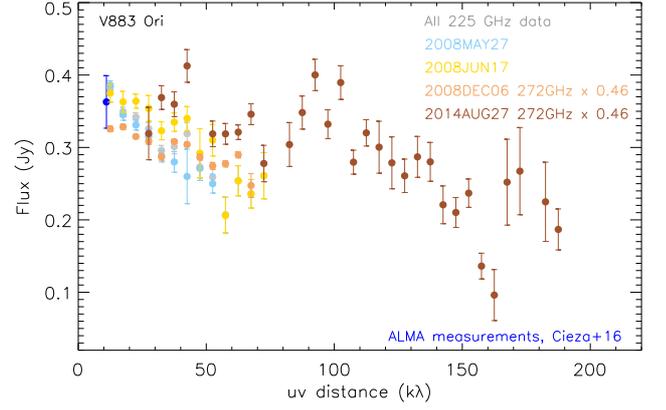}    \\
\end{tabular}
\vspace{-0.6cm}
\caption{\footnotesize{
Visibility amplitudes of V883\,Ori (similar to Figure \ref{fig:uvampCygnus}).
We scaled the observed amplitudes at frequencies higher than 225 GHz to the expected amplitudes at this frequency, by assuming the spectral index $\alpha$ $=$4.0.
We overplot the integrated flux measured at the same frequency by ALMA with 11 $k\lambda$ shortest {\it uv} distance, which were taken on December 12 of 2014 and April 05 of 2015 (Cieza et al. 2016). 
We assume a nominal 10\% error bar for the presented ALMA measurement.
There were ALMA observations taken with the more extended array configuration in August and October of 2015. 
We omit presenting measurements from the extended ALMA array observations since they observed quasars as absolute flux references, which are not ideal for the purpose of studying flux variability.
}}
\label{fig:uvampv883}
\end{figure}

We followed the standard data calibration strategy of SMA.
The application of system temperature (T$_{\mbox{\tiny sys}}$) information and the absolute flux, passband, and gain calibrations were carried out using the MIR IDL software package (Qi 2003). 
The absolute flux scalings were derived by comparing the visibility amplitudes of the gain calibrators with those of the absolute flux standard sources of SMA, which are planets or planet moons (i.e., the `Flux cal.' column in Table \ref{tab:obs}).
After calibration, the zeroth-order fitting of continuum levels from line-free channels and the joint weighted imaging of all continuum data were performed using the Miriad software package (Sault et al. 1995). 
We performed zeroth-order multi-frequency imaging combining the upper- and lower-sideband data, to produce sensitive continuum image at the central observing frequency (i.e., the local oscillator frequency).
We used a 0\farcs05 cell size and 2048 pixels in each spatial dimension when creating all images.
The cell size is representative to the fit errors of any coordinates or image component sizes in general.
Primary beam attenuation was corrected using the {\tt linmos} task of the Miriad software package.

In most of our flux measurements, we quote a nominal 10\% absolute flux error in the case that it is larger than the 1-$\sigma$ thermal noise (in terms of Jy\,beam$^{-1}$), and quote the 1-$\sigma$ thermal noise in the rest of the cases.
However, for some observations which were carried out in relatively poor weather conditions (e.g., $\tau_{\mbox{\tiny 225 GHz}}$ $\gtrsim$ 0.2, or in conditions with high phase dispersion), we quote up to 25\% absolute flux error in the case that it is larger than the 1-$\sigma$ thermal noise according to our experiences.
We also note that for observations in high phase dispersion weather conditions, for example, some filler time observations at $\sim$6-7 pm Hawaii time, the baseline length dependent phase de-coherence (for more discussion see P\'erez et al. 2010) can artificially cause the azimuthally-averaged visibility amplitude decrease with {\it uv} distance.
In most of our observations, the effect of phase de-coherence is not significant in the short baselines, and therefore does not significantly bias our flux measurements.
However, the degraded azimuthally-averaged visibility amplitudes can artificially make the sources appear slightly spatially resolved both in the image and in the visibility domain.
In the following sections, we will clarify our concern about data quality and potential defects whenever they are relevant.


\section{Results}\label{sec:results}
We present in Figure \ref{fig:poststamp1} and \ref{fig:poststamp2} the SMA images of the 1.3 or 1.1 mm continuum emission of the observed sources.
For FU\,Ori, Z\,CMa and V1057\,Cyg, we provide a comparison of our available millimeter images with the previously reported Subaru-HiCIAO near infrared coronagraphic polarization intensity images (Liu et al. 2016b) in Figure \ref{fig:hiciao}.
Details of the resolved continuum emission structures are described in Section \ref{sub:structure}.
The continuum flux measurements for the circumstellar material are introduced in Section \ref{sub:mmflux}.
Our flux measurements are summarized in Table \ref{tab:summary}.
Some of our target sources were observed at the same or similar frequencies in multiple  epochs.
We discuss their millimeter flux variability/stability in Section \ref{sub:mmvariability} and Appendix \ref{appendix:variability}.
The visibility amplitudes for sources which are not discussed in the text are provided in Appendix \ref{appendix:uvamp}.

\begin{figure}
\begin{tabular}{ c }
\includegraphics[width=9.5cm]{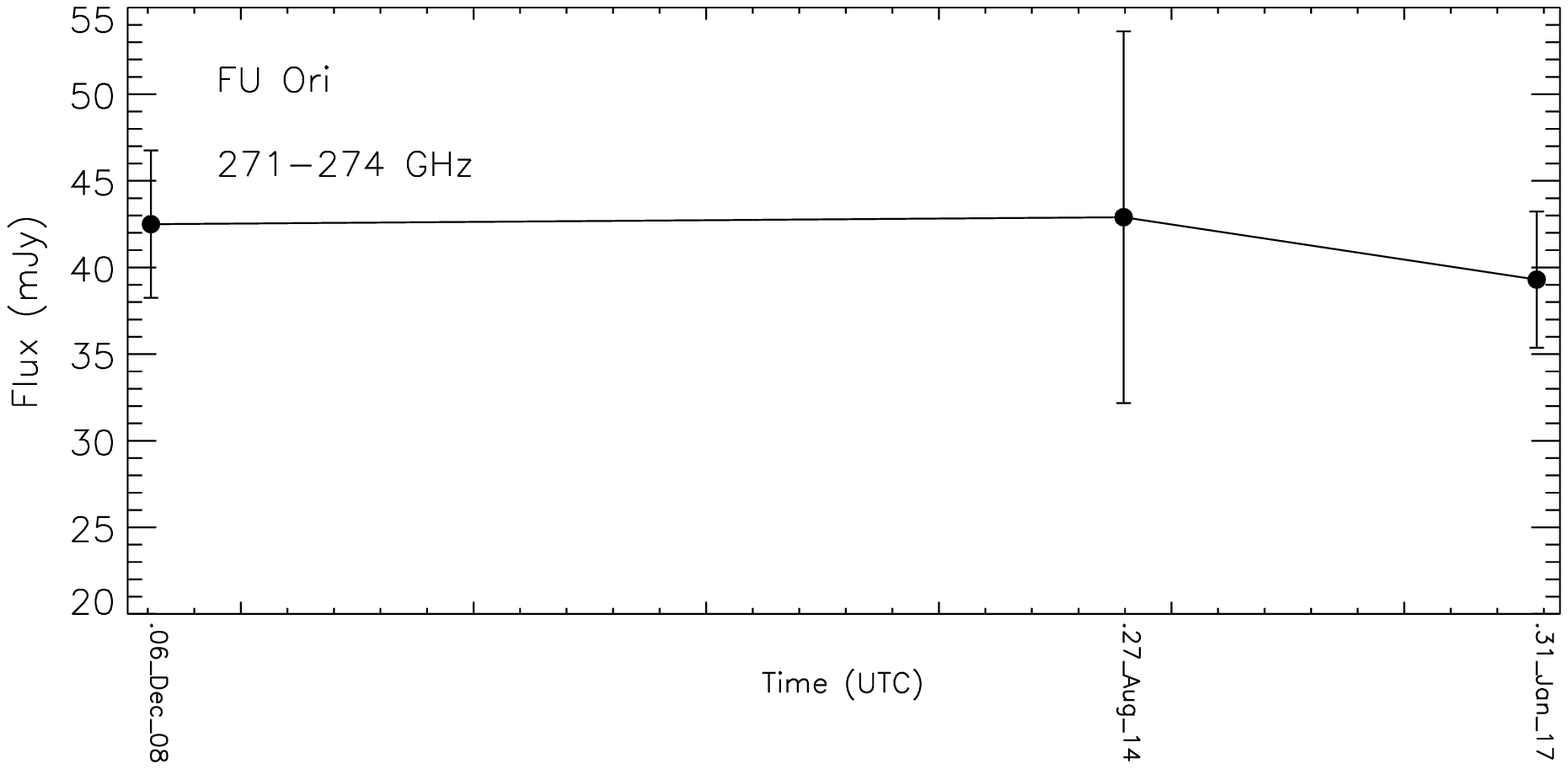} \\
\includegraphics[width=9.5cm]{0530135_fluxtime_270.eps} \\
\includegraphics[width=9.5cm]{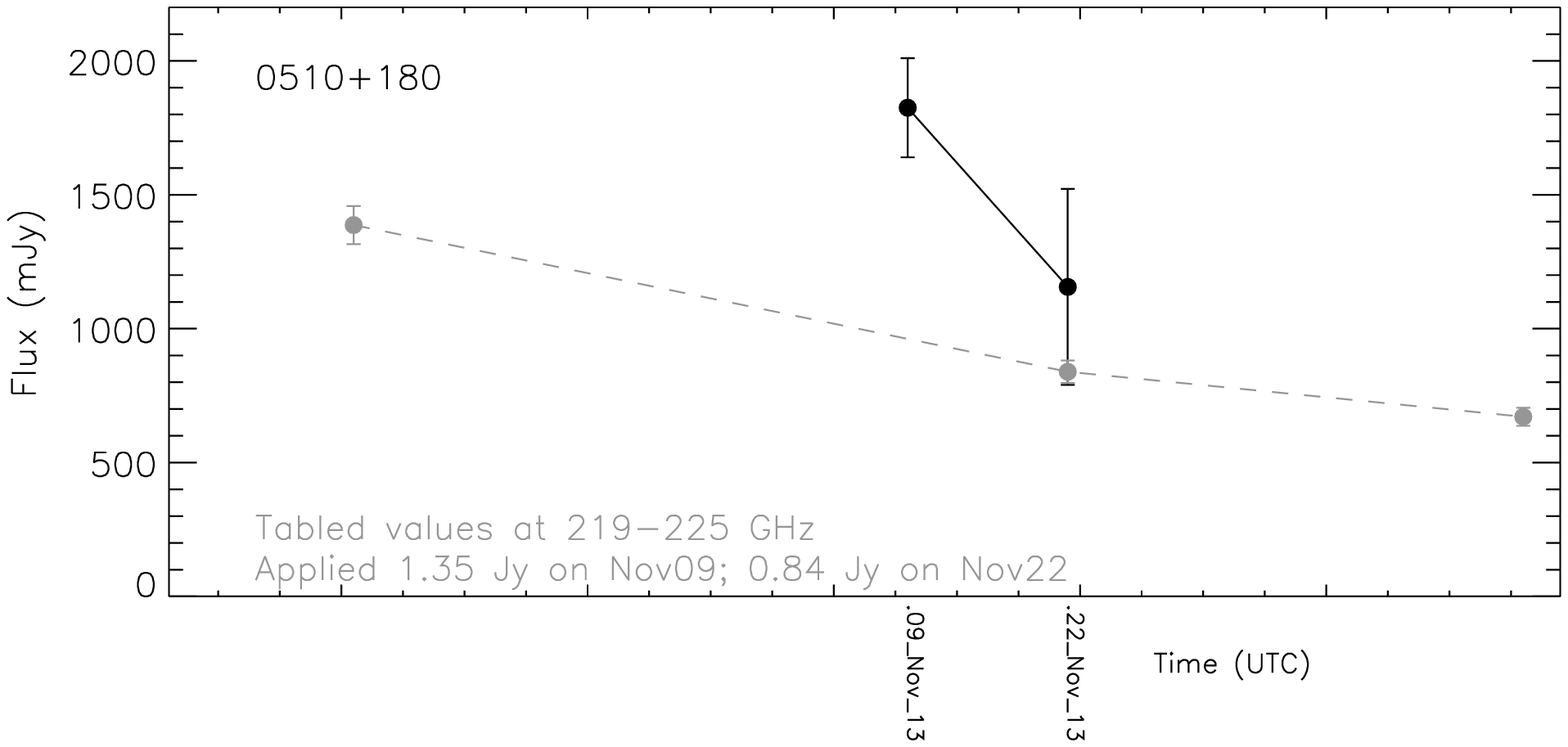} \\
\end{tabular}
\caption{\footnotesize{
Similar to Figure \ref{fig:fluxvarCygnus}, however, for the target source FU\,Ori and is presenting the measurements at 271-274 GHz.
Horizontal axes of the top and middle panels are presented on the same scale.
There were only two epochs of 225 GHz observations on FU\,Ori.
However, one of them (November 09, 2013) was taken at an extremely poor weather condition ($\tau_{\mbox{{\tiny 225 GHz}}}$ $\sim$ 0.4), such that the absolute flux calibration was very uncertain.
Middle and bottom panels present the gain calibrators for the 271-274 GHz observations (0530+135) and the 225 GHz observations (0510+180), respectively.
}}
\label{fig:fluxvarfuori}
\end{figure}

\begin{figure}
\begin{tabular}{ c }
\includegraphics[width=9.5cm]{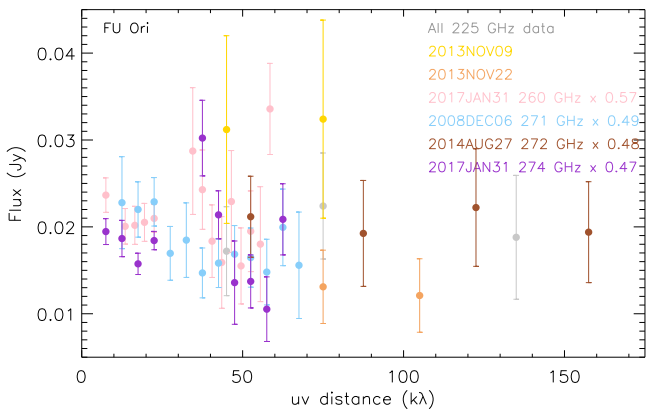} \\
\end{tabular}
\vspace{-0.6cm}
\caption{\footnotesize{
Visibility amplitudes of FU\,Ori (similar to Figure \ref{fig:uvampCygnus}).
We scaled the observed amplitudes at frequencies higher than 225 GHz to the expected amplitudes at this frequency, by assuming the spectral index $\alpha$ $\sim$3.8 (see discussion of Liu et al. 2017).
}}
\label{fig:uvampfuori}
\end{figure}

\subsection{Continuum structures}\label{sub:structure}
XZ\,Tau was observed with a short integration time thus was limited by the poor {\it uv} coverage.
Since a very bright millimeter emission source HL\,Tau was located at the edge of the SMA field of view when we observed XZ\,Tau, our image of XZ\,Tau is very seriously confused with imaging defects.
From Figure \ref{fig:poststamp1} we see that XZ\,Tau\,A and B are enclosed in a 3-$\sigma$ contour.
We tentatively consider that XZ\,Tau\,A and B are detected in the SMA observations, although the reported flux in Table \ref{tab:summary} should be regarded as an upper limit due to the confusion.
Osorio et al. (2016) reported that the 1.3 mm flux from the inner 3 AU radius around XZ\,Tau\,B is 7$\pm$2 mJy, although this source is located outside of the primary beam of their reported ALMA observations.
The sources RNO\,1B and 1C are located in a very condensed low-mass cluster-forming environment (Anglada et al. 1994; Quanz et al. 2007a; Ju\'{a}rez et al. in prep.)
Our SMA observations towards these sources show that RNO\,1C is associated with a $\sim$5000 AU scale clumpy gas/dust toroid; and RNO\,1B is associated with a $\sim$5000 AU scale gas/dust arm which connects to the clumpy toroid from the south (Figure \ref{fig:poststamp1}).
Due to the limited angular resolution of our SMA observations, we cannot separate the millimeter emission of the cirucmstellar material of RNO\,1B and 1C from these extended structures.
Therefore, their millimeter fluxes are considered as upper limits.
We do not detect millimeter continuum emission from VY\,Tau, V1143\,Ori, AR\,6a, AR\,6b, HBC\,722, IRAS\,20588+5215N, V1735\,Cyg, and V733\,Cep.

For the rest of the observed sources, SVS\,13A, PP\,13S, Haro\,5a/6a\,IRS, V2775\,Ori, NGC\,2071\,MM3, and Z\,CMa are resolved to be associated with $\gtrsim$1000 AU scales, spatially   asymmetric dust/gas structures, which may be the innermost parts of their residual circumstellar envelope.
This is consistent with the fact that these sources were previously considered as Class\,I YSOs (Table \ref{tab:summary}).
It is worth noting that such extended arm-like feature was also spatially resolved in the  the Atacama Large Millimeter Array (ALMA) observations of $^{13}$CO 2-1 towards the nearby low-mass YSO, HL\,Tau (Yen et al. 2017; see also Welch et al. 2000).
L1551\,IRS\,5 is associated with a $>$1000 AU scale envelope (e.g., Saito et al. 1996; Looney et al. 1997; Momose et al. 1998; Chou et al. 2014), which however was largely filtered out in our SMA continuum observations due to missing short-spacing data.
The millimeter emission of the rest of the detected sources appears to be dominated by spatially compact (e.g., disk-like) structures; some of them are spatially marginally resolved (V883\,Ori, NY\,Ori, V1057\,Cyg).
V2494\,Cyg and V2495\,Cyg present spatially unresolved compact sources immediately surrounded by spatially asymmetric or clumpy structures.
The higher angular resolution (0\farcs05-0\farcs5) millimeter continuum observations of FU\,Ori, V883\,Ori, V2775\,Ori, XZ\,Tau and PP\,13S have been reported by Hales et al. (2015), Liu et al. (2017), Cieza et al. (2016), Zurlo et al. (2017), Osorio et al. (2016), and P\'erez et al. (2010).

All objects with elongated emission appear to be of Class I type, implying possible interactions with parental clouds/filaments. 
Whether these elongated structures present injections of gas/dust from surroundings or ejections from the disk (e.g., Vorobyov 2016) remains to be understood (Akiyama et al. 2017, in prep.)
In the SMA image created at lower angular resolution (Figure \ref{fig:hiciao}), the extended millimeter emission around Z\,CMa appears more significant and presents a $\gtrsim$4000 AU scale arm-like structure extending from Z\,CMa toward the south.
The previously reported arm-like structure around Z\,CMa in the near infrared polarization coronagraphic image (Canovas et al. 2015; Liu et al. 2016b) is projectedly the innermost part of this $\gtrsim$4000 AU scale arm-like structure.
The previous JCMT-SCUBA 450 $\mu$m and 850 $\mu$m continuum observations show that these structures are further connected with a $\sim$2$'$ ($\sim$10$^{5}$ AU) scale elongated gas structure which has an east-west alignment (Sandell \& Weintraub 2001).
The lower angular resolution SMA image of Haro\,5a/6a\,IRS also revealed a $>$3000 AU scale arm-like structure which extends from Haro\,5a/6a\,IRS to the south (Figure \ref{fig:haro5a6a}).
The dominant millimeter emission source around V1057\,Cyg is spatially compact in our SMA observations.
However, we detected several millimeter emission clumps south of it (Figure \ref{fig:hiciao}).
The most significant one is marginally spatially resolved, which may have $\sim$500-1000 AU size scales. 
The previous JCMT-SCUBA 450 $\mu$m and 850 $\mu$m continuum observations show that V1057\,Cyg is associated with a $\gtrsim$2$'$ ($\sim$7.2 $\times$ 10$^{4}$ AU) scale elongated gas structure which has a north-south alignment (Sandell \& Weintraub 2001).
SVS\,13A and PP\,13S are also connected with exterior gas filaments (Sandell \& Knee 2001; Sandell \& Weintraub 2001).
Our SMA observations do not have fine enough {\it uv} coverages to address the detailed geometry of the extended emission on a few arcsecond scales around SVS\,13A and PP\,13S (Figure \ref{fig:poststamp1}).




\subsection{Millimeter fluxes}\label{sub:mmflux}

We measured millimeter fluxes of the detected sources in the image domain by performing two-dimensional Gaussian fits.
The synthesized beams of individual SMA images, and the obtained millimeter source sizes, fluxes, and positions, are summarized in Table \ref{tab:summary}.
We note that fitting fluxes from visibility amplitudes cannot be systematically applied to all of our targets, since some structures around them can significantly contribute to the visibility amplitudes, and it is not easy to subtract them off due to our limited {\it uv} coverage.
Based on previous radio observations towards a subset of our sample (Rodriguez et al. 1990; Liu et al. 2017), and the previous surveys of radio emission from Class 0-III YSOs (e.g., Liu et al. 2014; Dzib et al. 2015), we consider the millimeter flux contributed from free-free emission to be negligible.

We measured the flux of the central compact source using the peak intensity per synthesized beam (i.e., the  `peak intensity' column from Table \ref{tab:summary}), as these values are approximately the same (as long as the spatial extension is absent or marginal), but the peak intensity measurements suffer less from contaminating emission from more extended components.
Figure \ref{fig:fluxsummary} compares the fluxes of the central compact objects embedded in the observed sources.
The binary or multiple systems which cannot be spatially resolved with our SMA observations are presented with only one symbol.
The integrated fluxes derived from two-dimensional Gaussian fits can be significantly biased by the extended structures, which also depend on the angular resolution of our images (for some discussion see Dunham et al. 2014).
On the other hand, we found that the `peak intensity' returned from the two-dimensional Gaussian fit tends to be very close to the maximum pixel value in the image, which is less or not sensitive to angular resolution in most of our observed cases.
However, for the case of L1551\,IRS\,5, we quoted its `integrated flux' instead since it is located at a nearer distance ($\sim$140 pc) than most of the observed sources.
The shortest {\it uv} distance for L1551\,IRS\,5 ($\sim$30 $k\lambda$) traces a similar spatial scale as the $\sim$100 $k\lambda$ baseline observations on sources in the Orion molecular cloud ($d\sim$420 pc).
Quoting the 'peak intensity' for L1551\,IRS\,5 will reduce its flux by 30\%, which will not significantly affect our scientific discussion. 
For the other marginally spatially resolved sources V1057\,Cyg, Z\,CMa, NGC\,2071\,MM3, V2775\,Ori, V883\,Ori, NY\,Ori, Haro\,5a/6a\,IRS, SVS\,13A, and PP\,13S (Table \ref{tab:summary}), quoting `peak intensity' potentially leads to an underestimate of the millimeter flux of their circumstellar material by $\sim$20\%.
Given the widely spread millimeter luminosity of our observed sample over a range of $\gtrsim$3 orders of magnitude, such 20\% underestimates will not significantly bias the presentation of Figure \ref{fig:fluxsummary} and our following discussion.

When generating Figure \ref{fig:fluxsummary}, we have scaled the observed flux of Parsamian\,21 from 272 GHz to 225 GHz by assuming a spectral index $\alpha$ $=$ $3.8$.
Our assumption of $\alpha$ is motivated by the previous observations towards FU\,Ori (Liu et al. 2017; for a review of dust opacity see Draine 2003).
For scaling 272 GHz measurement to 225 GHz measurement, assuming the extreme $\alpha$ values 4.0 and 2.0, the scaling factors are 0.47 and 0.68, respectively.
This is merely a $\pm$10\% error thanks to the small frequency range we probed,
and is not significant as compared with our assumed uncertainties.
In addition, we scaled the plotted flux of each source by ($d$ [pc] / 353 [pc])$^{2}$, where $d$ is the source distance and 353 pc is the parallax distance of FU\,Ori (Gaia Collaboration 2016).
The vertical error bars incorporate a normal 10\% distance error for Taurus sources and 20\% for the rest of the sources, and the measurement errors (Section \ref{sec:observations}).
For a sense of the distance uncertainty we note that prior to the Gaia data release, FU\,Ori was usually considered to be at a 450 pc distance (Auddard et al. 2014) and that the updated distance by Gaia is $\sim$78\% of it.
We denote the rescaled flux by $F_{\mbox{\tiny 1.33 mm}}^{\mbox{\tiny 353 pc}}$.
We may potentially underestimate the distance of Z\,CMa by a factor of $\sim$2 (Audard et al. 2014; Gramajo et al. 2014), and thereby underestimate its $F_{\mbox{\tiny 1.33 mm}}^{\mbox{\tiny 353 pc}}$ by a factor of $\sim$4.

From Figure \ref{fig:fluxsummary}, we see that the millimeter emission of the archetype long duration outburst source, FU\,Ori, is in fact relatively faint as compared with the rest of the detected FUor or FUor-like objects.
The non-detections of V1143\,Ori and VY\,Ori constrained their millimeter luminosity to be considerably lower than that of FU\,Ori.
Our 3-$\sigma$ upper limits for HBC\,722 (see Dunham et al. 2012; K\'osp\'al et al. 2016; Dunham et al. in prep.), V733\,Cep, V1735\,Cyg, AR\,6a, AR\,6b, and IRAS\,20588+5125N are still consistent with the objects having a millimeter luminosity similar to that of FU\,Ori but are presently not yet detected either due to their larger distances or were observed with higher thermal noise levels. 
The previously reported silicate-emission type objects (FU\,Ori, Parsamian\,21, V1057\,Cyg, V1515\,Cyg, V1647\,Ori, XZ\,Tau) appear systematically less luminous than the silicate-absorption type objects (Z\,CMa, L1551\,IRS\,5, RNO\,1B, RNO\,1C, V1735\,Cyg, V883\,Ori) at 1\,mm, although their millimeter luminosity distributions overlap (for more discussion see Quanz et al. 2007b).

In Figure \ref{fig:fluxsummary}, we also quote the formulation that Ansdell et al. (2017) used to convert 1.33\,mm flux to the dust mass of circumstellar disk based on an optically thin assumption and an assumption of a fixed dust temperature $T_{\mbox{\scriptsize dust}}=$20 K.
We quote the derived dust masses from the ALMA surveys towards the Taurus, Lupus, Cha\,I, and Upper Sco molecular clouds from Pascucci et al. (2016), which show that the dust masses of the Class\,II YSO disks are mostly in the range of 1-100 $M_\oplus$.
In Figure \ref{fig:fluxsummary}, we limit the presentation of those ALMA surveys to sources for which the stellar mass is higher than 0.3 $M_{\odot}$.
Class II YSOs with $<$0.3 $M_{\odot}$ stellar mass tend to have lower millimeter luminosity, and therefore the ALMA samples were rather incomplete.
On the other hand, the previously estimated stellar masses for FUors/EXors are mostly higher than 0.3 $M_{\odot}$ although these estimates are very uncertain (for more discussion see Gramajo et al. 2014, and references therein).
We note that Ansdell et al. (2017) adopted the dust opacity $\kappa_{\mbox{\scriptsize 1000 GHz}}=$10 cm$^{2}$\,g$^{-1}$ and the opacity spectral index $\beta=$1.0, which corresponds to $\kappa_{\mbox{\scriptsize 230 GHz}}=$2.3 cm$^{2}$\,g$^{-1}$.
At 230 GHz, the dust opacity adopted by Pascucci et al. (2016) was the same with that adopted by Ansdell et al. (2017), in spite that Pascucci et al. (2016) adopted $\beta=$0.4.
In addition, Pascucci et al. (2016) has used self-consistent inputs and evolutionary models to estimate disk dust masses and stellar masses.
The millimeter luminosity of our observed accretion outburst YSOs appear systematically brighter than the average Class\,II YSOs.
We note that following the $\beta$=1 (i.e., $\alpha$=3) assumption of Ansdell et al. (2017) will infer still lower 225 GHz fluxes for the Class II YSOs which were observed by ALMA only at 345 GHz.

We caution that the dominant 1.33\,mm emission sources in some of our targets can be optically very thick, such that adopting an optically thin assumption can lead to an underestimate of dust mass by one order of magnitude (e.g., FU\,Ori, Liu et al. 2017; see also Dunham et al. 2014, Osorio et al. 2016, and Evans et al. 2017).
This effect is subtle, and may not be limited only to the case of accretion outburst YSOs.
The 0.85-1.3\,mm SED of FU\,Ori is indeed consistent with optically thin dust emission, whereas its condensed and optically very thick inner few AU scale disk may still hide $\sim$90\% of dust mass since it cannot significantly contribute to the overall $<$1 mm flux (Liu et al. 2017).
The previous Plateau de Bure Interferometry (PdBI) $\theta_{\mbox{\tiny maj}}$ $\times$ $\theta_{\mbox{\tiny min}}$ $=$2$\farcs$7$\times$2$\farcs$2 continuum observations on V1057\,Cyg at 2.75 mm from March 28 to June 22 of 2012 detected integrated flux and peak intensity of 4.9 mJy and 2.6 mJy\,beam$^{-1}$, respectively (Feh\'er et al. submitted).
If we combine all of our SMA 1.33 mm observations towards V1057\,Cyg, the image tappered  to the same angular resolution as that of the PdBI 2.75 mm image has a peak intensity of 18$\pm$2 mJy\,beam$^{-1}$.
Comparing the peak intensities at 1.33 mm and 2.75 mm implies a spectral index $\alpha$ $=$2.5-2.8 at this specific wavelength range. 
The previous PdBI $\theta_{\mbox{\tiny maj}}$ $\times$ $\theta_{\mbox{\tiny min}}$ $=$2$\farcs$4$\times$2$\farcs$2 continuum observations on V1735\,Cyg at 2.75 mm on April 05 and June 25 of 2014 detected integrated flux and peak intensity of 2.3 mJy and 1.8 mJy\,beam$^{-1}$, respectively (Feh\'er et al. 2017).
Comparing with our 3$\sigma$ detection limit of V1735\,Cyg at 1.33 mm implies an upper limit of spectral index $\alpha$ $=$1.7-2.0.
We cannot rule out the possibility that the derived very low spectral indices of V1057\,Cyg and V1735\,Cyg at 1.33-2.75 mm were confused by millimeter flux variability (more discussion see Section \ref{sub:mmvariability}).
Otherwise, the low spectral indices at 1.33-2.75 mm can be interpreted by the (partly) obscured, very optically thick hot inner disk of a few AU scales (e.g., Zhu et al. 2007) which is heated by the outbursts or outburst triggering (or related) mechanisms.
Hot inner disk presents in FU\,Ori (Liu et al. 2017) and may be common in Class 0/I stages (I-Hsiu Li et al. 2017).
Yet another possible interpretation for the low spectral indices is dust grain growth (Draine et al. 2006).
In the case that there are embedded hot inner disks, our reported $\sim$1\,mm fluxes approximately trace the dust mass outside of the millimeter photosphere of each YSO, which is also true for the case of FU\,Ori.
Without a spatially resolved image or a well sampled millimeter SED, the interpretation of the dust mass outside of the millimeter photosphere is inevitably model dependent.
More related discussion will be provided in Section \ref{sec:discussion}.
In the case of grain growth, dust opacity (Draine et al. 2006) will likely be a few times lower than what was assumed in Ansdell et al. (2017), which will lead to underestimates of the dust masses of our samples in Figure \ref{fig:fluxsummary}.
We note that the possibilities to interpret the low spectral indices of V1057\,Cyg and V1735\,Cyg are not mutually exclusive.
The few AU scales hot inner disk is possible to present flux variability on the orbital timescales of from a few months to few years, for example, if its density and/or thermal structures are altered by stellar accretion, or if it is heated by the time varying spiral shocks, viscous heating, or compression work.

\subsection{Millimeter variability/stability}\label{sub:mmvariability}
Some of our target sources were observed in multiple time epochs.
For not spatially resolved targets, or for resolved sources which were covered by observations with significantly overlapping {\it uv} coverages, we can constrain the variability/stability of their millimeter emission.
Since the observations were not specifically designed to quantify variability, the potential calibration or measurement errors require to be addressed in detail, which will be provided in Sections \ref{sub:cygnusvariability}-\ref{sub:FUorivariability}, and in Appendix \ref{sub:v1647variability}-\ref{sub:extendedvariability} for sources which are less well observed.

As a summary, we found that the 225\,GHz fluxes of V1057\,Cyg may have up to $\sim$20\% time variability; the 225\,GHz fluxes of V2494\,Cyg and V2495\,Cyg may have up to 30\%-60\% time variability from June 2013 to August 2014.
From May 2008 to October 2015, the 225\,GHz flux of V883\,Ori may vary by less than $\sim$10\%.
The 272 GHz flux of FU\,Ori is consistent with less than $\sim$10\% variability from December 2008 to January 2017.
V1647\,Ori is consistent with less than $\sim$10\% of 225\,GHz flux variability from November 2013 to October 2015.
NY\,Ori is consistent with $<$ $\pm$15\% of 225\,GHz flux variability from April 02 to 08 of 2013.
For NGC\,2071\,MM3 and Haro\,5a/6a\,IRS which were observed with limited {\it uv} coverages and were confused by extended emission, we tentatively consider that the former has less than $\pm$ $\sim$20\% millimeter flux variation in March 2014, and the latter has less than 10\% millimeter flux variation from March 2014 to September 2015.
For the rest of the observed sources, we do not have sufficient time sampling or data quality to address flux variability.

\subsubsection{V1057\,Cyg, V2494\,Cyg, and V2495\,Cyg}\label{sub:cygnusvariability}
V1057\,Cyg, V2494\,Cyg, and V2495\,Cyg were observed in the same tracks on June 17 of 2013, and August 17 and 20 of 2014. 
In addition, V1057\,Cyg and V2494\,Cyg were observed in a track on June 04 of 2014.
The gain phase and amplitude calibrator for all these sources in all tracks was 3C418, which was bright (0.8-1 Jy) at 225 GHz during these observations.
3C418 is 7.8$^{\circ}$, 3.3$^{\circ}$, and 6.0$^{\circ}$ separated from V1057\,Cyg, V2494\,Cyg, and V2494\,Cyg, respectively.
The weather conditions on June 17 of 2013 and June 04 of 2014 were the averaged condition  for the SMA operations at 225 GHz; the weather conditions on August 17 and 20 of 2014 were excellent (Table \ref{tab:obs}).

For each of these three sources, we imaged each epoch of observations separately but selected an identical shortest {\it uv} distance.
In Figure \ref{fig:fluxvarCygnus} we plot the fluxes measured in the image domain via Gaussian fits.
In addition, we shifted the phase referencing centers of our observations to the locations of the millimeter continuum emission peaks in post processing using the {\tt uvedit} task of the Miriad software package, and then measured the azimuthally vector-averaged visibility amplitudes using the {\tt uvamp} task of Miriad.
Since the phase referencing centers for the observations of these three sources, the stellar positions, are very close to the millimeter emission peaks (Figure \ref{fig:poststamp1}, \ref{fig:poststamp2}), whether or not we shift the phase referencing centers in post processing will not significantly change the measured visibility amplitudes. 
The obtained visibility amplitudes are presented in Figure \ref{fig:uvampCygnus}.
The visibility amplitudes of V1057\,Cyg are rather complicated, in particular, at $<$50 $k\lambda$ {\it uv} distances, due to the contribution from the millimeter emission clumps adjacent to it (Figure \ref{fig:hiciao}).

From Figure \ref{fig:fluxvarCygnus}, we see that the measured fluxes on August 17 and 20 of 2014 are very well consistent, in spite of that these two tracks observed different absolute flux standard sources (Table \ref{tab:obs}).
The visibility amplitudes of V2494\,Cyg taken on June 17 of 2013 are only slightly higher than those taken on June 04 of 2014 at the overlapping {\it uv} distance range (Figure \ref{fig:uvampCygnus}).
However, the observed fluxes of V2494\,Cyg in August 2014 are approximately 1.4 times the observed flux on June 17 of 2013.
The fluxes of V2495\,Cyg in August 2014 are 1.6 times those on June 17 of 2013.
The fluxes of V1057\,Cyg in August 2014 may be 1.1 times the observed flux on June 17 of 2013, although the difference is consistent with our assumed measurement/calibration errors; its flux may drop by $\sim$20\% on June 04 of 2014, although $\sim$10\% of it may be attributed to calibration errors.

The observed flux variations of V2494\,Cyg and V2495\,Cyg are larger than what can be caused by the typical absolute flux calibration uncertainty of SMA at 225 GHz.
Our measured and applied flux values of 3C418 on June 04 and August 17 and 20 of 2014 are also very well consistent with the flux monitoring results of SMA (Figure \ref{fig:fluxvarCygnus}, bottom panel).
Figure \ref{fig:fluxvarCygnus} also shows that the fluxes of 3C418 varies no more than $\sim$10\% over the timescales of from a few days to few months. Therefore, our flux measurements are unlikely to be largely confused by the time variability of 3C418.
For spatially resolved sources, the differences in {\it uv} sampling can bias the observed fluxes.
However, the millimeter emission sources associated with V1057\,Cyg, V2494\,Cyg, and V2495\,Cyg are all spatially relatively compact.
Moreover, we in fact detect higher fluxes from observations with the more extended array configurations, which cannot be explained by the effect of missing short-spacing data. 
The effect of atmospheric phase decoherence (e.g., P\'erez et al. 2010) will also more likely degrade the fluxes observed from longer baselines.
Due to the consistent observations and calibrations for these three sources, the flux variations are hard to be fully attributed to calibration errors.
Therefore, we consider that millimeter flux variability was defected from at least two of these three sources.
We are also open to the possibility that there are subtle instrumentation issues (e.g., unexpected antenna pointing errors for target sources but not calibrators) that were not identified or cannot be diagnosed by us, which might lead to larger flux uncertainties than what we assumed.

\subsubsection{V883\,Ori}\label{sub:v883variability}
We performed analyses which are similar to those outlined in Section \ref{sub:cygnusvariability}, and present the results in Figure \ref{fig:fluxvarv883ori} and \ref{fig:uvampv883}.
The measured fluxes on May 27 and June 17 of 2008 are consistent to within our assumed measurement errors.
In addition, the visibility amplitudes measured with SMA are also very well consistent with the integrated fluxes taken by ALMA at the same frequency on December 12 of 2014 and April 05 of 2015.
Moreover, we scaled the observed fluxes on December 06 of 2008 and August 27 of 2014 at 272 GHz to 225 GHz by assuming the spectral index $\alpha$=4.0 (see Cieza et al. 2016), and found that they are also consistent with the 225 GHz measurements to within the assumed measurement error.
These observations constrain the millimeter flux variability of V883\,Ori from 2008 to 2015 to less than $\sim$10\%.

\subsubsection{FU\,Ori}\label{sub:FUorivariability}
We present the observed fluxes and visibility amplitudes of FU\,Ori in Figure \ref{fig:fluxvarfuori} and \ref{fig:uvampfuori}.
(Sub)millimeter emission around FU\,Ori cannot be spatially resolved by any of the presented SMA observations (see Hales et al. 2015 and Liu et al. 2017 for higher angular resolution (sub)millimeter images).
Therefore, we imaged individual epochs of observations without limiting the {\it uv} distance range, and then measured fluxes in the image domain by performing two-dimensional Gaussian fits.

There were two epochs of 225 GHz observations.
However, the first epoch was taken at poor weather conditions with $\tau_{\mbox{\tiny 225 GHz}}$  $\sim$0.4 (Table \ref{tab:obs}), thus the absolute flux scaling cannot be derived reliably.
Therefore, our analysis of variability focuses on the 272 GHz fluxes measured on December 06 of 2008, August 27 of 2014, and January 31 of 2017, which are 43, 43 and 39 mJy, respectively.
All of these measurements are consistent within our assumed measurement error.
We constraint the 272 GHz flux variability of FU\,Ori to at most $\sim$10\%.
Our result is consistent with Green et al. (2016b) who claimed that FU\,Ori does not present infrared variability over the last decade.

\begin{figure*}[!]
\hspace{-0.7cm}
\begin{tabular}{ p{8.5cm} p{8.5cm} }
\includegraphics[width=10.5cm]{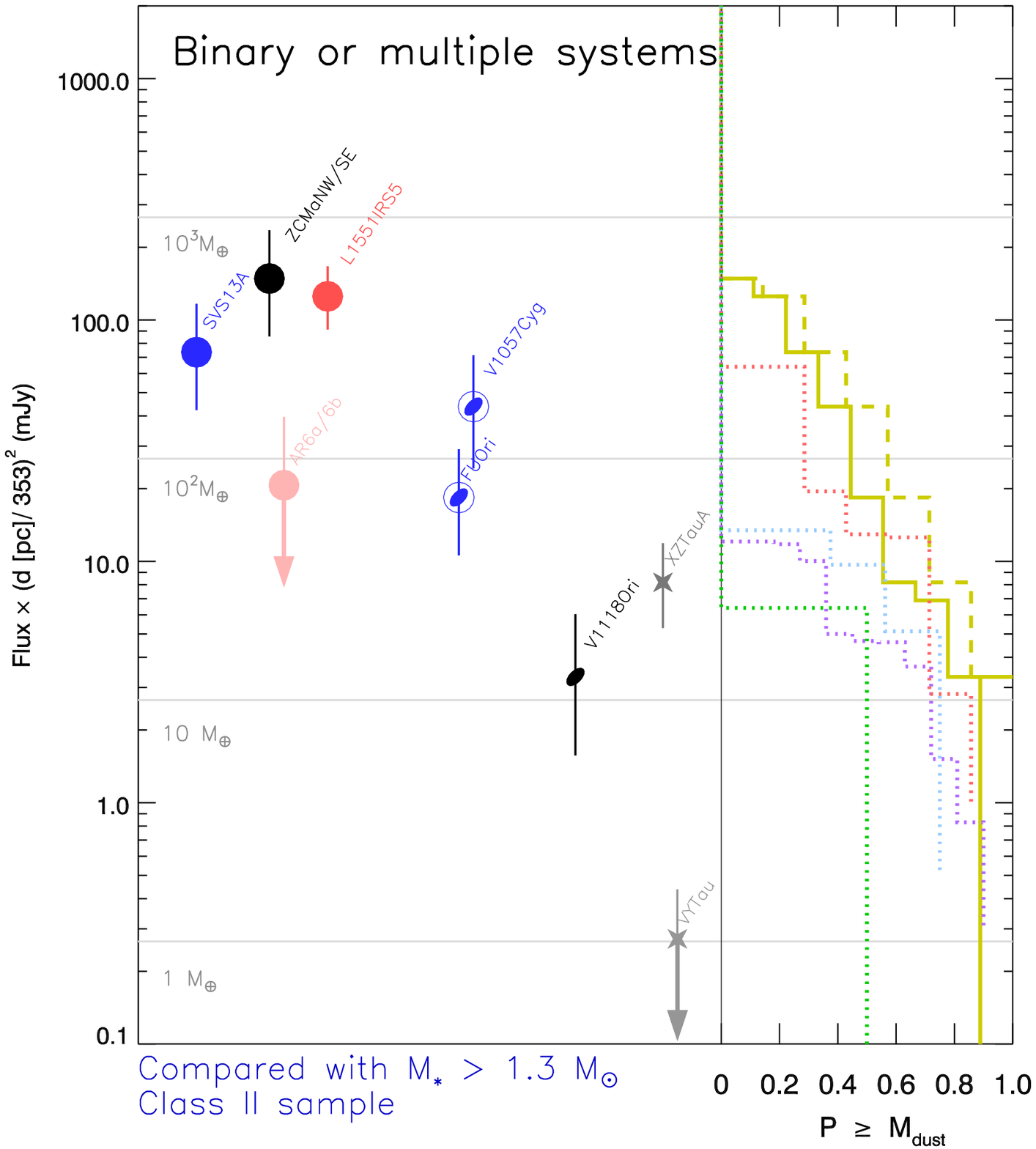} &
\includegraphics[width=10.5cm]{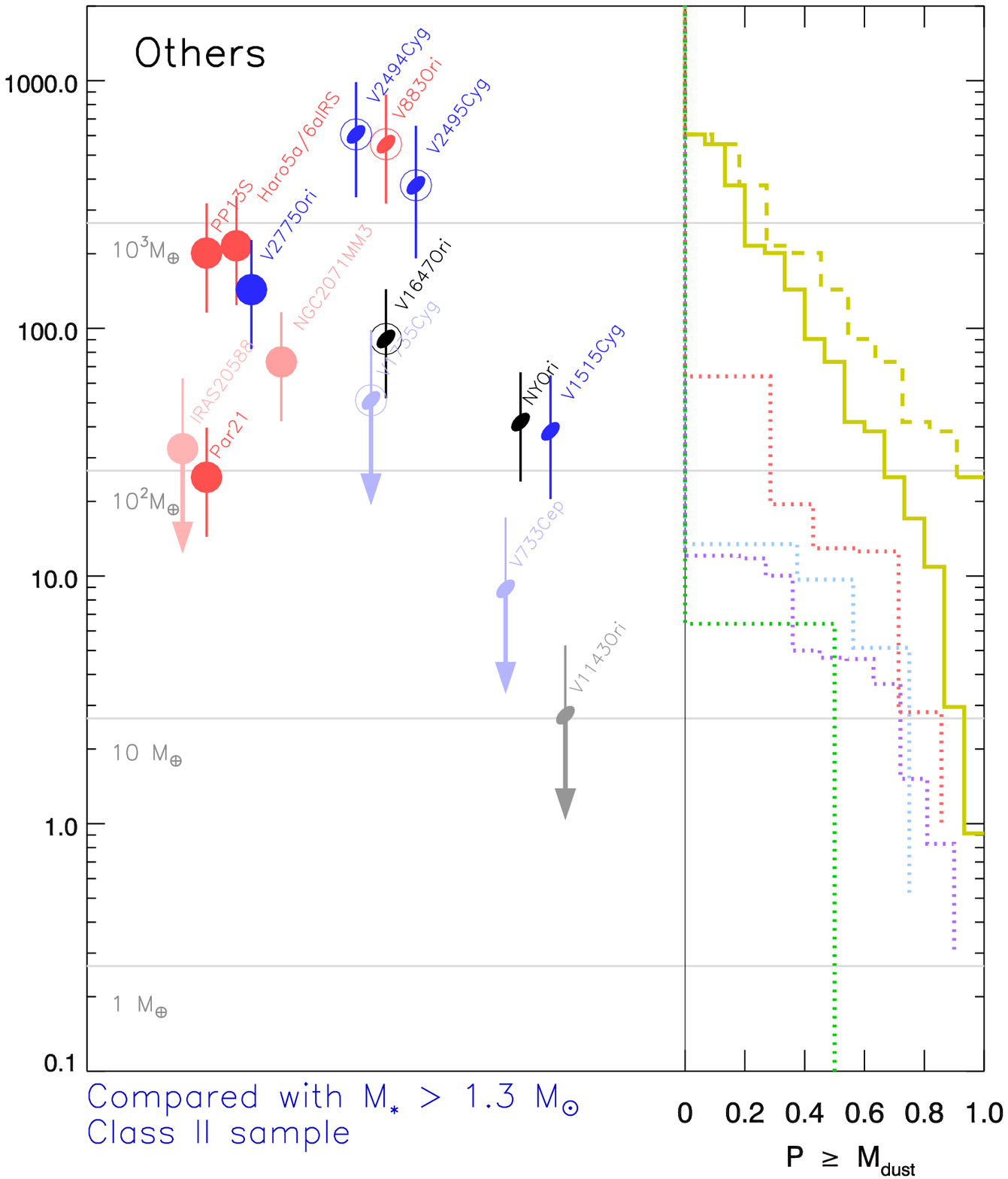} \\
\end{tabular}
\caption{\footnotesize{
Similar to Figure \ref{fig:fluxsummary}. In this figure, we plot the confirmed binary or multiple sources in the left panel, and plot the rest of the sources in the right panel. 
We exclude RNO\,1B, RNO\,1C, and HBC\,722 which are located in condensed stellar cluster forming regions (Figure \ref{fig:poststamp1}; see also Dunham et al. 2012).
Purple, light blue, red, and green dotted lines present the $M_{*}>$1.3 $M_{\odot}$ Class II objects in the Taurus, Lupus, Chameleon I, and Upper Sco regions, respectively (quoted from Pascucci et al. 2016; the original observations can be found in Andrews et al. 2013, Carpenter et al. 2014, Ansdell et al. 2016, Barenfeld et al. 2016, and Pascucci et al. 2016).
}}
\label{fig:fluxcomparison}
\end{figure*}

\section{Discussion}\label{sec:discussion}
To understand better the observed distribution of $F_{\mbox{\tiny 1.33 mm}}^{\mbox{\tiny 353 pc}}$, in Figure \ref{fig:fluxcomparison} we plot the confirmed binary or triple systems separately from the rest of the sources.
In addition, we exclude the sources which are in very condensed cluster-forming environments (RNO\,1B, RNO\,1C, HBC\,722).
We found that millimeter emission of the confirmed binary or multiple systems may be systematically lower than that of the rest of the sources. 
It is also possible that there is an additional, millimeter bright population in the later samples, for example, V883\,Ori, V2494\,Cyg and V2495\,Cyg.
We note that the samples presented in the right panel of Figure \ref{fig:fluxcomparison} may contain unresolved binary or multiple systems, which follow a $F_{\mbox{\tiny 1.33 mm}}^{\mbox{\tiny 353 pc}}$ distribution similar with what is presented in the left panel.
We refer to Harris et al. (2012) who reported that multiple YSO systems with $>$300 AU separations show similar (sub-)millimeter emission properties with single ones; multiple YSO systems with $<$300 AU separations are in general considerably fainter at (sub-)millimeter bands, although there are exceptions which posses millimeter bright circumbinary disks.

The different $F_{\mbox{\tiny 1.33 mm}}^{\mbox{\tiny 353 pc}}$ distributions of the two samples can be seen more clearly in the comparison with the cumulative mass distribution derived from the ALMA surveys of Class II disks (Figure \ref{fig:fluxcomparison}).
The cumulative mass distribution function of the samples of confirmed binary or multiple systems resembles that of the $M_{*}>$1.3 $M_{\odot}$ Class II objects in the Cham\,I region.
We note that the $M_{*}>$1.3 $M_{\odot}$ Class II objects are among the highest mass but rare populations of the samples previously surveyed by ALMA. They tend to have higher millimeter luminosity than the lower stellar mass ones.
Pascucci et al. (2016) and Ansdell et al. (2017) based on the ALMA surveys towards Taurus, Lupus, Upper Sco, $\sigma$ Orionis, and Cham\,I regions suggested that $F_{\mbox{\tiny 1.33 mm}}^{\mbox{\tiny 353 pc}}$ and $M_{*}$ may be correlated, and may be approximately described by the power laws $F_{\mbox{\tiny 1.33 mm}}^{\mbox{\tiny 353 pc}}$[mJy]$\propto$($M_{*}$ [$M_{\odot}$])$^{\gamma}$, where $\gamma$ may be in the range of $\sim$1.5-3.
The masses of the host protostars in our FUor(-like) and EXor samples are not well determined due to that spectral-typing for the protostars is seriously confused by the emission from the hot inner disk (e.g., Zhu et al. 2007).
We cannot know whether or not the millimeter bright sources in our samples possess higher mass host YSOs.
We also do not yet know whether or not the correlation of $F_{\mbox{\tiny 1.33 mm}}^{\mbox{\tiny 353 pc}}$ and $M_{*}$ should hold in the FUor(-like) and EXor samples.
The protostellar mass of the millimeter brightest FUor-like object, V883\,Ori, was constrained to be $\sim$1.3 $M_{\odot}$ by the Keplerian rotation curve of CO gas (Cieza et al. 2016), which may be a relatively massive YSO in our sample but not extreme.
However, we caution that interpreting the higher millimeter luminosity of the observed FUors/EXors, compared to the $M_{*}>$1.3 $M_{\odot}$ Class II objects, as a consequence of a higher average stellar mass of the FUors/EXors sample, may require extrapolating the correlation between $F_{\mbox{\tiny 1.33 mm}}^{\mbox{\tiny 353 pc}}$ and $M_{*}$ to a stellar mass beyond the range for which this correlation was measured (c.f., Boissier et al. 2011; Andrews et al. 2013).

Although many of the seven observed EXors show lower values of $F_{\mbox{\tiny 1.33 mm}}^{\mbox{\tiny 353 pc}}$ (Figure \ref{fig:fluxcomparison}), it is not yet clear to us how the outburst nature (e.g., long duration, short duration and repetitive, FUor-like) is related to the value of $F_{\mbox{\tiny 1.33 mm}}^{\mbox{\tiny 353 pc}}$.
They may not be related, which may imply that the outburst nature is determined by the physical mechanisms on unresolved spatial scales (e.g., magnetosphere or inner circumstellar disk).
The millimeter flux variability that we potentially have detected from V1057\,Cyg, V2494\,Cyg, and V2495\,Cyg may indicate that the thermal or density structures of the dusty disks in the inner AU scale regions around these sources are perturbed.
We refer to Johnstone et al.(2013) and Yoo et al. (2017) who argued that the millimeter flux variability can also be due to the variation of protostellar irradiation.
K{\'o}sp{\'a}l et al. (2011a) reported the 1.25 mm flux variability over a timescale of a few days of the EXor source UZ\,Tau\,E, which is binary.
However, it is not clear whether the 1.25 mm flux is dominated by dust emission.

There are four sources (VY\,Tau, XZ\,Tau, V1118\,Ori, V1143\,Ori) which our SMA observations have constrained to have considerably lower $F_{\mbox{\tiny 1.33 mm}}^{\mbox{\tiny 353 pc}}$ values than FU\,Ori (Figure \ref{fig:fluxcomparison}).
All of them are EXors.
In particular, VY\,Tau may be a Class\,III source which possess considerably lower disk mass than the rest of our samples (Liu et al. 2016a).
The accretion outbursts may only require some mass to be concentrated immediately around the host YSOs instead of requiring the bulk of circumstellar disk to be very massive, although the presence of a very massive disk naturally make it easier to fulfill the required mass concentration to feed accretion outbursts.
We emphasize that the observed $F_{\mbox{\tiny 1.33 mm}}^{\mbox{\tiny 353 pc}}$ traces dust mass outside of the millimeter photosphere (e.g., the $\tau=$1 boundary).
The sources which have low $F_{\mbox{\tiny 1.33 mm}}^{\mbox{\tiny 353 pc}}$ values can still possess massive but compact and optically thick disks (e.g., XZ\,Tau\,B, see Osorio et al. 2016; see also Figure 12 of Zhu et al. 2010).
In addition, it becomes rather difficult to detect such compact disks unless they are very significantly heated, either by protostellar irradiation or mechanic works.
Capturing exterior, infalling gas arms/streams/clumps may replenish the circumstellar disk mass, which is conducive to triggering accretion outbursts.
Either disk gravitational instability (e.g., Vorobyov \& Basu 2010; Vorobyov \& Basu 2015; Mercer \& Stamatellos 2017; Tsukamoto et al. 2017) or binary interaction (e.g., Bonnell \& Bastien 1992; Pfalzner 2008; Nayakshin \& Lodato 2012) can help pile up gas and dust to small spatial scales.

Moreover, the sources which are resolved to be connected with dense, $\ge$10$^{3}$ AU scale millimeter emission filament(s) (e.g., SVS\,13, PP\,13S, NGC\,2071\,MM3) or arm(s) (e.g., Haro\,5a/6a\,IRS, ZCMa) are not necessarily particularly bright in our measured $F_{\mbox{\tiny 1.33 mm}}^{\mbox{\tiny 353 pc}}$.
On the other hand, many sources which show high $F_{\mbox{\tiny 1.33 mm}}^{\mbox{\tiny 353 pc}}$ values (e.g., V2494\,Cyg, V2495\,Cyg, V883\,Ori, V2775\,Ori, PP\,13S, Haro\,5a/6a\,IRS) are surrounded by circumstellar disk structures (or inner accreting envelope) that have spatial scales larger than few tens of AU (Figure \ref{fig:poststamp1}, \ref{fig:poststamp2}, \ref{fig:uvampCygnus}, \ref{fig:uvampv883}, \ref{fig:uvamp_many}; see also  P\'erez et al. 2010; Cieza et al. 2016), which are very different from the case of FU\,Ori which presents spatially compact millimeter emission (Figure \ref{fig:hiciao}; Hales et al. 2015; Liu et al. 2017).
For confirmed binary or triple systems, the compact emission may be explained by several physical mechanisms which are not mutually exclusive.
During the earlier evolutionary stage, members of the binary  may preferentially capture the infalling gas stream which has less relative motion with respect to them.
In other words, part of the initial angular momentum budget of the parental core goes to the orbital motion of the binary components.
Therefore, the circumstellar disks formed around each of the individual members may be initially small due to the lower accumulated angular momentum.
In addition, binary interaction may truncate large disks or may induce efficient inward migration of dust.
A disk will become faint in $F_{\mbox{\tiny 1.33 mm}}^{\mbox{\tiny 353 pc}}$ once the dust grains are migrated inside the millimeter photosphere.
On the other hand, if the $F_{\mbox{\tiny 1.33 mm}}^{\mbox{\tiny 353 pc}}$ bright sources are indeed isolated YSOs (V883\,Ori is very likely the case according to the ALMA image presented in Cieza et al. 2016) or very compact binary systems, then the brighter $F_{\mbox{\tiny 1.33 mm}}^{\mbox{\tiny 353 pc}}$ emission observed around the higher mass YSOs may be naturally understood by that a certain fraction of the accreted gas and dust to form YSOs is yet centrifugally supported at extended (e.g., $\sim$100 AU) regions (some theoretical discussion is presented in Vorobyov 2013, Kuffmeier et al. 2017, and references therein).
Accreting gas from large to small scales and onto protostars requires either the rotational motion of extended circumstellar disk or the orbital motion of the binary/multiple components to serve as reservoirs of angular momentum.
More insight can be gained by combining radiative transfer modeling and hydrodynamic simulations, which is not in the scope of the present manuscript.


\begin{table*}[h]{\scriptsize
\caption{\footnotesize{Summary for individual target sources}}\label{tab:summary}
      \begin{tabular}{  l  p{3.1cm} p{3.1cm} p{3.1cm} p{3.1cm}   }\hline\hline
Source name	&           
   RNO\,1B (V710\,Cas) & 	
   RNO\,1C  & 
   SVS\,13A (Per-emb-44 A/B) &
   PP\,13S  
   \\
   \hline
Evolutionary Class	    &
  Class I/II (cluster)   &
  Class I   (cluster) &
  Class I (binary)  &
  Class I   
  \\
stellar R.A. (J2000)	&
  00$^{\mbox{\scriptsize{h}}}$36$^{\mbox{\scriptsize{m}}}$46$^{\mbox{\scriptsize{s}}}$.05 &
  00$^{\mbox{\scriptsize{h}}}$36$^{\mbox{\scriptsize{m}}}$46$^{\mbox{\scriptsize{s}}}$.65 &
  03$^{\mbox{\scriptsize{h}}}$29$^{\mbox{\scriptsize{m}}}$03$^{\mbox{\scriptsize{s}}}$.759  &
  04$^{\mbox{\scriptsize{h}}}$10$^{\mbox{\scriptsize{m}}}$41$^{\mbox{\scriptsize{s}}}$.119     
   \\
stellar Decl. (J2000)	&
  +63$^{\circ}$28$'$53$\farcs$29 &
  +63$^{\circ}$28$'$57$\farcs$90  &
  $+$31$^{\circ}$16$'$03$\farcs$99 &
  $+$38$^{\circ}$07$'$54$\farcs$41    
  \\
Spectral type		    &
  F8   &
  M (?)   &
  late-F to early-G (unknown type companion separated from $\sim$0$\farcs$3)   &
  mid-K (?)   
  \\
Onset (yr)				&
  1978   &
  FUor-like   &
  $>$1988, $<$1990   &
  FUor-like   
  \\
Outburst duration	 (yr)			&
  $\gtrsim$40   &
  $\cdots$   &
  $\gtrsim$30   &
  $\cdots$   
  \\
%
Assumed distance (pc)  &
   929  &
   929  &
   235  &
   350  
  \\
  \hline
Synthesized beam &
   1\farcs3$\times$1\farcs1; -25$^{\circ}$  &
   1\farcs3$\times$1\farcs1; -25$^{\circ}$  &
   1\farcs5$\times$0\farcs88; 43$^{\circ}$  &
   1\farcs7$\times$1\farcs0; 54$^{\circ}$  
  \\
 ($\theta_{\mbox{\scriptsize{maj}}}$ $\times$ $\theta_{\mbox{\scriptsize{min}}}$; P.A.)	 &
       &
       &
   [31-140 $k\lambda$]    &
       
    \\
Image RMS (mJy\,beam$^{-1}$)		&
  1.2   &
  1.2   &
  10.6   &
  7.0   
  \\
  \hline
mm R.A. (J2000)	&
  confused   &
  confused   &
  03$^{\mbox{\scriptsize{h}}}$29$^{\mbox{\scriptsize{m}}}$03$^{\mbox{\scriptsize{s}}}$.75   &
  04$^{\mbox{\scriptsize{h}}}$10$^{\mbox{\scriptsize{m}}}$41$^{\mbox{\scriptsize{s}}}$.12   
  \\
mm Decl. (J2000)	&
 confused    &
 confused    &
 $+$31$^{\circ}$16$'$03$\farcs$7    &
 $+$38$^{\circ}$07$'$54$\farcs$5    
  \\
Image component size  & 
 $\cdots$    &
 $\cdots$    &
 $1\farcs 7$ $\times$ $1\farcs 1$; 39$^{\circ}$    &
 $1\farcs 8$ $\times$ $1\farcs 1$; 49$^{\circ}$    
  \\
($FWHM{\mbox{\scriptsize{maj}}}$ $\times$ $FWHM{\mbox{\scriptsize{min}}}$; P.A.) &
     &
     &
  lightly confused   &
     
  \\
Peak intensity (mJy\,beam$^{-1}$)	&
 $\lesssim$6    &
 $\lesssim$12   &
 162$\pm$16    &
 200$\pm$20    
  \\
Peak S/N		&
 $\lesssim$5    &
 $\lesssim$10   &
 15    &
 29    
  \\
Integrated 1.3 mm Flux (mJy)	&
 $\cdots$    &
 $\cdots$    &
 225$\pm$23    &
 229$\pm$23    
  \\
  \hline
      \end{tabular}

\vspace{1cm}
      \begin{tabular}{  l  p{3.1cm} p{3.1cm} p{3.1cm} p{3.1cm}   }\hline\hline
Source name	&           
  L1551\,IRS5  & 	
  XZ\,Tau\,A      & 
  VY\,Tau      &
  V1118\,Ori  
   \\
   \hline
Evolutionary Class	    &
 Class I (binary)    &
 Class III (binary)  &
 Class III (binary) &
 Class II  (binary)  
  \\
stellar R.A. (J2000)	&
 04$^{\mbox{\scriptsize{h}}}$31$^{\mbox{\scriptsize{m}}}$34$^{\mbox{\scriptsize{s}}}$.08     &
 04$^{\mbox{\scriptsize{h}}}$31$^{\mbox{\scriptsize{m}}}$40$^{\mbox{\scriptsize{s}}}$.095     &
 04$^{\mbox{\scriptsize{h}}}$39$^{\mbox{\scriptsize{m}}}$17$^{\mbox{\scriptsize{s}}}$.412     &
 05$^{\mbox{\scriptsize{h}}}$34$^{\mbox{\scriptsize{m}}}$44$^{\mbox{\scriptsize{s}}}$.747     
   \\
stellar Decl. (J2000)	&
 $+$18$^{\circ}$08$'$04$\farcs$90     &
 $+$18$^{\circ}$13$'$56$\farcs$71     &
 +22$^{\circ}$47$'$53$\farcs$40     &
 $-$05$^{\circ}$33$'$42$\farcs$26     
  \\
Spectral type		    &
   Both mid-K ? (separated by 0$\farcs$3)  &
   M3 (one M0 comparion at $\sim$42 AU separation, and potentially an unknown type one at 13 AU separation) &
   M0 (with a M2-M4 companion, orbital period $>$350 yr)  &
   M2-M3 (with a companion of unclear spectral type, separation$\sim$76 AU)  
  \\
Onset (yr)				&
   FUor-like  &
   many  &
   many (1900-1970, 2013-present)  &
   many  
  \\
Outburst duration	 (yr)			&
  $\cdots$   &
  a few   &
  0.5-2   &
  $\sim$1.2   
  \\
Assumed distance (pc)  &
  140   &
  140   &
  140   &
  420
  \\
  \hline
Synthesized beam &
  1\farcs2$\times$0\farcs90; -85$^{\circ}$   &
  0\farcs54$\times$0\farcs39; 86$^{\circ}$   &
  $0\farcs59$ $\times$ $0\farcs40$; 75$^{\circ}$   &
  $0\farcs61$ $\times$ $0\farcs47$; 70$^{\circ}$   
  \\
 ($\theta_{\mbox{\scriptsize{maj}}}$ $\times$ $\theta_{\mbox{\scriptsize{min}}}$; P.A.)	 &
       &
       &
       &
       
    \\
Image RMS (mJy\,beam$^{-1}$)		&
  3.0   &
  17.0  &
  0.55   &
  0.6   
  \\
  \hline
mm R.A. (J2000)	&
  04$^{\mbox{\scriptsize{h}}}$31$^{\mbox{\scriptsize{m}}}$34$^{\mbox{\scriptsize{s}}}$.16   &
  $\cdots$   &
  $\cdots$   &
  05$^{\mbox{\scriptsize{h}}}$34$^{\mbox{\scriptsize{m}}}$44$^{\mbox{\scriptsize{s}}}$.75   
  \\
mm Decl. (J2000)	&
  $+$18$^{\circ}$08$'$04$\farcs$6   &
  $\cdots$   &
  $\cdots$   &
  $-$05$^{\circ}$33$'$42$\farcs$3   
  \\
Image component size  & 
  $1\farcs3$ $\times$ $1\farcs1$; 137$^{\circ}$   &
  $\cdots$   &
  $\cdots$   &
  $0\farcs7$ $\times$ $0\farcs37$; 17$^{\circ}$   
  \\
($FWHM{\mbox{\scriptsize{maj}}}$ $\times$ $FWHM{\mbox{\scriptsize{min}}}$; P.A.) &
     &
     &
     &
     
  \\
Peak intensity (mJy\,beam$^{-1}$)	&
  488$\pm$49   &
  3$\sigma$ $<$51   &
  3$\sigma$ $<$1.7   &
  2.3$\pm$0.6   
  \\
Peak S/N		&
  150   &
  $\cdots$   &
  $\cdots$   &
  3.8   
  \\
Integrated 1.3 mm Flux (mJy)	&
  780$\pm$78   &
  3$\sigma$ $<$51   &
  3$\sigma$ $<$1.7   &
  2.3$\pm$0.6   
  \\
  \hline
      \end{tabular}

\vspace{1cm}
      \begin{tabular}{  l  p{3.1cm} p{3.1cm} p{3.1cm} p{3.1cm}   }\hline\hline
Source name	&           
  Haro\,5a/6a\,IRS  & 	
  NY\,Ori           & 
  V1143\,Ori         &
  V883\,Ori  
   \\
   \hline
Evolutionary Class	    &
   Class I    &
   Class II   &
   Class II  &
   Class I/II   
  \\
stellar R.A. (J2000)	&
   05$^{\mbox{\scriptsize{h}}}$35$^{\mbox{\scriptsize{m}}}$26$^{\mbox{\scriptsize{s}}}$.560   &
   05$^{\mbox{\scriptsize{h}}}$35$^{\mbox{\scriptsize{m}}}$36$^{\mbox{\scriptsize{s}}}$.011   &
   05$^{\mbox{\scriptsize{h}}}$38$^{\mbox{\scriptsize{m}}}$03$^{\mbox{\scriptsize{s}}}$.896   &
   05$^{\mbox{\scriptsize{h}}}$38$^{\mbox{\scriptsize{m}}}$18$^{\mbox{\scriptsize{s}}}$.1   
   \\
stellar Decl. (J2000)	&
   $-$05$^{\circ}$03$'$55$\farcs$08   &
   $-$05$^{\circ}$12$'$25$\farcs$32   &
   $-$04$^{\circ}$16$'$42$\farcs$85   &
   $-$07$^{\circ}$02$'$26$''$   
  \\
Spectral type		    &
  uncertain   &
  mid-G to early-K   &
  M2   &
  F5   
  \\
Onset (yr)				&
  FUor-like   &
  many   &
  many   &
  FUor-like   
  \\
Outburst duration	 (yr)			&
  $\cdots$   &
  $>$0.3   &
  $\sim$1  &
  $\cdots$   
  \\
%
Assumed distance (pc)  &
   450  &
   420  &
   420  &
   460
  \\
  \hline
Synthesized beam &
   1\farcs2$\times$0\farcs88; 90$^{\circ}$  &
   $2\farcs0$ $\times$ $0\farcs86$; 64$^{\circ}$  &
   $0\farcs58$ $\times$ $0\farcs47$; 65$^{\circ}$  &
   2\farcs4$\times$2\farcs2; -55$^{\circ}$  
  \\
 ($\theta_{\mbox{\scriptsize{maj}}}$ $\times$ $\theta_{\mbox{\scriptsize{min}}}$; P.A.)	 &
       &
       &
       &
 [271 GHz]      
    \\
Image RMS (mJy\,beam$^{-1}$)		&
   2.4  &
   1.7  &
   0.62  &
   7.6  
  \\
  \hline
mm R.A. (J2000)	&
   05$^{\mbox{\scriptsize{h}}}$35$^{\mbox{\scriptsize{m}}}$26$^{\mbox{\scriptsize{s}}}$.56  &
   05$^{\mbox{\scriptsize{h}}}$35$^{\mbox{\scriptsize{m}}}$36$^{\mbox{\scriptsize{s}}}$.01  &
   $\cdots$  &
   05$^{\mbox{\scriptsize{h}}}$38$^{\mbox{\scriptsize{m}}}$18$^{\mbox{\scriptsize{s}}}$.11 
  \\
mm Decl. (J2000)	&
   $-$05$^{\circ}$03$'$55$\farcs$2  &
   $-$05$^{\circ}$12$'$25$\farcs$3  &
   $\cdots$  &
   $-$07$^{\circ}$02$'$25$\farcs$9  
  \\
Image component size  & 
   $1\farcs 3$ $\times$ $1\farcs 08$; 78$^{\circ}$  &
   $2\farcs 0$ $\times$ $0\farcs94$; 66$^{\circ}$  &
   $\cdots$  &
   $2\farcs 5$ $\times$ $2\farcs 3$; 123$^{\circ}$
  \\
($FWHM{\mbox{\scriptsize{maj}}}$ $\times$ $FWHM{\mbox{\scriptsize{min}}}$; P.A.) &
     &
     &
     &
     
  \\
Peak intensity (mJy\,beam$^{-1}$)	&
  132$\pm$13   &
  29$\pm$2.9   &
  $\cdots$   &
  664$\pm$66   
  \\
Peak S/N		&
  55   &
  17   &
  $\cdots$   &
  87   
  \\
Integrated 1.3 mm Flux (mJy)	&
  171$\pm$17   &
  32$\pm$3.2   &
  3$\sigma$ $<$1.9   &
  704$\pm$70
  \\
  \hline
      \end{tabular}
   }
\end{table*}

\begin{table*}[h]{\scriptsize
{\bf Table \ref{tab:summary}}. (Continued)

      \begin{tabular}{  l  p{3.1cm} p{3.1cm} p{3.1cm} p{3.1cm}   }\hline\hline
Source name	&           
  V2775\,Ori ([ctf93] 216-2)   & 	
  FU\,Ori       & 
  V1647\,Ori    &
  NGC\,2071\,MM3  
   \\
   \hline
Evolutionary Class	    &
   Class I  &
   Class I/II      &
   Class I/II  &
   Class I  
  \\
stellar R.A. (J2000)	&
   05$^{\mbox{\scriptsize{h}}}$42$^{\mbox{\scriptsize{m}}}$48$^{\mbox{\scriptsize{s}}}$.488   &
   05$^{\mbox{\scriptsize{h}}}$45$^{\mbox{\scriptsize{m}}}$22$^{\mbox{\scriptsize{s}}}$.368      &
   05$^{\mbox{\scriptsize{h}}}$46$^{\mbox{\scriptsize{m}}}$13$^{\mbox{\scriptsize{s}}}$.135   &
   05$^{\mbox{\scriptsize{h}}}$47$^{\mbox{\scriptsize{m}}}$36$^{\mbox{\scriptsize{s}}}$.6   
   \\
stellar Decl. (J2000)	&
   $-$08$^{\circ}$16$'$34$\farcs$74   &
   $+$09$^{\circ}$04$'$12$\farcs$25      &
   $-$00$^{\circ}$06$'$04$\farcs$82   &
   $+$00$^{\circ}$20$'$06
  \\
Spectral type		    &
  M5   &
  early-K to mid-M (with a $\sim$1 $M_{\odot}$ companion, $\sim$0$\farcs$5 separation)  &
  M0   &
  uncertain   
  \\
Onset (yr)				&
  $>$2005, $<$2007   &
  1936   &
  many   &
  FUor-like   
  \\
Outburst duration	 (yr)			&
  $>$5   &
  $>$80   &
  a few   &
  $\cdots$   
  \\
%
Assumed distance (pc)  &
   420  &
   353      &
   400  &
   400
  \\
  \hline
Synthesized beam &
   1\farcs1$\times$1\farcs0; 73$^{\circ}$  &
   1\farcs7$\times$0\farcs77; 67$^{\circ}$     &
   1\farcs2$\times$1\farcs0; $-$82$^{\circ}$  &
   3\farcs9$\times$2\farcs9; 86$^{\circ}$  
  \\
 ($\theta_{\mbox{\scriptsize{maj}}}$ $\times$ $\theta_{\mbox{\scriptsize{min}}}$; P.A.)	 &
       &
       &
       &
       
    \\
Image RMS (mJy\,beam$^{-1}$)		&
   1.9  &
   1.8     &
   0.66  &
   4.0  
  \\
  \hline
mm R.A. (J2000)	&
   05$^{\mbox{\scriptsize{h}}}$42$^{\mbox{\scriptsize{m}}}$48$^{\mbox{\scriptsize{s}}}$.49  &
   05$^{\mbox{\scriptsize{h}}}$45$^{\mbox{\scriptsize{m}}}$22$^{\mbox{\scriptsize{s}}}$.38     &
   05$^{\mbox{\scriptsize{h}}}$46$^{\mbox{\scriptsize{m}}}$13$^{\mbox{\scriptsize{s}}}$.14  &
   05$^{\mbox{\scriptsize{h}}}$47$^{\mbox{\scriptsize{m}}}$36$^{\mbox{\scriptsize{s}}}$.59  
  \\
mm Decl. (J2000)	&
   $-$08$^{\circ}$16$'$34$\farcs$7    &
   $+$09$^{\circ}$04$'$12$\farcs$3     &
   $-$00$^{\circ}$06$'$04$\farcs$9    &
   $+$00$^{\circ}$20$'$06$\farcs$1  
  \\
Image component size  & 
   $1\farcs 1$ $\times$ $1\farcs 1$; 76$^{\circ}$  &
   $1\farcs 3$ $\times$ $0\farcs 83$; 69$^{\circ}$  &
   $1\farcs 3$ $\times$ $1\farcs 1$; 98$^{\circ}$  &
   $4\farcs 2$ $\times$ $3\farcs 4$; 78$^{\circ}$   
  \\
($FWHM{\mbox{\scriptsize{maj}}}$ $\times$ $FWHM{\mbox{\scriptsize{min}}}$; P.A.) &
     &
     &
     &
  confused   
  \\
Peak intensity (mJy\,beam$^{-1}$)	&
     99$\pm$10  &
     18$\pm$1.8     &
     69$\pm$6.9 &
     55$\pm$5.5
  \\
Peak S/N		&
     52  &
     9.3 &
     105 &
     14
  \\
Integrated 1.3 mm Flux (mJy)	&
     115$\pm$12  &
     18$\pm$1.8 &
     74$\pm$7.4 &
     $\lesssim$70
  \\
  \hline
      \end{tabular}

\vspace{1cm}
      \begin{tabular}{  l  p{3.1cm} p{3.1cm} p{3.1cm} p{3.1cm}   }\hline\hline
Source name	&           
  AR\,6A/6B    & 	
  Z\,CMa\,NW/SE       & 
  Parsamian\,21 (HBC\,687)    &
  V1515\,Cyg 
   \\
   \hline
Evolutionary Class	    &
  Class I (binary)   &
  Class I (binary)  &
  Class I   &
  Class II   
  \\
stellar R.A. (J2000)	&
 06$^{\mbox{\scriptsize{h}}}$40$^{\mbox{\scriptsize{m}}}$59$^{\mbox{\scriptsize{s}}}$.31/59$^{\mbox{\scriptsize{s}}}$.31    &
 07$^{\mbox{\scriptsize{h}}}$03$^{\mbox{\scriptsize{m}}}$43$^{\mbox{\scriptsize{s}}}$.164     &
 19$^{\mbox{\scriptsize{h}}}$29$^{\mbox{\scriptsize{m}}}$00$^{\mbox{\scriptsize{s}}}$.86     &
 20$^{\mbox{\scriptsize{h}}}$23$^{\mbox{\scriptsize{m}}}$48$^{\mbox{\scriptsize{s}}}$.010     
   \\
stellar Decl. (J2000)	&
  $+$09$^{\circ}$35$'$52$\farcs$0/49$\farcs$2    &
  $-$11$^{\circ}$33$'$06$\farcs$22    &
  $+$09$^{\circ}$38$'$42.9     &
  $+$42$^{\circ}$12$'$25$\farcs$70    
  \\
Spectral type		    &
  early-G to late-K/ Unknown  (separated by 2$\farcs$8) &
  B8-outbursting / F5-FUor-like (separation $\sim$0$\farcs$1)   &
  A5e   &
  G2-G5   
  \\
Onset (yr)				&
  FUor-like/FUor-like   &
  many (1987, 2000, 2004, 2008, etc) / $\cdots$  &
  FUor-like   &
  1950   
  \\
Outburst duration	 (yr)			&
  $\cdots$ / $\cdots$  &
  a few / $\cdots$  &
  $\cdots$   &
  $\sim$30   
  \\
Assumed distance (pc)  &
   800  &
   930  &
   400  &
   1000
  \\
  \hline
Synthesized beam &
   4\farcs2$\times$1\farcs4; 37$^{\circ}$   &
   1\farcs1$\times$0\farcs93; 42$^{\circ}$  &
   1\farcs2$\times$0\farcs83; 40$^{\circ}$  &
   1\farcs7$\times$1\farcs3; 78$^{\circ}$	  
  \\
 ($\theta_{\mbox{\scriptsize{maj}}}$ $\times$ $\theta_{\mbox{\scriptsize{min}}}$; P.A.)	 &
       &
  [35-162 $k\lambda$]     &
  [266 GHz]     &
  [220-277 GHz]     
    \\
Image RMS (mJy\,beam$^{-1}$)		&
   1.3  &
   1.3  &
   1.1  &
   1.0  
  \\
  \hline
mm R.A. (J2000)	&
   $\cdots$   &
   07$^{\mbox{\scriptsize{h}}}$03$^{\mbox{\scriptsize{m}}}$43$^{\mbox{\scriptsize{s}}}$.16  &
   19$^{\mbox{\scriptsize{h}}}$29$^{\mbox{\scriptsize{m}}}$00$^{\mbox{\scriptsize{s}}}$.86  &
   20$^{\mbox{\scriptsize{h}}}$23$^{\mbox{\scriptsize{m}}}$48$^{\mbox{\scriptsize{s}}}$.02  
  \\
mm Decl. (J2000)	&
  $\cdots$   &
  $-$11$^{\circ}$33$'$06$\farcs$0   &
  $+$09$^{\circ}$38$'$42$\farcs$5   &
  $+$42$^{\circ}$12$'$25$\farcs$8
  \\
Image component size  & 
  $\cdots$    &
  $1\farcs4$ $\times$ $0\farcs 90$; 38$^{\circ}$   &
  $1\farcs1$ $\times$ $0\farcs 78$; 37$^{\circ}$   &
  $1\farcs7$ $\times$ $1\farcs 3$; 77$^{\circ}$
  \\
($FWHM{\mbox{\scriptsize{maj}}}$ $\times$ $FWHM{\mbox{\scriptsize{min}}}$; P.A.) &
     &
 confused    &
     &
     
  \\
Peak intensity (mJy\,beam$^{-1}$)	&
   3$\sigma$ $<$3.9  &
   21$\pm$2.1  &
   37$\pm$3.7  &
   6.0$\pm$1.0  
  \\
Peak S/N		&
   $\cdots$  &
   16  &
   34  &
   6  
  \\
Integrated 1.3 mm Flux (mJy)	&
   3$\sigma$ $<$3.9  &
   27$\pm$2.7  &
   37$\pm$3.7  &
   6.0$\pm$1.0 
  \\
  \hline
      \end{tabular}

\vspace{1cm}
      \begin{tabular}{  l  p{3.1cm} p{3.1cm} p{3.1cm} p{3.1cm}   }\hline\hline
Source name	&           
  HBC\,722  (V2493\,Cyg or PTF\,10qpf)    &
  V2494\,Cyg  (HH\,381\,IRS)   & 
  V1057\,Cyg    &
  IRAS\,20588+5215N 
   \\
   \hline
Evolutionary Class	    &
  Class II   &
  Class I/II   &
  Class I/II (binary)   &
  Class I   
  \\
stellar R.A. (J2000)	&
  20$^{\mbox{\scriptsize{h}}}$58$^{\mbox{\scriptsize{m}}}$17$^{\mbox{\scriptsize{s}}}$.029   &
  20$^{\mbox{\scriptsize{h}}}$58$^{\mbox{\scriptsize{m}}}$21$^{\mbox{\scriptsize{s}}}$.09    &
  20$^{\mbox{\scriptsize{h}}}$58$^{\mbox{\scriptsize{m}}}$53$^{\mbox{\scriptsize{s}}}$.72    &
  21$^{\mbox{\scriptsize{h}}}$00$^{\mbox{\scriptsize{m}}}$21$^{\mbox{\scriptsize{s}}}$.40    
   \\
stellar Decl. (J2000)	&
  $+$43$^{\circ}$53$'$43$\farcs$29    &
  $+$52$^{\circ}$29$'$27$\farcs$70    &
  $+$44$^{\circ}$15$'$28$\farcs$3     &
  $+$52$^{\circ}$27$'$09$\farcs$40    
  \\
Spectral type		    &
  K7-M0   &
  G   &
  F7-G3 I-II (with a M-type companion, separation $\sim$30 AU)   &
  uncertain   
  \\
Onset (yr)				&
  2010   &
  $>$1983, $<$1989   &
  1970   &
  Actively accreting,    
  \\
Outburst duration	 (yr)			&
  $>$6   &
  $\gtrsim$20   &
  10-30   &
  but not necessarily a FUor or EXor   
  \\
Assumed distance (pc)  &
  520   &
  800   &
  600   &
  800
  \\
  \hline
Synthesized beam &
   3\farcs3$\times$3\farcs0; $-$14$^{\circ}$  &
   0\farcs86$\times$0\farcs53; 74$^{\circ}$   &
   0\farcs87$\times$0\farcs50; 76$^{\circ}$   &
   1\farcs2$\times$1\farcs0; 85$^{\circ}$  
  \\
 ($\theta_{\mbox{\scriptsize{maj}}}$ $\times$ $\theta_{\mbox{\scriptsize{min}}}$; P.A.)	 &
       &
       &
       &
       
    \\
Image RMS (mJy\,beam$^{-1}$)		&
   2.4  &
   1.0  &
   0.7  &
   2.1  
  \\
  \hline
mm R.A. (J2000)	&
   $\cdots$  &
   20$^{\mbox{\scriptsize{h}}}$58$^{\mbox{\scriptsize{m}}}$21$^{\mbox{\scriptsize{s}}}$.10  &
   20$^{\mbox{\scriptsize{h}}}$58$^{\mbox{\scriptsize{m}}}$53$^{\mbox{\scriptsize{s}}}$.73  &
   $\cdots$  
  \\
mm Decl. (J2000)	&
   $\cdots$  &
   $+$52$^{\circ}$29$'$27$\farcs$5  &
   $+$44$^{\circ}$15$'$28$\farcs$4  &
   $\cdots$  
  \\
Image component size  & 
   $\cdots$   &
   $0\farcs87$  $\times$ $0\farcs53$ ; 75$^{\circ}$   &
   $1\farcs0$  $\times$ $0\farcs59$ ; 84$^{\circ}$   &
   $\cdots$  
  \\
($FWHM{\mbox{\scriptsize{maj}}}$ $\times$ $FWHM{\mbox{\scriptsize{min}}}$; P.A.) &
     &
  variable   &
     &
     
  \\
Peak intensity (mJy\,beam$^{-1}$)	&
   3$\sigma$ $<$7.2  &
   117$\pm$15  &
   15$\pm$2.0  &
   3$\sigma$ $<$6.3  
  \\
Peak S/N		&
   $\cdots$  &
   117  &
   21  &
   $\cdots$  
  \\
Integrated 1.3 mm Flux (mJy)	&
   3$\sigma$ $<$7.2  &
   117$\pm$15  &
   19$\pm$2.0  &
   3$\sigma$ $<$6.3  
  \\
  \hline
      \end{tabular}
   }
\end{table*}

\begin{table*}[h]{\scriptsize
   
{\bf Table \ref{tab:summary}.} (Continued)

      \begin{tabular}{  l  p{3.1cm} p{3.1cm} p{3.1cm} p{3.1cm}   }\hline\hline
Source name	&   
  V2495\,Cyg    & 
  V1735\,Cyg (Elias\,1-12)   &
  V733\,Cep (Persson's star)    &      
   \\
   \hline
Evolutionary Class	    &
  Class I/II   &
  Class I/II   &
  Class II   &
     
  \\
stellar R.A. (J2000)	&
   21$^{\mbox{\scriptsize{h}}}$00$^{\mbox{\scriptsize{m}}}$25$^{\mbox{\scriptsize{s}}}$.25   &
   21$^{\mbox{\scriptsize{h}}}$47$^{\mbox{\scriptsize{m}}}$20$^{\mbox{\scriptsize{s}}}$.65   &
   22$^{\mbox{\scriptsize{h}}}$53$^{\mbox{\scriptsize{m}}}$33$^{\mbox{\scriptsize{s}}}$.24   &
      
   \\
stellar Decl. (J2000)	&
   $+$52$^{\circ}$30$'$17$\farcs$0   &
   $+$47$^{\circ}$32$'$03$\farcs$8   &
   $+$62$^{\circ}$32$'$23$\farcs$8   &
      
  \\
Spectral type		    &
  K5-M1 (uncertain)   &
  F0 II-G0 II   &
  early- to mid-G   &
     
  \\
Onset (yr)				&
  $\sim$1999   &
  $>$1957 and $<$1965   &
  $>$1953 and $<$1984   &
     
  \\
Outburst duration	 (yr)			&
  $>$8   &
  $\gtrsim$30   &
  $\gtrsim$30   &
     
  \\
%
Assumed distance (pc)  &
   800  &
   900  &
   800  &
  \\
  \hline
Synthesized beam &
  0\farcs78$\times$0\farcs46; 74$^{\circ}$    &
  1\farcs2$\times$0\farcs97; 85$^{\circ}$     &
  3\farcs3$\times$2\farcs9; $-$84$^{\circ}$   &
     
  \\
 ($\theta_{\mbox{\scriptsize{maj}}}$ $\times$ $\theta_{\mbox{\scriptsize{min}}}$; P.A.)	 &
       &
       &
       &
       
    \\
Image RMS (mJy\,beam$^{-1}$)		&
   1.0  &
   2.6  &
   0.59  &
     
  \\
  \hline
mm R.A. (J2000)	&
   21$^{\mbox{\scriptsize{h}}}$00$^{\mbox{\scriptsize{m}}}$25$^{\mbox{\scriptsize{s}}}$.24  &
   $\cdots$  &
   $\cdots$  &
     
  \\
mm Decl. (J2000)	&
   $+$52$^{\circ}$30$'$16$\farcs$9  &
   $\cdots$  &
   $\cdots$  &
     
  \\
Image component size  & 
   $0\farcs78$ $\times$ $0\farcs49$; 75$^{\circ}$  &
   $\cdots$  &
   $\cdots$  &
     
  \\
($FWHM{\mbox{\scriptsize{maj}}}$ $\times$ $FWHM{\mbox{\scriptsize{min}}}$; P.A.) &
   variable  &
     &
     &
     
  \\
Peak intensity (mJy\,beam$^{-1}$)	&
  73$\pm$15   &
  3$\sigma$ $<$7.8   &
  3$\sigma$ $<$1.7   &
     
  \\
Peak S/N		&
   73  &
   $\cdots$  &
   $\cdots$  &
     
  \\
Integrated 1.3 mm Flux (mJy)	&
   89$\pm$15  &
   3$\sigma$ $<$7.8  &
   3$\sigma$ $<$1.7  &
     
  \\
  \hline
      \end{tabular}
   }

\vspace{0.6cm}  
{\bf
Notes.
}
{\footnotesize
$^{\mbox{\tiny a}}$ Except for FU\,Ori of which the parallax distance has been provided by Gaia data releases (Gaia Collaboration 2016), we quoted the target source distances mostly from Audard et al. (2014). We note that the distance of Z\,CMa may be as high as 1800 pc, which is approximately two times higher than the value we quote here. 
$^{\mbox{\tiny b}}$ Spectral types of our target sources may be uncertain. We quote them mainly based on the discussion in Gramajo et al. (2014), Audard et al. (2014) and references therein. We provide the list of references for the sake of crediting the original works, on the other hand not duplicating the related description given that we are not providing any better observational constraints. The protostellar mass of V883\,Ori has been constrained by the observations of Keplerian rotation curve with ALMA (Cieza et al. 2016). We do not find dedicated discussion about the protostellar masses of Haro\,5a/6a\,IRS, NGC\,2071\,MM3, and IRAS\,20588+5215N.
$^{\mbox{\tiny c}}$ We specify whether a target source is a confirmed visual or spectroscopic binary or multiple systems in this table. However, we cannot rule out the possibility that some unspecified sources are not yet resolved close binaries. We refer to Reipurth \& Aspin (2004b) and references therein for more discussion about multiplicity of these objects.
} 

\vspace{0.2cm}
{\bf
References. 
}
{\scriptsize
$\bullet$ {\'A}brah{\'a}m et al. (2004) 
$\bullet$ Andrews et al. (2004)  
$\bullet$ Anglada et al. (2004)  
$\bullet$ Antoniucci et al. (2016) 
$\bullet$ Aspin \& Sandell (1994) 
$\bullet$ Aspin \& Reipurth (2003) 
$\bullet$ Aspin et al. (2008)  
$\bullet$ Aspin et al. (2009) 
$\bullet$ Beck \& Aspin(2012) 
$\bullet$ Beljawsky (1928) 
$\bullet$ Bonnefoy et al. (2017) 
$\bullet$ Brice{\~n}o et al. (2004)  
$\bullet$ Caratti o Garatti et al. (2011) 
$\bullet$ Caratti o Garatti et al. (2012) 
$\bullet$ Carrasco-Gonz{\'a}lez et al. (2009) 
$\bullet$ Chen et al. (2016) 
$\bullet$ Cieza et al. (2016) 
$\bullet$ Cohen \& Kuhi (1979) 
$\bullet$ Dibai (1969) 
$\bullet$ Dodin et al. (2016) 
$\bullet$ Eisloeffel et al. (1991) 
$\bullet$ Elias (1978)  
$\bullet$ Fischer et al. (2012)  
$\bullet$ Forgan et al. 2014; 
$\bullet$ Gramajo et al. (2014)  
$\bullet$ Green et al. (2006)  
$\bullet$ Green et al. (2013)  
$\bullet$ Green et al. (2016a) 
$\bullet$ Haas et al. (1990) 
$\bullet$ Hartmann et al. (1989) 
$\bullet$ Hartmann \& Kenyon (1996) 
$\bullet$ Herbig (1977) 
$\bullet$ Herbig et al. (2003)  
$\bullet$ Herbig (2008) 
$\bullet$ Hillenbrand (1997) 
$\bullet$ Hodapp \& Chini (2014) 
$\bullet$ Kenyon et al. (1988) 
$\bullet$ Kenyon et al. (1989) 
$\bullet$ Kenyon et al. (1991)  
$\bullet$ Kenyon et al. (1993) 
$\bullet$ Kolotilov \& Petrov (1983) 
$\bullet$ Koresko et al. (1991)  
$\bullet$ Kospal et al. (2005) 
$\bullet$ Kopatskaya et al. (2013)  
$\bullet$ Krist et al. (1997) 
$\bullet$ Lim et al. (2016) 
$\bullet$ Lodato \& Bertin (2001)  
$\bullet$ Lodato \& Bertin (2003)  
$\bullet$ Magakian et al. (2013)  
$\bullet$ Miller et al. (2011) 
$\bullet$ Movsessian et al. (2006) 
$\bullet$ Munari et al.(2010) 
$\bullet$ Peneva et al. (2010) 
$\bullet$ Persson (2004)       
$\bullet$ Pueyo et al. (2012)  
$\bullet$ Quanz et al. (2007a) 
$\bullet$ Reipurth \& Aspin (1997) 
$\bullet$ Reipurth et al.(2004) 
$\bullet$ Reipurth \& Aspin(2004) 
$\bullet$ Reipurth \& Aspin (2004b) 
$\bullet$ Reipurth et al. (2007a) 
$\bullet$ Reipurth et al. (2007b) 
$\bullet$ Reipurth \& Aspin (2010) 
$\bullet$ Rodr{\'{\i}}guez et al. (2003) 
$\bullet$ Sandell \& Aspin (1998) 
$\bullet$ Semkov et al. (2010) 
$\bullet$ Semkov et al. (2012) 
$\bullet$ Staude \& Neckel (1991) 
$\bullet$ Staude \& Neckel (1992) 
$\bullet$ Strom \& Strom (1993) 
$\bullet$ Szeifert et al. (2010)  
$\bullet$ The et al. (1994) 
$\bullet$ van den Ancker et al. (2004)  
$\bullet$ Wang et al. (2004) 
$\bullet$ Welin (1971) 
$\bullet$ Zhu et al. (2007) 
}
\end{table*}


\section{Conclusions}\label{sec:conclusion}
We have observed a sample of 29 FUors, EXors, and FUor-like objects using the SMA, at $\sim$1.1-1.3 mm wavelengths. 
Most of the detected sources, which are located at distances of $\sim$140-1000 parsec, were observed with $\sim$1$''$ angular resolution.
Therefore, our observations trace the circumstellar material which is feeding the outburst YSOs.
Our main findings are:

\begin{itemize}
\item[$\bullet$] We detected asymmetric, 10$^{3}$ AU scales millimeter continuum emission structures around Haro\,5a/6a\,IRS, NGC\,2071\,MM3, and Z\,CMa. We detected asymmetric or clumpy, few hundreds AU scales structures immediately around SVS\,13, PP\,13S, V2775\,Ori, V2494\,Cyg, and V2495\,Cyg. V1057\,Cyg is surrounded by a few 500-1000 AU scale millimeter emission clumps which are at a few 10$^{3}$ AU projected separations. RNO\,1B, RNO\,1C, and HBC\,722 are located in crowded low-mass cluster forming environments. The other detected sources are  spatially compact, which were marginally resolved or were not resolved by our SMA observations.\\
\item[$\bullet$] Among our observed sources, the confirmed binary or multiple systems show a systematically lower $F_{\mbox{\tiny 1.33 mm}}^{\mbox{\tiny 353 pc}}$ distribution than the rest of the sources. Unfortunately, the presently large uncertainty in the spectral types and the target source distances prohibit addressing whether or not the difference is related to a different protostellar mass distribution. Finally, our samples show a systematically higher $F_{\mbox{\tiny 1.33 mm}}^{\mbox{\tiny 353 pc}}$ distribution than that of the sample of Class\,II YSOs summarized by Pascucci et al. (2016). \\
%
\item[$\bullet$] All detected sources which are confirmed to have lower $F_{\mbox{\tiny 1.33 mm}}^{\mbox{\tiny 353 pc}}$ than FU\,Ori, are EXors (XZ\,Tau, VY\,Tau, V1143\,Ori, V1118\,Ori). The $F_{\mbox{\tiny 1.33 mm}}^{\mbox{\tiny 353 pc}}$ values of the longer duration outburst sources and FUor-like objects are widely spread. The lower bound of their 1.33 mm fluxes will only become certain once the future, deeper observations recover those presently non-detected sources.\\
%
\item[$\bullet$] The distribution of $F_{\mbox{\tiny 1.33 mm}}^{\mbox{\tiny 353 pc}}$ for those targets which exhibit 10 $\mu$m silicate emission (FU\,Ori, Parsamian\,21, V1057\,Cyg, V1515\,Cyg, V1647\,Ori, and XZ\,Tau) is systematically lower than that of the silicate absorption objects (Z\,CMa, L1551\,IRS\,5, RNO\,1B, RNO\,1C, V1735\,Cyg, and V883\,Ori), although these two distributions partly overlap.
This may be consistent with the interpretation of Quanz et al. (2007b) that the silicate absorption objects are embedded by denser residual circumstellar envelopes as compared with the silicate emission ones. \\
\item[$\bullet$] We may have detected the millimeter flux variability of a few tens percents from V2494\,Cyg and V2495\,Cyg. The millimeter emission of FU\,Ori and V883\,Ori appears stationary over the approximately one decade timescales.  Millimeter emission of V1057\,Cyg, V1647\,Ori and Haro\,5a/6a\,IRS may be stationary over a 1-2 years timescale. Over the timescale of several days, we constrained the millimeter flux variability of NGC\,2071\,MM3 and NY\,Ori to be less than $\pm$10\%-15\%, which should be regarded as upper limits given the relatively limited data quality of these two sources. Given the high accretion rate of FU\,Ori over the last $\sim$80 years ($\sim$10$^{-4}$ $M_{\odot}$\,yr$^{-1}$), the stability of its millimeter emission implies that the inner few AU region of its circumstellar disk is very massive such that its density and thermal structures are not yet significantly perturbed due to the massive inflow motion. This is consistent with the mass derivation of Liu et al. (2017). \\
\item[$\bullet$] The high $F_{\mbox{\tiny 1.33 mm}}^{\mbox{\tiny 353 pc}}$ values of our observed sources may imply that they possess relatively massive circumstellar disks. This may be conducive to triggering accretion outbursts. \\
\end{itemize}

\begin{acknowledgements}
We acknowledge Dr. Mark Gurwell for compiling the SMA Calibrator List (http://sma1.sma.hawaii.edu/callist/callist.html). 
The presented calibrator flux density data was obtained at the Submillimeter Array (SMA) near the summit of Mauna Kea (Hawaii). 
3C418, 0530+135, 0510+180, and 0607-085 are included in an ongoing monitoring program at the SMA to determine the fluxes of compact extragalactic radio sources that can be used as calibrators at millimeter wavelengths (Gurwell et al. 2007).  
Observations of available potential calibrators are from time to time observed for 3 to 5 minutes, and the measured source signal strength calibrated against known standards, typically solar system objects (Titan, Uranus, Neptune, or Callisto).  
Data from this program are updated regularly and are available at the SMA website
HBL thanks Dr. Cornelis Dullemond for very insightful discussion.
E. Vorobyov acknowledges support from the RFBR grant 17-02-00644.
This work was supported by the Momentum grant of the MTA CSFK Lend\"ulet Disk Research Group.
\end{acknowledgements}


\appendix




\section{Observational details}\label{appendix:obs} 

We summarize information about each track of the presented SMA observations in Table \ref{tab:obs}, including the starting date, array configuration, number of available antenna, approximated {\it uv} distance range, atmospheric optical depth $\tau$ at 225 GHz ($\tau_{\mbox{\tiny{225 GHz}}}$), intermediate frequency (IF) coverage, central observing frequency (i.e., local oscillator frequency), observed target sources, and absolute flux reference source.
We note that for track sharing observations which covered multiple target sources, the actually sampled {\it uv} distance ranges of individual target sources may slightly vary from the listed values.

\begin{figure}
\begin{tabular}{ c }
\includegraphics[width=9.5cm]{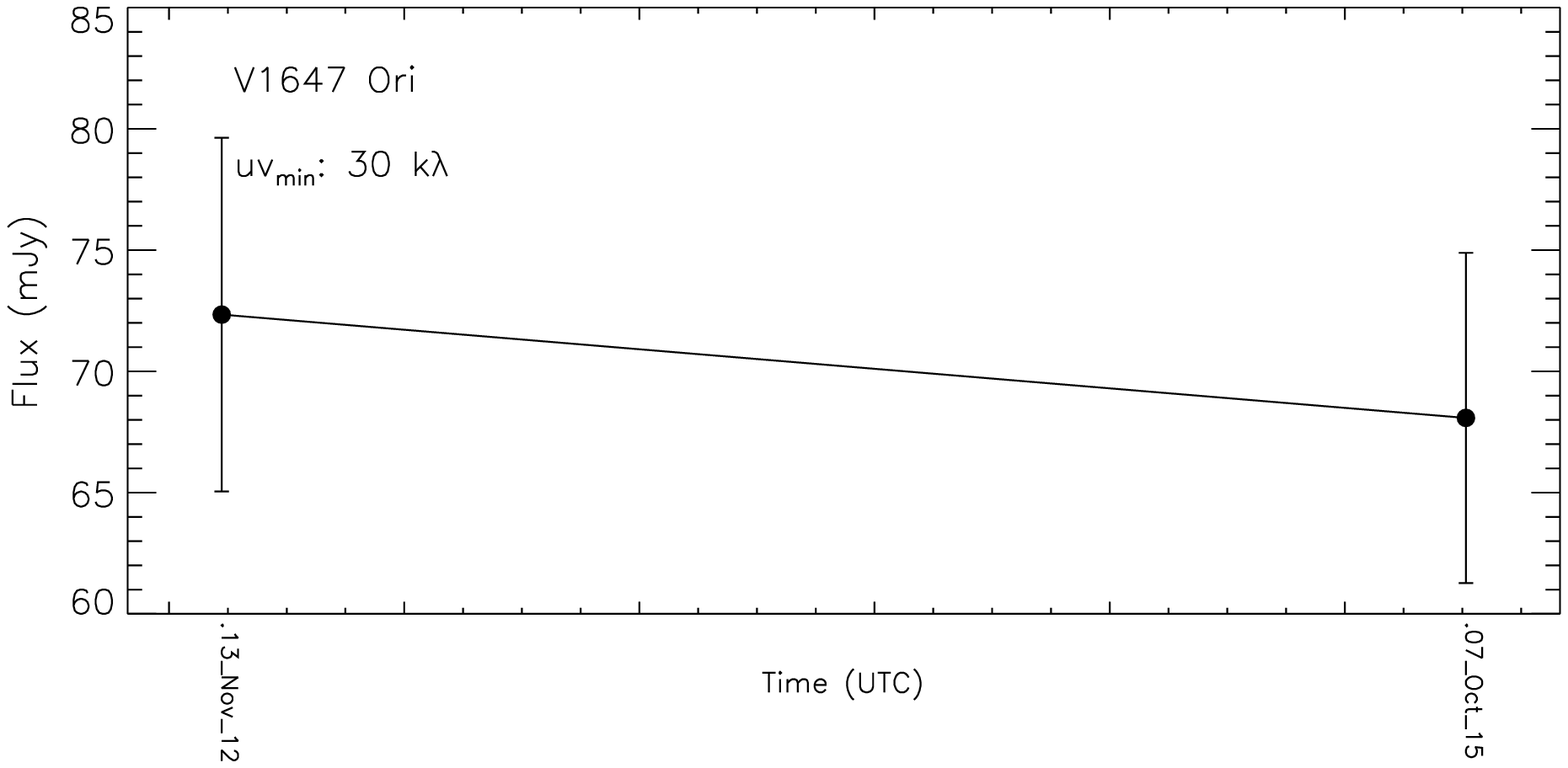} \\
\includegraphics[width=9.5cm]{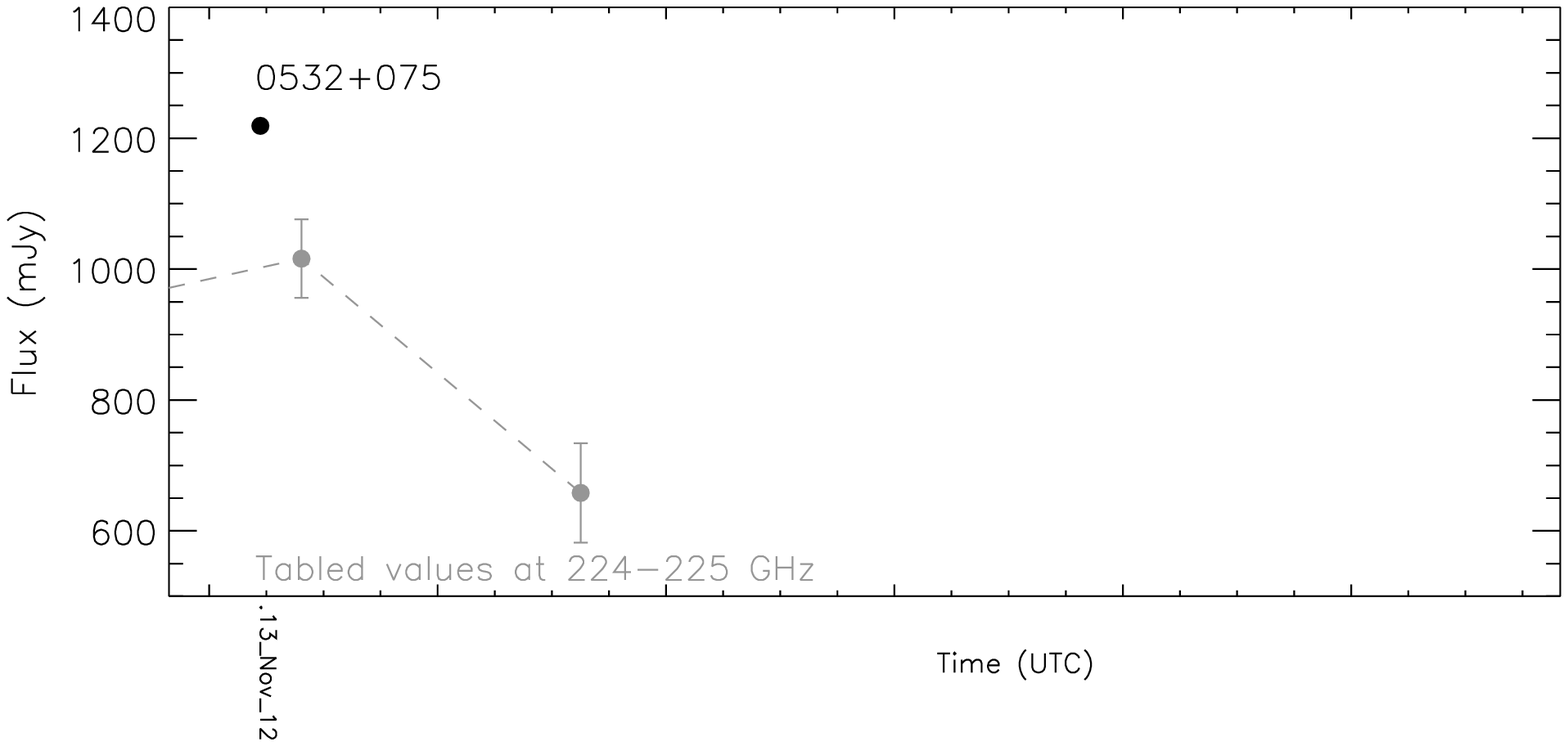} \\
\includegraphics[width=9.5cm]{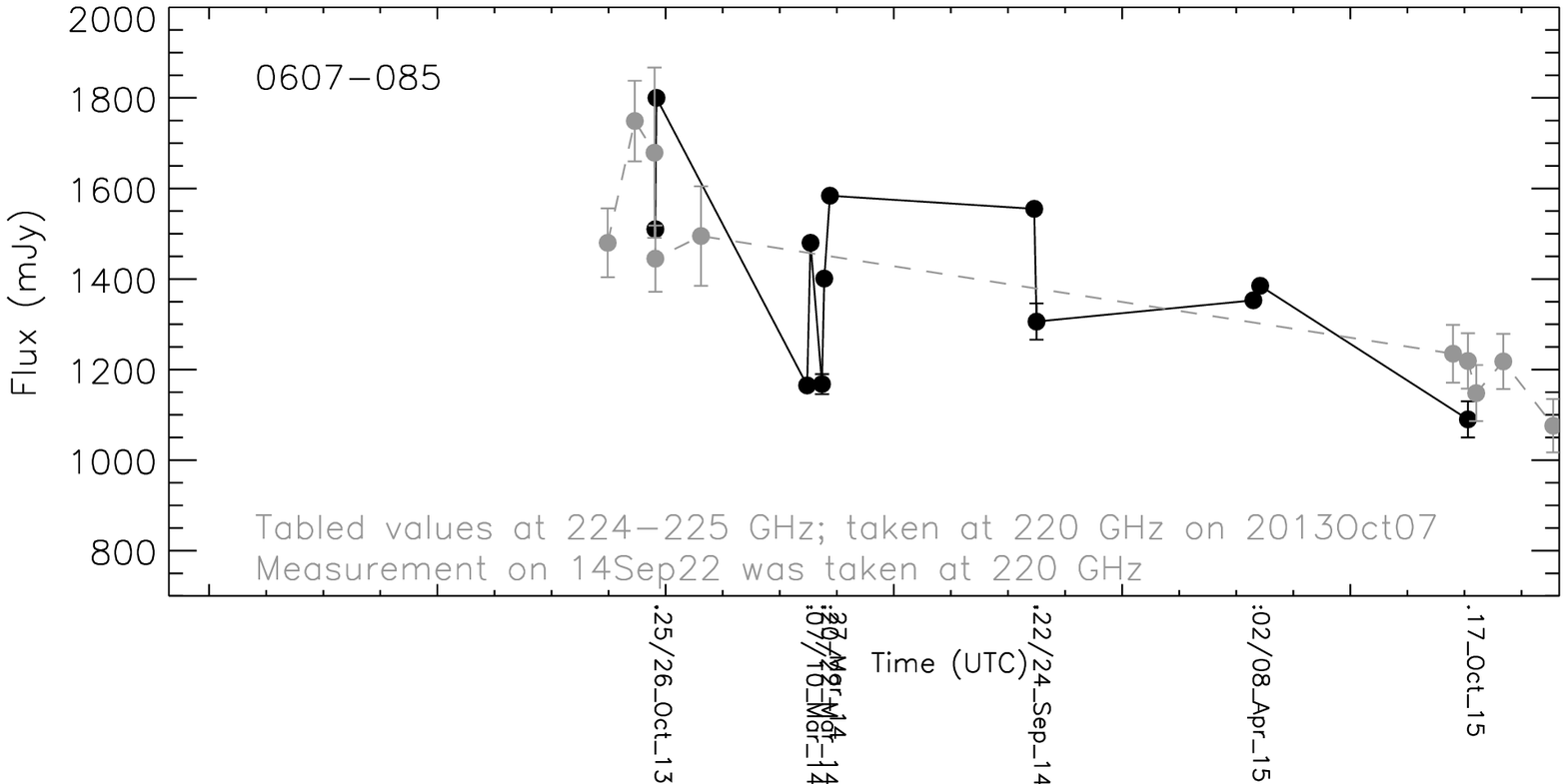} \\
\end{tabular}
\caption{\footnotesize{
Similar to Figure \ref{fig:fluxvarCygnus}, however, for the target source V1647\,Ori.
Horizontal axes of all three panels are presented on the same scale.
The gain calibrator was 0532+075 on November 13 of 2012, and was 0607-085 on October 17 of 2015.
}}
\label{fig:fluxvarv1647ori}
\end{figure}

\begin{figure}
\begin{tabular}{ c }
\includegraphics[width=9.5cm]{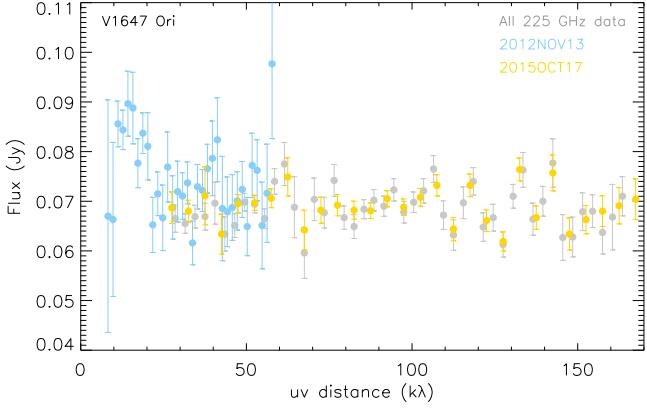} \\
\end{tabular}
\vspace{-0.6cm}
\caption{\footnotesize{
Visibility amplitudes of V1647\,Ori (similar to Figure \ref{fig:uvampCygnus}).
}}
\label{fig:uvampv1647ori}
\end{figure}

\section{Millimeter variability/stability of V1647\,Ori, NY\,Ori, NGC\,2071\,MM3 and Haro\,5a/6a\,IRS}\label{appendix:variability}

\subsection{V1647\,Ori}\label{sub:v1647variability}
We present the observed fluxes and visibility amplitudes of V1647\,Ori in Figure \ref{fig:fluxvarv1647ori} and \ref{fig:uvampv1647ori}.
There were two epochs of SMA observations at 225 GHz towards V1647\,Ori.
Figure \ref{fig:uvampv1647ori} show that this source was spatially compact.
We imaged individual epochs of observations limiting the {\it uv} distance range to $>$30 $k\lambda$, and then measured fluxes in the image domain by performing two-dimensional Gaussian fits.
The detected fluxes were 73 and 68 mJy on November 13 of 2012 and October 07 of 2015, respectively.
Figure \ref{fig:fluxvarv1647ori} shows that the fluxes in the overlapping baseline range agree very well at the two different observing epochs.
We conclude that V1647\,Ori has less than 10\% millimeter flux variability over this time period.
It is worthy of noting that the observed gain quasars, in particular, 0607-085 (Figure \ref{fig:fluxvarv1647ori}) presented large flux variability on as short as a few days timescales.
To constrain millimeter flux variability to less than 10\% precision, it is necessary to observe a stationary absolute flux standard source instead of adopting the tabulated flux values of gain calibration quasars.

\subsection{NY\,Ori}\label{sub:nyvariability}
NY\,Ori was observed on April 02 and 08 of 2014.
We present the observed fluxes and visibility amplitudes of NY\,Ori in Figure \ref{fig:fluxvarnyori} and \ref{fig:uvampnyori}.
The April 02 epoch was taken at poor weather condition with $\tau_{\mbox{\tiny 225 GHz}}$  $\sim$0.3-0.4 (Table \ref{tab:obs}).
The thermal noise of the April 02 observations was larger than the assumed 10\% flux error.
However, due to the poor weather on this date, we may underestimated the potential absolute flux error.
The derived fluxes on these two dates are 29 and 22 mJy, respectively.
We consider the upper limit of the millimeter flux variability of NY\,Ori over this time period to be $\pm$15\%.

\begin{figure}
\begin{tabular}{ c }
\includegraphics[width=9.5cm]{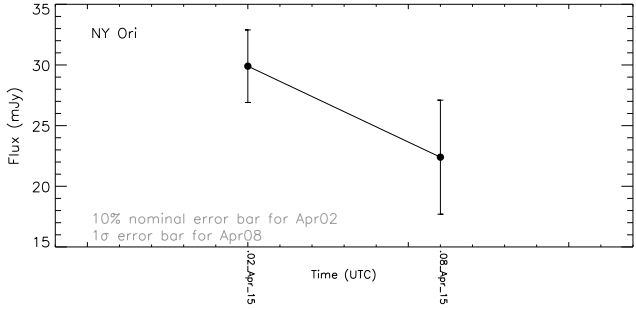} \\
\end{tabular}
\caption{\footnotesize{
Similar to Figure \ref{fig:fluxvarv1647ori}, however, for the target source NY\,Ori.
Gain calibrator of these observations was 0607-085, of which the measured fluxes are presented in Figure \ref{fig:fluxvarv1647ori}.
}}
\label{fig:fluxvarnyori}
\end{figure}

\begin{figure}
\begin{tabular}{ c }
\includegraphics[width=9.5cm]{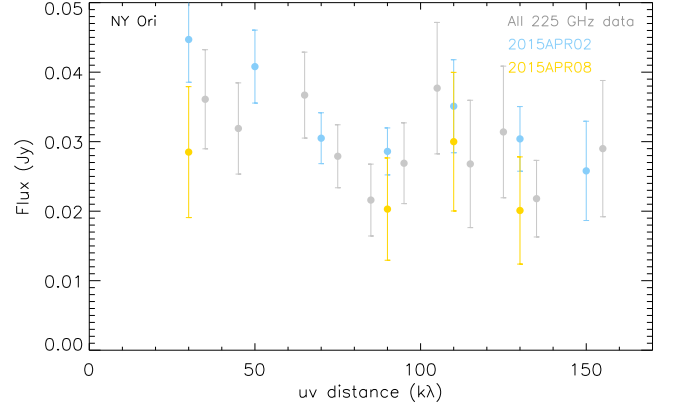} \\
\end{tabular}
\vspace{-0.6cm}
\caption{\footnotesize{
Visibility amplitudes of NY\,Ori (similar to Figure \ref{fig:uvampCygnus}).
}}
\label{fig:uvampnyori}
\end{figure}

\subsection{NGC\,2071\,MM3 and Haro\,5a/6a\,IRS}\label{sub:extendedvariability}
The millimeter emission of dust around NGC\,2071\,MM3 and Haro\,5a/6a\,IRS are spatially extended as compared with the angular scale of the SMA primary beam (Figure \ref{fig:poststamp1}).
Limited by {\it uv} coverages of the observations towards these sources, we were not able to produce images from individual epochs of observations.
Based on the comparison of the visibility amplitudes (Figure \ref{fig:uvampngc2071}-\ref{fig:uvampharo5a6a}), we tentatively consider that NGC\,2071\,MM3 has less than $\pm$ $\sim$20\% millimeter flux variation in March 2014, and Haro\,5a/6a\,IRS has less than 10\% millimeter flux variation from March 2014 to September 2015.

\begin{figure}
\begin{tabular}{ c }
\includegraphics[width=9.5cm]{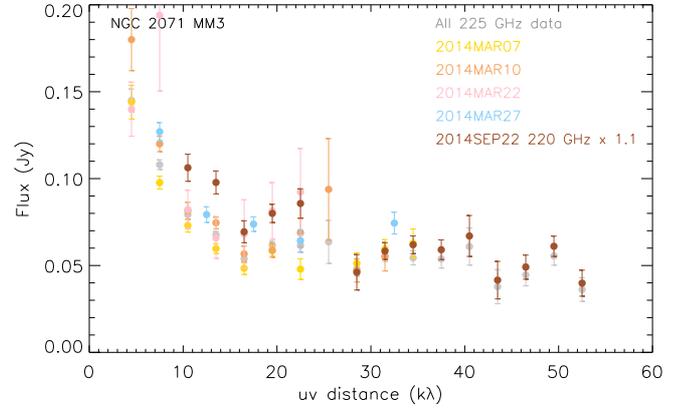} \\
\includegraphics[width=9.5cm]{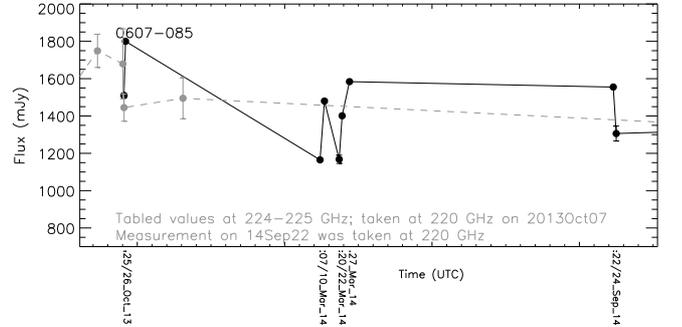} \\
\end{tabular}
\caption{\footnotesize{
Top and bottom panels show the visibility amplitudes of NGC\,2071\,MM3, and the measured fluxes of the gain calibrator 0607-085.
We scaled the observed amplitudes at frequencies lower than 225 GHz to the expected amplitudes at this frequency, by assuming the spectral index $\alpha$ $=$4.0.
}}
\label{fig:uvampngc2071}
\end{figure}

\begin{figure}
\begin{tabular}{ c }
\includegraphics[width=9.5cm]{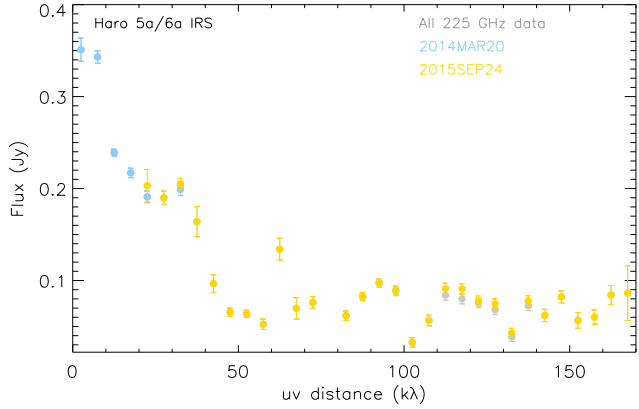} \\
\end{tabular}
\vspace{-0.6cm}
\caption{\footnotesize{
Visibility amplitudes of Haro\,5a/6a\,IRS. 
Gain calibrator of these observations was 0607-085, of which the measured fluxes are presented in Figure \ref{fig:fluxvarv1647ori} and \ref{fig:uvampngc2071}.
}}
\label{fig:uvampharo5a6a}
\end{figure}

\section{Visibility amplitudes}\label{appendix:uvamp}
We summarize the visibility amplitudes of SVS\,13A, PP\,13S, L1551\,IRS\,5, Haro\,5a/6a\,IRS, and Parsamian\,21 in Figure \ref{fig:uvamp_many}.

We scaled the 341 GHz amplitudes of Haro\,5a/6a\,IRS to an observational frequency of 225 GHz based on the assumption of spectral index $\alpha=3.0$.
The assumed $\alpha$ appears too small at $\lesssim$15 $k\lambda$ {\it uv} distance, and is too large at $>$20 $k\lambda$ {\it uv} distance.
This may imply that the 341 GHz emission of Haro\,5a/6a\,IRS is becoming optically thick on $\lesssim$10$''$ angular scales ($\sim$4500 AU).

\begin{table*}{\scriptsize
\begin{center}
\caption{\footnotesize{Summary of the SMA observations.}}
\label{tab:obs}
\hspace{0cm}
\begin{tabular}{  p{1.5cm}  p{2.0cm} p{1.5cm} p{1.5cm} p{1.5cm} p{1.5cm} p{1.5cm} p{2.4cm} p{1.6cm}  }\hline\hline
Dates & Array Config. & \# of Antennas & $uv$ range   & $\tau_{\mbox{\tiny{225 GHz}}}$ & IFs & Central Freq. & Targets & Flux cal. \\[15pt]
(UTC)                &                     &                    & ($k\lambda$) &  & (GHz) & (GHz) & \\\hline
2008May27  & SUB  & 8    & 10-90  & 0.1-0.15 & 4-6 & 225 & V883\,Ori & Uranus \\[16pt]
2008Jun17  & COM  & 7    & 10-90  & 0.1-0.15 & 4-6 & 225 & V883\,Ori & Uranus \\[16pt]
2008Dec06  & COM  & 8     & 10-70  & 0.1-0.15  & 4-6 &  272 & FU\,Ori, V883\,Ori   & Callisto  \\[16pt]
2011May20  & COM  & 7     & 8-52   & 0.2-0.25  & 4-8  & 225  & HBC\,722   & Saturn   \\[16pt]
2012Nov13  & COM  & 7    & 8-58  & 0.05-0.1 & 4-8 & 225 & V1647\,Ori  & Callisto \\[16pt]
2013Jun17  & EXT  & 6     & 40-140  & $\sim$0.2  &  4-8 & 225  & IRAS\,20588+5215N, V1057\,Cyg, V1735\,Cyg, V2494\,Cyg, V2495\,Cyg  &  Neptune \\[56pt]
2013Aug02  & COM  & 5    & 8-52  & 0.1-0.15 & 4-8 & 224 & RNO\,1B/1C  & Uranus \\[16pt]
2013Oct24  & EXT  & 6    & 23-138  & 0.2-0.3 & 4-8 & 224 & RNO\,1B/1C  & Uranus \\[16pt]
2013Oct25  & EXT  & 5    & 20-158  & 0.1-0.2 & 4-8 & 224 & RNO\,1B/1C, V2775\,Ori, Z\,CMa  & Uranus \\[35pt]
2013Oct26  & EXT  & 6    & 30-140  & 0.2-0.25 & 4-8 & 224 & PP\,13S, SVS\,13, Z\,CMa, V2775\,Ori  & Neptune \\[35pt]
2013Nov09  & EXT  & 7    & 30-170  & 0.4 & 4-8 & 224 & VY\,Tau, FU\,Ori  & Uranus \\[16pt]
2013Nov22  & EXT  & 7    & 25-175  & 0.1-0.4 & 4-8 & 224 & VY\,Tau, FU\,Ori  & Callisto \\[16pt]
2014Feb10  & SUB  & 6    & 4-19  & 0.15 & 4-8 & 224 & RNO\,1B/1C  & Callisto \\[16pt]
2014Mar07  & SUB  & 7    & 5-35  & 0.2-0.3 & 4-8 & 225 &  NGC2071\,MM3, AR\,6A/6B & Callisto \\[16pt]
2014Mar10  & SUB  & 7    & 5-35  & 0.1-0.25 & 4-8 & 225 &  NGC2071\,MM3, AR\,6A/6B & Titan \\[16pt]
2014Mar20  & SUB-N  & 7    &  5-140/7-52 & 0.07  & 4-6 & 225/342 (dual RX) & Haro\,5a6a\,IRS, Z\,CMa & Callisto \\[20pt]
2014Mar22  & SUB  & 7    & 3-35  & $\sim$0.2 & 4-8 & 225 &  NGC2071\,MM3, AR\,6A/6B &  Callisto \\[16pt]
2014Mar27  & SUB-N  & 5    & 5-140  & $\sim$0.1 & 4-8 & 225 & NGC2071\,MM3, AR\,6A/6B  & Callisto \\[16pt]
2014Apr08  & EXT  & 6    & 35-160  & 0.05 & 4-8 & 266 & Parsamian\,21 &  Titan \\[16pt]
2014Apr22  & COM-N  & 7    & 15-90  & 0.1-0.2 & 4-6 & 225 & V1515\,Cyg, V733\,Cep  & Neptune \\[16pt]\hline
\end{tabular}
\end{center}
\vspace{0.7cm}
}
\end{table*}
\normalsize{}

\begin{table*}{\scriptsize
{\bf Table \ref{tab:obs}.} (Continued)
\begin{center}
\hspace{0cm}
\begin{tabular}{  p{1.5cm}  p{2.0cm} p{1.5cm} p{1.5cm} p{1.5cm} p{1.5cm} p{1.5cm} p{2.4cm} p{1.6cm}  }\hline\hline
Dates & Array Config. & \# of Antennas & $uv$ range   & $\tau_{\mbox{\tiny{225 GHz}}}$ & IFs & Central Freq. & Targets & Flux cal. \\[15pt]
(UTC)                &                     &                    & ($k\lambda$) &  & (GHz) & (GHz) & \\\hline
2014Jun04  & COM  & 7    & 10-60  & 0.15-0.2 & 4-8 & 225 &  V1057\,Cyg, V2494\,Cyg & Neptune \\[16pt]
2014Jun12  & COM  & 8    & 10-79  & 0.15 & 4-6 & 272 & V1515\,Cyg, V2495\,Cyg  & Neptune \\[16pt]
2014Jun13  & COM  & 8    & 8-52  & 0.1 & 4-8 & 225 & V733\,Cep  & Neptune \\[16pt]
2014Jul25  & SUB  & 6    & 5-32  & 0.1-0.15 & 4-8 & 225 & RNO\,1B/1C  & Neptune \\[16pt]
2014Aug17  & VEX  & 6    & 30-320  & 0.05 & 4-8 & 225 & V1057\,Cyg, V2494\,Cyg, V2495\,Cyg  & Titan \\[24pt]
2014Aug20  & VEX  & 7    & 30-320  & 0.1-0.15 & 4-8 & 225 & V1057\,Cyg, V2494\,Cyg, V2495\,Cyg  & Neptune \\[24pt]
2014Aug27  & EXT  & 8    & 30-200  & 0.15 & 4-8  & 272 &  V883\,Ori, FU\,Ori & Uranus \\[16pt]
2014Sep02  & EXT  & 7     & 30-150  & 0.1  & 4-6 &  236 & V1515\,Cyg, V733\,Cep   &  Neptune \\[24pt]
2014Sep22  & COM  & 7    & 10-55  & 0.15-0.2 & 4-8 & 220 & NGC2071\,MM3, AR\,6A/6B  & Uranus \\[16pt]
2015Jan22  & VEX  & 7   & 25-390   &  0.07   & 4-8, 8-9.5, 10.5-12  &  225 & XZ\,Tau, VY\,tau, V1118\,Ori, V1143\,Ori & Callisto \\[34pt]
2015Jan26  & VEX  & 5   & 25-350   &  $\lesssim$0.1   & 4-8, 8-9.5, 10.5-12  &  225 & VY\,tau, V1118\,Ori, V1143\,Ori & Callisto \\[24pt]
2015Jan27  & VEX  & 6   & 25-350   &  0.07   & 4-8, 8-9.5, 10.5-12  &  225 & VY\,tau, V1118\,Ori, V1143\,Ori & Callisto \\[24pt]
2015Feb03  & EXT  & 6   & 25-175   &  0.3   & 4-8 &  225 & VY\,tau, V1118\,Ori, V1143\,Ori & Callisto \\[24pt]
2015Mar08  & EXT  & 6   & 25-175   &  0.1   & 4-8 &  225 & VY\,tau, V1118\,Ori, V1143\,Ori & Ganymede \\[24pt]
2015Apr02  & EXT  & 6   & 30-170   &  0.3-0.4   & 4-8 &  225 & NY\,Ori, V1118\,Ori, V1143\,Ori & Ganymede \\[24pt]
2015Apr08  & EXT  & 5   & 20-140   &  0.15   & 4-8 &  225 & NY\,Ori, V1118\,Ori, V1143\,Ori & Titan \\[24pt]
2015Sep18  & EXT  & 7   &  30-177  & 0.1   & 4-8, 8-9.5, 10.5-12 & 225           & L1551\,IRS5 & Uranus \\[14pt]
2015Sep24  & EXT  & 7   &  24-167  & 0.2   & 4-8, 8-9.5, 10.5-12 & 225           & Haro\,5a/6a\,IRS & Uranus \\[14pt]
2015Oct07  & EXT  & 7   &  25-175  & 0.07   & 4-8, 8-9.5, 10.5-12 & 225           & V1647\,Ori & Uranus \\[14pt]
2017Jan30  & SUB  & 7   &  7.5-62/8-64  &  0.08 & 4-12 & 259/275 (dual RX) & FU\,Ori & Uranus \\[14pt]\hline
\end{tabular}
\end{center}
\vspace{0.7cm}
}
\end{table*}
\normalsize{}

\clearpage
\begin{figure*}
\hspace{0.0cm}
\begin{tabular}{ p{8.5cm} p{8.5cm} }
\includegraphics[width=9.5cm]{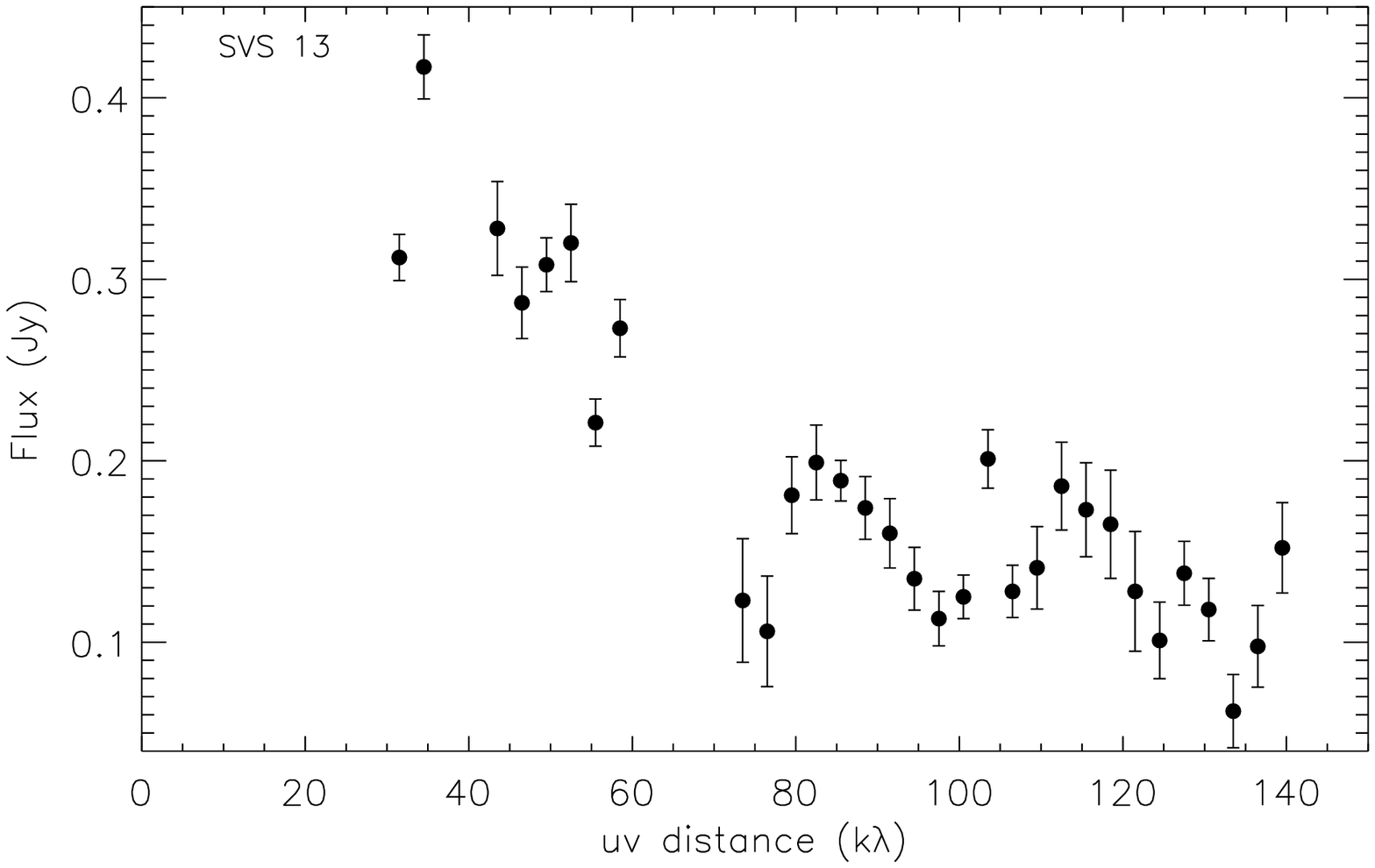} &
\includegraphics[width=9.5cm]{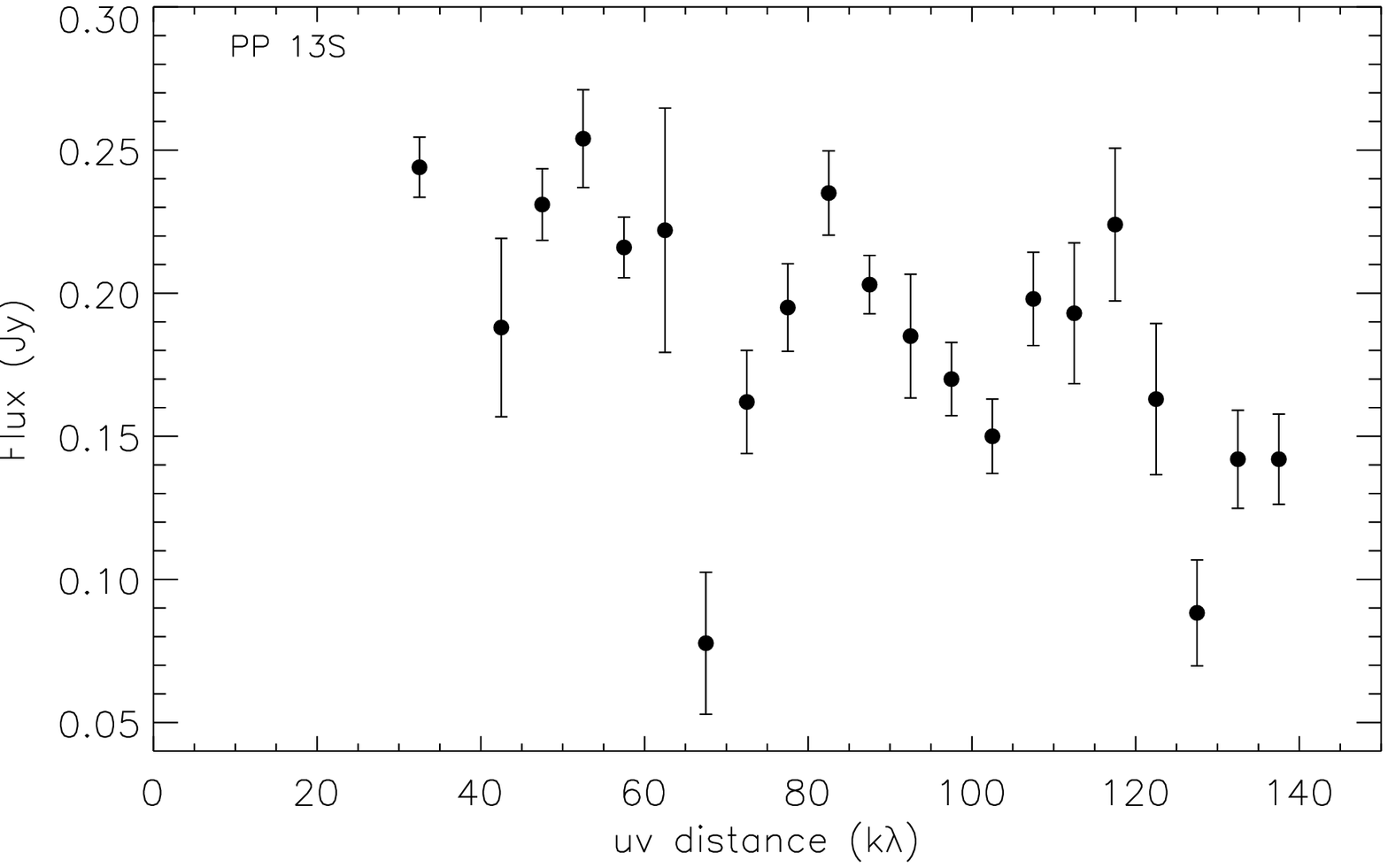} \\
\end{tabular}

\vspace{-0.6cm}
\begin{tabular}{ p{8.5cm} p{8.5cm} }
\includegraphics[width=9.5cm]{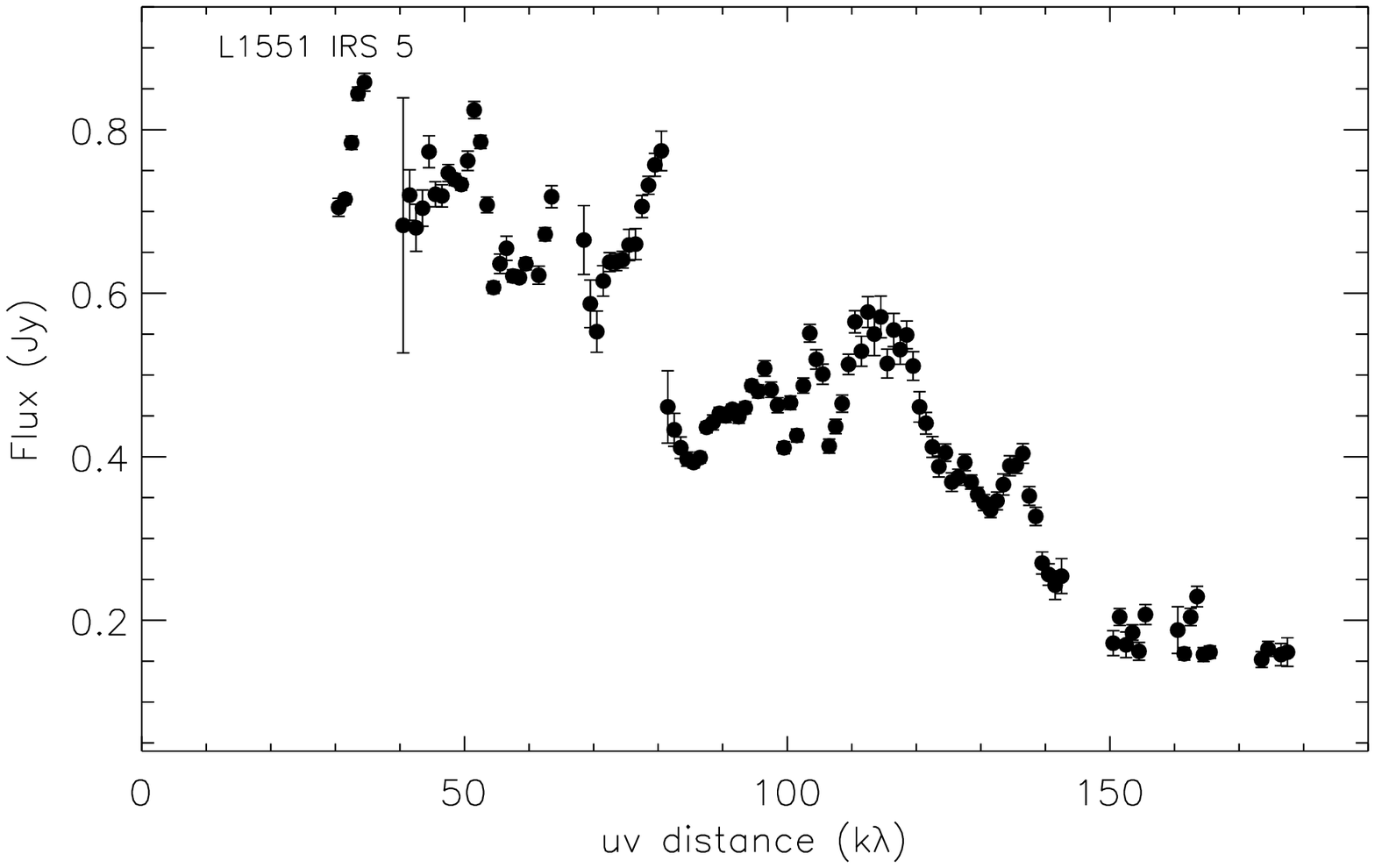} &
\includegraphics[width=9.5cm]{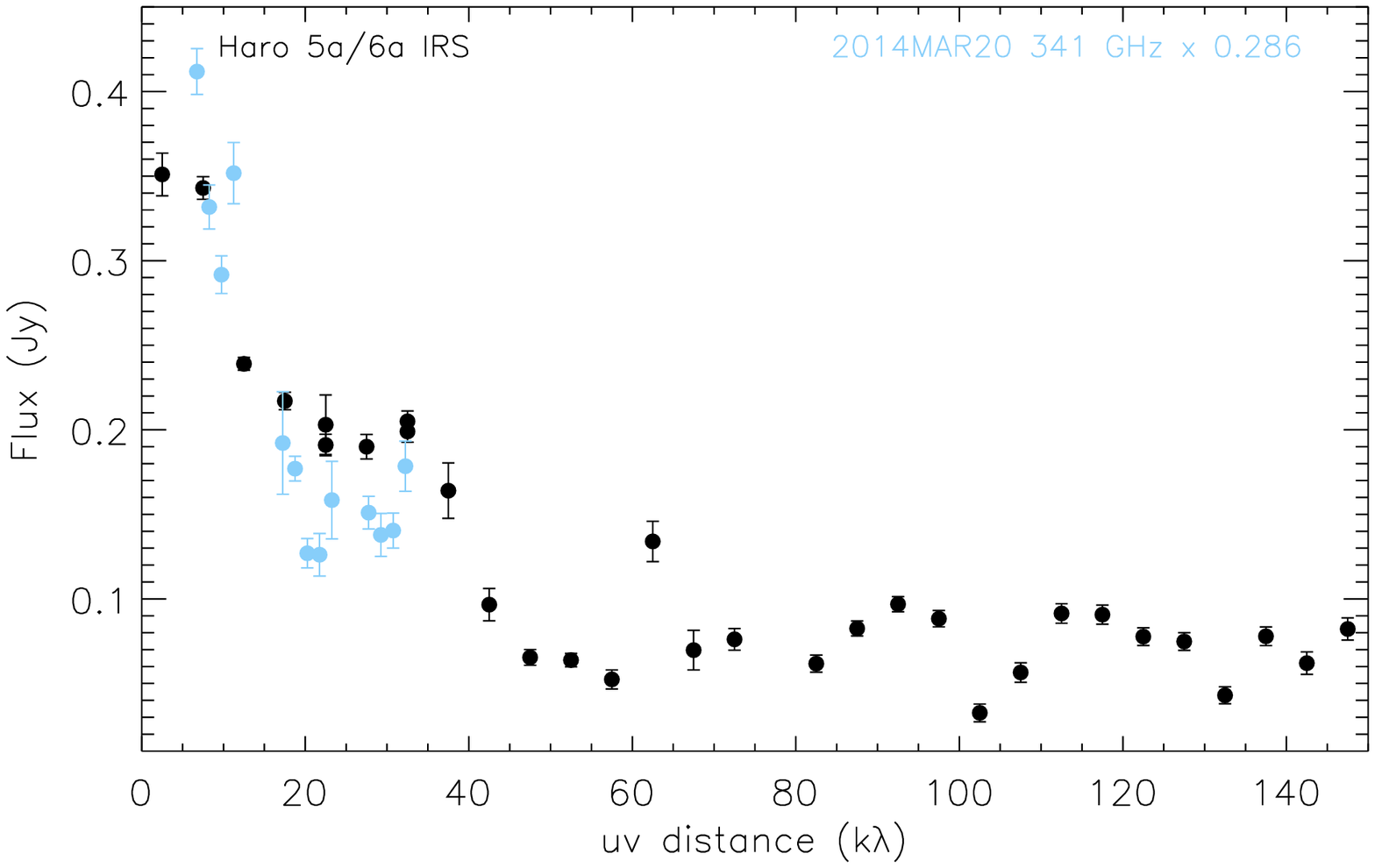} \\
\end{tabular}

\vspace{-0.6cm}
\begin{tabular}{ p{8.5cm} p{8.5cm} }
\includegraphics[width=9.5cm]{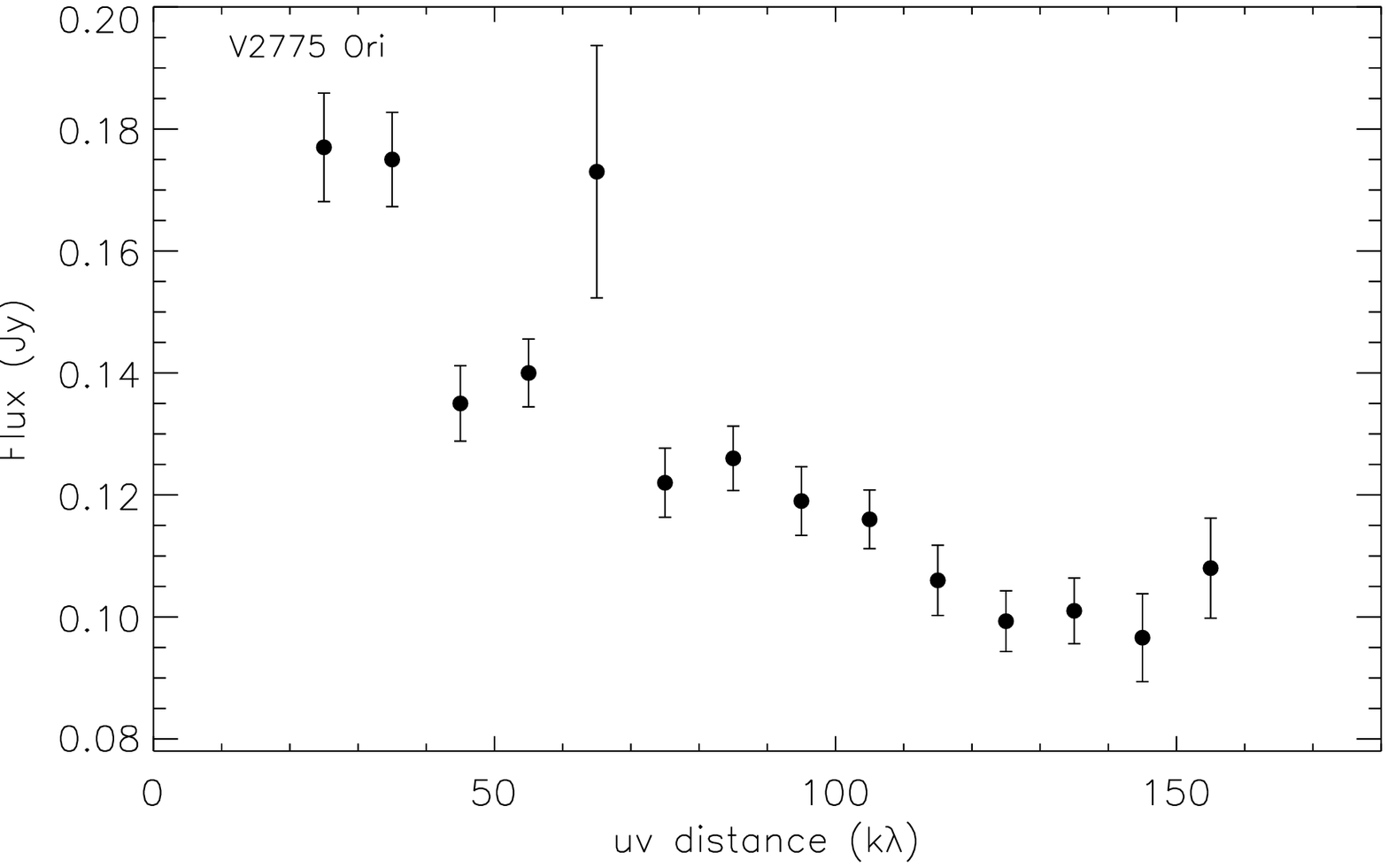} &
\includegraphics[width=9.5cm]{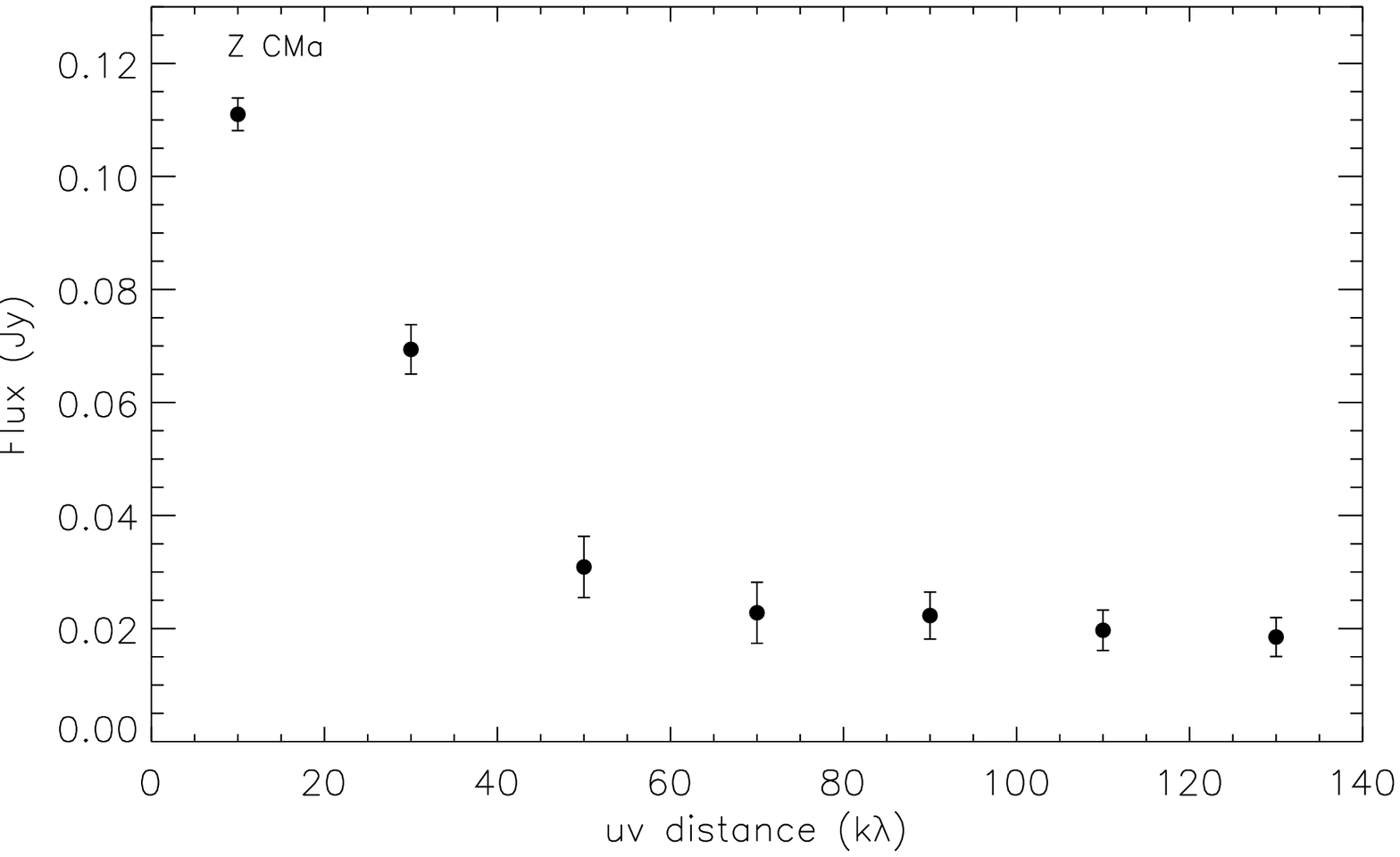} \\
\end{tabular}

\vspace{-0.6cm}
\begin{tabular}{ p{8.5cm} p{8.5cm} }
\includegraphics[width=9.5cm]{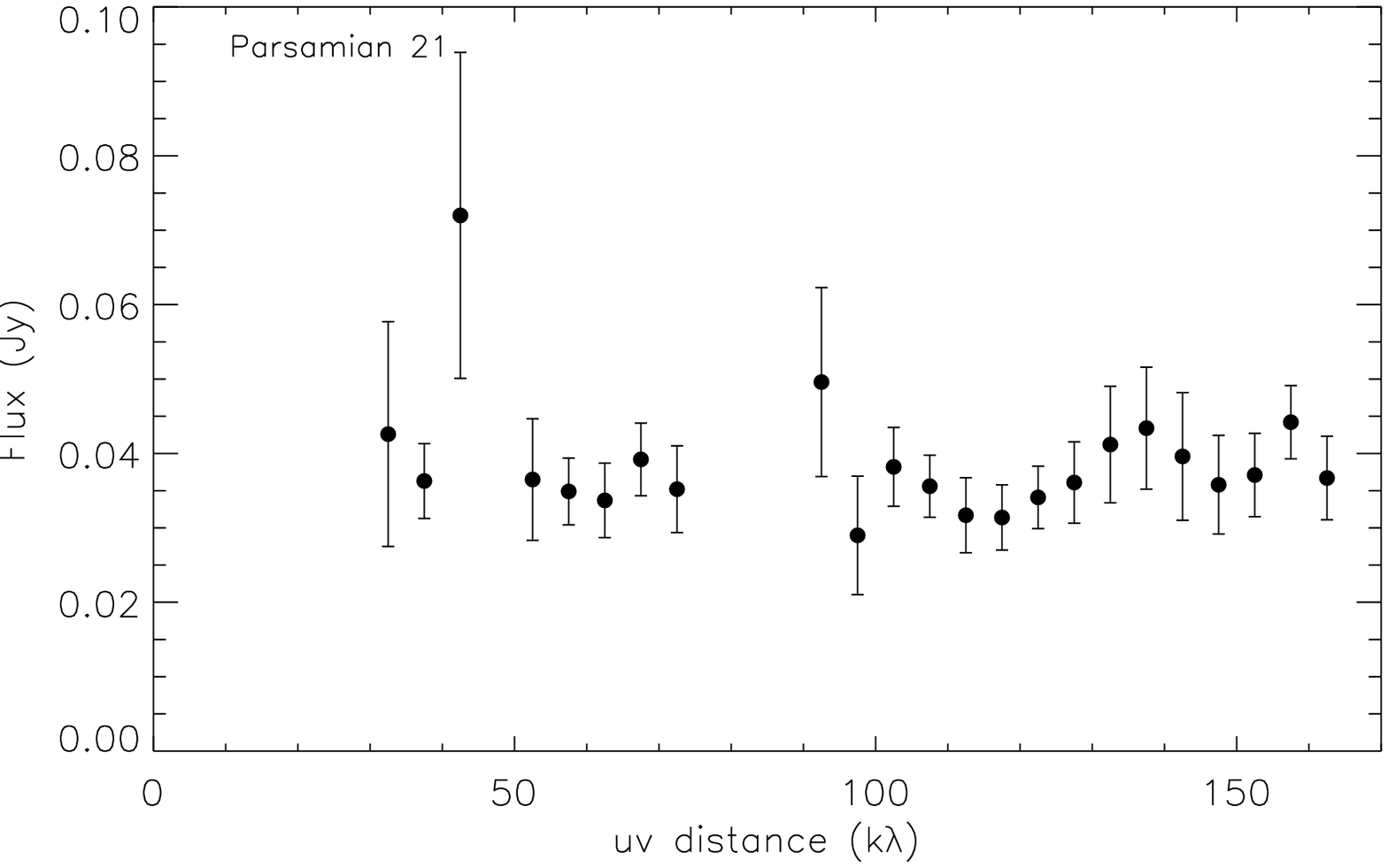}  & \\
\end{tabular}
\vspace{-0.6cm}
\caption{\footnotesize{
Observed visibility amplitudes of SVS\,13A, PP\,13S, L1551\,IRS\,5, Haro\,5a/6a\,IRS, V2775\,Ori, Z\,CMa, and Parsamian\,21, which were measured using the {\tt uvamp} task of the Miriad software package.
The visibility amplitudes of SVS\,13 is partly contributed from the adjacent YSOs and the parent molecular cloud structures.
Black curves are observations at $\sim$225 GHz.
We combined the data of V2775\,Ori taken on October 25 and 26 of 2013; we combined the data of Z\,CMa taken on October 25 and 26 of 2013, and on March 20 of 2014.
Blue curve in the panel of Haro\,5a/6a\,IRS presents the 341 GHz observations taken on March 20, 2014.
We scaled the 341 GHz amplitudes to an observational frequency of 225 GHz based on the assumption of spectral index $\alpha=3.0$, which corresponds to a scaling factor of 0.286.
}}
\label{fig:uvamp_many}
\end{figure*}


\begin{thebibliography}{}

\bibitem[{\'A}brah{\'a}m et al.(2004)]{2004A&A...428...89A} {\'A}brah{\'a}m, P., K{\'o}sp{\'a}l, {\'A}., Csizmadia, S., et al.\ 2004, \aap, 428, 89 


\bibitem[Alonso-Albi et al.(2009)]{2009A&A...497..117A} Alonso-Albi, T., Fuente, A., Bachiller, R., et al.\ 2009, \aap, 497, 117 

\bibitem[Andrews et al.(2004)]{2004ApJ...610L..45A} Andrews, S.~M., Rothberg, B., \& Simon, T.\ 2004, \apjl, 610, L45 

\bibitem[Andrews et al.(2013)]{2013ApJ...771..129A} Andrews, S.~M., Rosenfeld, K.~A., Kraus, A.~L., \& Wilner, D.~J.\ 2013, \apj, 771, 129

\bibitem[Anglada et al.(1994)]{1994ApJ...420L..91A} Anglada, G., Rodriguez, L.~F., Girart, J.~M., Estalella, R., \& Torrelles, J.~M.\ 1994, \apjl, 420, L91 

\bibitem[Anglada et al.(2004)]{2004ApJ...605L.137A} Anglada, G., Rodr{\'{\i}}guez, L.~F., Osorio, M., et al.\ 2004, \apjl, 605, L137


\bibitem[Ansdell et al.(2016)]{2016ApJ...828...46A} Ansdell, M., Williams, J.~P., van der Marel, N., et al.\ 2016, \apj, 828, 46

\bibitem[Ansdell et al.(2017)]{2017AJ....153..240A} Ansdell, M., Williams, J.~P., Manara, C.~F., et al.\ 2017, \aj, 153, 240

\bibitem[Antoniucci et al.(2016)]{2016A&A...593L..13A} Antoniucci, S., Podio, L., Nisini, B., et al.\ 2016, \aap, 593, L13

\bibitem[Aspin \& Sandell(1994)]{1994A&A...288..803A} Aspin, C., \& Sandell, G.\ 1994, \aap, 288, 803 

\bibitem[Aspin \& Reipurth(2003)]{2003AJ....126.2936A} Aspin, C., \& Reipurth, B.\ 2003, \aj, 126, 2936 

\bibitem[Aspin et al.(2008)]{2008AJ....135..423A} Aspin, C., Beck, T.~L., \& Reipurth, B.\ 2008, \aj, 135, 423 

\bibitem[Aspin et al.(2009)]{2009AJ....137..431A} Aspin, C., Beck, T.~L., Pyo, T.-S., et al.\ 2009, \aj, 137, 431 


\bibitem[Audard et al.(2014)]{2014prpl.conf..387A} Audard, M., {\'A}brah{\'a}m, P., Dunham, M.~M., et al.\ 2014, Protostars and Planets VI, 387 


\bibitem[Barenfeld et al.(2016)]{2016ApJ...827..142B} Barenfeld, S.~A., Carpenter, J.~M., Ricci, L., \& Isella, A.\ 2016, \apj, 827, 142


\bibitem[Beljawsky(1928)]{1928AN....234...41B} Beljawsky, S.\ 1928, Astronomische Nachrichten, 234


\bibitem[Boissier et al.(2011)]{2011A&A...531A..50B} Boissier, J., Alonso-Albi, T., Fuente, A., et al.\ 2011, \aap, 531, A50 

\bibitem[Bonnell \& Bastien(1992)]{1992ApJ...401L..31B} Bonnell, I., \& Bastien, P.\ 1992, \apjl, 401, L31 

\bibitem[Bonnefoy et al.(2017)]{2017A&A...597A..91B} Bonnefoy, M., Chauvin, G., Dougados, C., et al.\ 2017, \aap, 597, A91 

\bibitem[Brice{\~n}o et al.(2004)]{2004ApJ...606L.123B} Brice{\~n}o, C., Vivas, A.~K., Hern{\'a}ndez, J., et al.\ 2004, \apjl, 606, L123 

\bibitem[Canovas et al.(2015)]{2015A&A...578L...1C} Canovas, H., Perez, S., Dougados, C., et al.\ 2015, \aap, 578, L1 

\bibitem[Caratti o Garatti et al.(2011)]{2011A&A...526L...1C} Caratti o Garatti, A., Garcia Lopez, R., Scholz, A., et al.\ 2011, \aap, 526, L1

\bibitem[Caratti o Garatti et al.(2012)]{2012A&A...538A..64C} Caratti o Garatti, A., Garcia Lopez, R., Antoniucci, S., et al.\ 2012, \aap, 538, A64

\bibitem[Carpenter et al.(2014)]{2014ApJ...787...42C} Carpenter, J.~M., Ricci, L., \& Isella, A.\ 2014, \apj, 787, 42

\bibitem[Carrasco-Gonz{\'a}lez et al.(2009)]{2009ApJ...693L..86C} Carrasco-Gonz{\'a}lez, C., Rodr{\'{\i}}guez, L.~F., Anglada, G., \& Curiel, S.\ 2009, \apjl, 693, L86

\bibitem[Cieza et al.(2016)]{2016Natur.535..258C} Cieza, L.~A., Casassus, S., Tobin, J., et al.\ 2016, \nat, 535, 258 

\bibitem[Chen et al.(2016)]{2016ApJ...824...72C} Chen, X., Arce, H.~G., Zhang, Q., Launhardt, R., \& Henning, T.\ 2016, \apj, 824, 72

\bibitem[Chou et al.(2014)]{2014ApJ...796...70C} Chou, T.-L., Takakuwa, S., Yen, H.-W., Ohashi, N., \& Ho, P.~T.~P.\ 2014, \apj, 796, 70 

\bibitem[Cohen \& Kuhi(1979)]{1979ApJS...41..743C} Cohen, M., \& Kuhi, L.~V.\ 1979, \apjs, 41, 743 

\bibitem[Dibai(1969)]{1969Ap......5..115D} Dibai, E.~A.\ 1969, Astrophysics, 5, 115 

\bibitem[Dodin et al.(2016)]{2016AstL...42...29D} Dodin, A.~V., Emelyanov, N.~V., Zharova, A.~V., et al.\ 2016, Astronomy Letters, 42, 29 


\bibitem[Draine(2003)]{2003ARA&A..41..241D} Draine, B.~T.\ 2003, \araa, 41, 241 

\bibitem[Draine(2006)]{2006ApJ...636.1114D} Draine, B.~T.\ 2006, \apj, 636, 1114 


\bibitem[Dunham et al.(2012)]{2012ApJ...755..157D} Dunham, M.~M., Arce, H.~G., Bourke, T.~L., et al.\ 2012, \apj, 755, 157 

\bibitem[Dunham \& Vorobyov(2012)]{2012ApJ...747...52D} Dunham, M.~M., \& Vorobyov, E.~I.\ 2012, \apj, 747, 52 

\bibitem[Dunham et al.(2014)]{2014MNRAS.444..887D} Dunham, M.~M., Vorobyov, E.~I., \& Arce, H.~G.\ 2014, \mnras, 444, 887 


\bibitem[Dzib et al.(2015)]{2015ApJ...801...91D} Dzib, S.~A., Loinard, L., Rodr{\'{\i}}guez, L.~F., et al.\ 2015, \apj, 801, 91

\bibitem[Eisloeffel et al.(1991)]{1991ApJ...383L..19E} Eisloeffel, J., Guenther, E., Hessman, F.~V., et al.\ 1991, \apjl, 383, L19

\bibitem[Elias(1978)]{1978ApJ...223..859E} Elias, J.~H.\ 1978, \apj, 223, 859 

\bibitem[Evans et al.(2017)]{2017arXiv170600254E} Evans, M.~G., Ilee, J.~D., Hartquist, T.~W., et al.\ 2017, arXiv:1706.00254

\bibitem[Feh{\'e}r et al.(2017)]{2017arXiv170907458F} Feh{\'e}r, O., K{\'o}sp{\'a}l, {\'A}., {\'A}brah{\'a}m, P., Hogerheijde, M.~R., \& Brinch, C.\ 2017, arXiv:1709.07458 

\bibitem[Fischer et al.(2012)]{2012ApJ...756...99F} Fischer, W.~J., Megeath, S.~T., Tobin, J.~J., et al.\ 2012, \apj, 756, 99

\bibitem[Forgan et al.(2014)]{2014MNRAS.439.4057F} Forgan, D., Ivison, R.~J., Sibthorpe, B., Greaves, J.~S., \& Ibar, E.\ 2014, \mnras, 439, 4057 

\bibitem[Gaia Collaboration et al.(2016)]{2016A&A...595A...2G} Gaia Collaboration, Brown, A.~G.~A., Vallenari, A., et al.\ 2016, \aap, 595, A2 




\bibitem[Gramajo et al.(2014)]{2014AJ....147..140G} Gramajo, L.~V., Rod{\'o}n, J.~A., \& G{\'o}mez, M.\ 2014, \aj, 147, 140 

\bibitem[Green et al.(2006)]{2006ApJ...648.1099G} Green, J.~D., Hartmann, L., Calvet, N., et al.\ 2006, \apj, 648, 1099

\bibitem[Green et al.(2013)]{2013ApJ...772..117G} Green, J.~D., Evans, N.~J., II, K{\'o}sp{\'a}l, {\'A}., et al.\ 2013, \apj, 772, 117 

\bibitem[Green et al.(2016a)]{2016ApJ...830...29G} Green, J.~D., Kraus, A.~L., Rizzuto, A.~C., et al.\ 2016a, \apj, 830, 29

\bibitem[Green et al.(2016b)]{2016ApJ...832....4G} Green, J.~D., Jones, O.~C., Keller, L.~D., et al.\ 2016b, \apj, 832, 4 

\bibitem[Gurwell et al.(2007)]{2007ASPC..375..234G} Gurwell, M.~A., Peck, A.~B., Hostler, S.~R., Darrah, M.~R., \& Katz, C.~A.\ 2007, From Z-Machines to ALMA: (Sub)Millimeter Spectroscopy of Galaxies, 375, 234 

\bibitem[Haas et al.(1990)]{1990A&A...230L...1H} Haas, M., Leinert, C., \& Zinnecker, H.\ 1990, \aap, 230, L1

\bibitem[Hales et al.(2015)]{2015ApJ...812..134H} Hales, A.~S., Corder, S.~A., Dent, W.~R.~D., et al.\ 2015, \apj, 812, 134 

\bibitem[Harris et al.(2012)]{2012ApJ...751..115H} Harris, R.~J., Andrews, S.~M., Wilner, D.~J., \& Kraus, A.~L.\ 2012, \apj, 751, 115 

\bibitem[Hartmann et al.(1989)]{1989ApJ...338.1001H} Hartmann, L., Kenyon, S.~J., Hewett, R., et al.\ 1989, \apj, 338, 1001 

\bibitem[Hartmann \& Kenyon(1996)]{1996ARA&A..34..207H} Hartmann, L., \& Kenyon, S.~J.\ 1996, \araa, 34, 207 

\bibitem[Herbig(1977)]{1977ApJ...217..693H} Herbig, G.~H.\ 1977, \apj, 217, 693 

\bibitem[Herbig et al.(2003)]{2003ApJ...595..384H} Herbig, G.~H., Petrov, P.~P., \& Duemmler, R.\ 2003, \apj, 595, 384 

\bibitem[Herbig(2008)]{2008AJ....135..637H} Herbig, G.~H.\ 2008, \aj, 135, 637



\bibitem[Hillenbrand(1997)]{1997AJ....113.1733H} Hillenbrand, L.~A.\ 1997, \aj, 113, 1733 

\bibitem[Ho et al.(2004)]{2004ApJ...616L...1H} Ho, P.~T.~P., Moran, J.~M., \& Lo, K.~Y.\ 2004, \apjl, 616, L1 

\bibitem[Hodapp \& Chini(2014)]{2014ApJ...794..169H} Hodapp, K.~W., \& Chini, R.\ 2014, \apj, 794, 169 



\bibitem[I-Hsiu Li et al.(2017)]{2017ApJ...840...72I} I-Hsiu Li, J., Liu, H.~B., Hasegawa, Y., \& Hirano, N.\ 2017, \apj, 840, 72 

\bibitem[Johnstone et al.(2013)]{2013ApJ...765..133J} Johnstone, D., Hendricks, B., Herczeg, G.~J., \& Bruderer, S.\ 2013, \apj, 765, 133

\bibitem[Kenyon et al.(1988)]{1988ApJ...325..231K} Kenyon, S.~J., Hartmann, L., \& Hewett, R.\ 1988, \apj, 325, 231

\bibitem[Kenyon et al.(1989)]{1989ApJ...344..925K} Kenyon, S.~J., Hartmann, L., Imhoff, C.~L., \& Cassatella, A.\ 1989, \apj, 344, 925 

\bibitem[Kenyon et al.(1991)]{1991PASP..103.1069K} Kenyon, S.~J., Hartmann, L.~W., \& Kolotilov, E.~A.\ 1991, \pasp, 103, 1069 

\bibitem[Kenyon et al.(1993)]{1993AJ....105.1505K} Kenyon, S.~J., Hartmann, L., Gomez, M., Carr, J.~S., \& Tokunaga, A.\ 1993, \aj, 105, 1505 


\bibitem[Kolotilov \& Petrov(1983)]{1983PAZh....9..171K} Kolotilov, E.~A., \& Petrov, P.~P.\ 1983, Pisma v Astronomicheskii Zhurnal, 9, 171 

\bibitem[Koresko et al.(1991)]{1991AJ....102.2073K} Koresko, C.~D., Beckwith, S.~V.~W., Ghez, A.~M., Matthews, K., \& Neugebauer, G.\ 1991, \aj, 102, 2073

\bibitem[Kospal et al.(2005)]{2005IBVS.5661....1K} Kospal, A., Abraham, P., Acosta-Pulido, J., et al.\ 2005, Information Bulletin on Variable Stars, 5661, 1 

\bibitem[K{\'o}sp{\'a}l et al.(2011)]{2011A&A...527A..96K} K{\'o}sp{\'a}l, {\'A}., Salter, D.~M., Hogerheijde, M.~R., Mo{\'o}r, A., \& Blake, G.~A.\ 2011a, \aap, 527, A96

\bibitem[K{\'o}sp{\'a}l et al.(2011)]{2011A&A...527A.133K} K{\'o}sp{\'a}l, {\'A}., {\'A}brah{\'a}m, P., Acosta-Pulido, J.~A., et al.\ 2011b, \aap, 527, A133 

\bibitem[K{\'o}sp{\'a}l(2011)]{2011A&A...535A.125K} K{\'o}sp{\'a}l, {\'A}.\ 2011b, \aap, 535, A125  

\bibitem[K{\'o}sp{\'a}l et al.(2016)]{2016A&A...596A..52K} K{\'o}sp{\'a}l, {\'A}., {\'A}brah{\'a}m, P., Acosta-Pulido, J.~A., et al.\ 2016, \aap, 596, A52 

\bibitem[K{\'o}sp{\'a}l et al.(2017)]{2017ApJ...843...45K} K{\'o}sp{\'a}l, {\'A}., {\'A}brah{\'a}m, P., Csengeri, T., et al.\ 2017, \apj, 843, 45 


\bibitem[Kopatskaya et al.(2013)]{2013MNRAS.434...38K} Kopatskaya, E.~N., Kolotilov, E.~A., \& Arkharov, A.~A.\ 2013, \mnras, 434, 38

\bibitem[Krist et al.(1997)]{1997ApJ...481..447K} Krist, J.~E., Burrows, C.~J., Stapelfeldt, K.~R., et al.\ 1997, \apj, 481, 447 

\bibitem[Kuffmeier et al.(2017)]{2017ApJ...846....7K} Kuffmeier, M., Haugb{\o}lle, T., \& Nordlund, {\AA}.\ 2017, \apj, 846, 7

\bibitem[Lim \& Takakuwa(2006)]{2006ApJ...653..425L} Lim, J., \& Takakuwa, S.\ 2006, \apj, 653, 425 

\bibitem[Lim et al.(2016)]{2016ApJ...826..153L} Lim, J., Yeung, P.~K.~H., Hanawa, T., et al.\ 2016, \apj, 826, 153 


\bibitem[Liu et al.(2014)]{2014ApJ...780..155L} Liu, H.~B., Galv{\'a}n-Madrid, R., Forbrich, J., et al.\ 2014, \apj, 780, 155 

\bibitem[Liu et al.(2016)]{2016ApJ...816L..29L} Liu, H.~B., Galv{\'a}n-Madrid, R., Vorobyov, E.~I., et al.\ 2016a, \apjl, 816, L29 

\bibitem[Liu et al.(2016)]{2016SciA....200875L} Liu, H.~B., Takami, M., Kudo, T., et al.\ 2016b, Science Advances, 2, e1500875 

\bibitem[Liu et al.(2017)]{2017A&A...602A..19L} Liu, H.~B., Vorobyov, E.~I., Dong, R., et al.\ 2017, \aap, 602, A19 

\bibitem[Lodato \& Bertin(2001)]{2001A&A...375..455L} Lodato, G., \& Bertin, G.\ 2001, \aap, 375, 455 

\bibitem[Lodato \& Bertin(2003)]{2003A&A...408.1015L} Lodato, G., \& Bertin, G.\ 2003, \aap, 408, 1015 

\bibitem[Looney et al.(1997)]{1997ApJ...484L.157L} Looney, L.~W., Mundy, L.~G., \& Welch, W.~J.\ 1997, \apjl, 484, L157 

\bibitem[Magakian et al.(2013)]{2013MNRAS.432.2685M} Magakian, T.~Y., Nikogossian, E.~H., Movsessian, T., et al.\ 2013, \mnras, 432, 2685 

\bibitem[Mercer \& Stamatellos(2017)]{2017MNRAS.465....2M} Mercer, A., \& Stamatellos, D.\ 2017, \mnras, 465, 2 


\bibitem[Miller et al.(2011)]{2011ApJ...730...80M} Miller, A.~A., Hillenbrand, L.~A., Covey, K.~R., et al.\ 2011, \apj, 730, 80 

\bibitem[Momose et al.(1998)]{1998ApJ...504..314M} Momose, M., Ohashi, N., Kawabe, R., Nakano, T., \& Hayashi, M.\ 1998, \apj, 504, 314 

\bibitem[Movsessian et al.(2006)]{2006A&A...455.1001M} Movsessian, T.~A., Khanzadyan, T., Aspin, C., et al.\ 2006, \aap, 455, 1001 

\bibitem[Munari et al.(2010)]{2010ATel.2808....1M} Munari, U., Milani, A., Valisa, P., \& Semkov, E.\ 2010, The Astronomer's Telegram, 2808


\bibitem[Pascucci et al.(2016)]{2016ApJ...831..125P} Pascucci, I., Testi, L., Herczeg, G.~J., et al.\ 2016, \apj, 831, 125 

\bibitem[Peneva et al.(2010)]{2010A&A...515A..24P} Peneva, S.~P., Semkov, E.~H., Munari, U., \& Birkle, K.\ 2010, \aap, 515, A24 

\bibitem[P{\'e}rez et al.(2010)]{2010ApJ...724..493P} P{\'e}rez, L.~M., Lamb, J.~W., Woody, D.~P., et al.\ 2010, \apj, 724, 493 

\bibitem[Persson(2004)]{2004IAUC.8441....2P} Persson, R.\ 2004, \iaucirc, 8441, 2 

\bibitem[Pfalzner(2008)]{2008A&A...492..735P} Pfalzner, S.\ 2008, \aap, 492, 735 

\bibitem[Pueyo et al.(2012)]{2012ApJ...757...57P} Pueyo, L., Hillenbrand, L., Vasisht, G., et al.\ 2012, \apj, 757, 57 

\bibitem[Nayakshin \& Lodato(2012)]{2012MNRAS.426...70N} Nayakshin, S., \& Lodato, G.\ 2012, \mnras, 426, 70 

\bibitem[Qi(2003)]{2003cdsf.conf..393Q} Qi, C.\ 2003, SFChem 2002: Chemistry as a Diagnostic of Star Formation, 393 

\bibitem[Osorio et al.(2016)]{2016ApJ...825L..10O} Osorio, M., Mac{\'{\i}}as, E., Anglada, G., et al.\ 2016, \apjl, 825, L10 

\bibitem[Quanz et al.(2007)]{2007ApJ...658..487Q} Quanz, S.~P., Henning, T., Bouwman, J., Linz, H., \& Lahuis, F.\ 2007a, \apj, 658, 487 

\bibitem[Quanz et al.(2007)]{2007ApJ...668..359Q} Quanz, S.~P., Henning, T., Bouwman, J., et al.\ 2007b, \apj, 668, 359 

\bibitem[Reipurth \& Aspin(1997)]{1997AJ....114.2700R} Reipurth, B., \& Aspin, C.\ 1997, \aj, 114, 2700

\bibitem[Reipurth et al.(2004)]{2004AJ....127.1736R} Reipurth, B., Rodr{\'{\i}}guez, L.~F., Anglada, G., \& Bally, J.\ 2004, \aj, 127, 1736 

\bibitem[Reipurth \& Aspin(2004)]{2004ApJ...606L.119R} Reipurth, B., \& Aspin, C.\ 2004a, \apjl, 606, L119 

\bibitem[Reipurth \& Aspin(2004)]{2004ApJ...608L..65R} Reipurth, B., \& Aspin, C.\ 2004b, \apjl, 608, L65 

\bibitem[Reipurth et al.(2007)]{2007AJ....133.1000R} Reipurth, B., Aspin, C., Beck, T., et al.\ 2007a, \aj, 133, 1000 

\bibitem[Reipurth et al.(2007)]{2007AJ....134.2272R} Reipurth, B., Guimar{\~a}es, M.~M., Connelley, M.~S., \& Bally, J.\ 2007b, \aj, 134, 2272

\bibitem[Reipurth \& Aspin(2010)]{2010vaoa.conf...19R} Reipurth, B., \& Aspin, C.\ 2010, Evolution of Cosmic Objects through their Physical Activity, 19

\bibitem[Rodriguez et al.(1990)]{1990PASP..102.1413R} Rodriguez, L.~F., Hartmann, L.~W., \& Chavira, E.\ 1990, \pasp, 102, 1413 

\bibitem[Rodr{\'{\i}}guez et al.(2003)]{2003ApJ...583..330R} Rodr{\'{\i}}guez, L.~F., Curiel, S., Cant{\'o}, J., et al.\ 2003, \apj, 583, 330 

\bibitem[Ru{\'{\i}}z-Rodr{\'{\i}}guez et al.(2017)]{2017MNRAS.466.3519R} Ru{\'{\i}}z-Rodr{\'{\i}}guez, D., Cieza, L.~A., Williams, J.~P., et al.\ 2017, \mnras, 466, 3519 

\bibitem[Saito et al.(1996)]{1996ApJ...473..464S} Saito, M., Kawabe, R., Kitamura, Y., \& Sunada, K.\ 1996, \apj, 473, 464

\bibitem[Sandell \& Aspin(1998)]{1998A&A...333.1016S} Sandell, G., \& Aspin, C.\ 1998, \aap, 333, 1016

\bibitem[Sandell \& Knee(2001)]{2001ApJ...546L..49S} Sandell, G., \& Knee, L.~B.~G.\ 2001, \apjl, 546, L49

\bibitem[Sandell \& Weintraub(2001)]{2001ApJS..134..115S} Sandell, G., \& Weintraub, D.~A.\ 2001, \apjs, 134, 115 

\bibitem[Sault et al.(1995)]{1995ASPC...77..433S} Sault, R.~J., Teuben, P.~J., \& Wright, M.~C.~H.\ 1995, Astronomical Data Analysis Software and Systems IV, 77, 433

\bibitem[Semkov et al.(2010)]{2010A&A...523L...3S} Semkov, E.~H., Peneva, S.~P., Munari, U., Milani, A., \& Valisa, P.\ 2010, \aap, 523, L3

\bibitem[Semkov et al.(2012)]{2012A&A...542A..43S} Semkov, E.~H., Peneva, S.~P., Munari, U., et al.\ 2012, \aap, 542, A43 

\bibitem[Staude \& Neckel(1991)]{1991A&A...244L..13S} Staude, H.~J., \& Neckel, T.\ 1991, \aap, 244, L13

\bibitem[Staude \& Neckel(1992)]{1992ApJ...400..556S} Staude, H.~J., \& Neckel, T.\ 1992, \apj, 400, 556

\bibitem[Strom \& Strom(1993)]{1993ApJ...412L..63S} Strom, K.~M., \& Strom, S.~E.\ 1993, \apjl, 412, L63

\bibitem[Szeifert et al.(2010)]{2010A&A...509L...7S} Szeifert, T., Hubrig, S., Sch{\"o}ller, M., et al.\ 2010, \aap, 509, L7 

\bibitem[The et al.(1994)]{1994A&AS..104..315T} The, P.~S., de Winter, D., \& Perez, M.~R.\ 1994, \aaps, 104,  


\bibitem[Tsukamoto et al.(2017)]{2017ApJ...838..151T} Tsukamoto, Y., Okuzumi, S., \& Kataoka, A.\ 2017, \apj, 838, 151 

\bibitem[van den Ancker et al.(2004)]{2004MNRAS.349.1516V} van den Ancker, M.~E., Blondel, P.~F.~C., Tjin A Djie, H.~R.~E., et al.\ 2004, \mnras, 349, 1516 


\bibitem[Vorobyov \& Basu(2010)]{2010ApJ...719.1896V} Vorobyov, E.~I., \& Basu, S.\ 2010, \apj, 719, 1896 

\bibitem[Vorobyov(2013)]{2013A&A...552A.129V} Vorobyov, E.~I.\ 2013, \aap, 552, A129 



\bibitem[Vorobyov \& Basu(2015)]{2015ApJ...805..115V} Vorobyov, E.~I., \& Basu, S.\ 2015, \apj, 805, 115 

\bibitem[Vorobyov(2016)]{2016A&A...590A.115V} Vorobyov, E.~I.\ 2016, \aap, 590, A115 

\bibitem[Wang et al.(2004)]{2004ApJ...601L..83W} Wang, H., Apai, D., Henning, T., \& Pascucci, I.\ 2004, \apjl, 601, L83 

\bibitem[Welch et al.(2000)]{2000ApJ...540..362W} Welch, W.~J., Hartmann, L., Helfer, T., \& Brice{\~n}o, C.\ 2000, \apj, 540, 362

\bibitem[Welin(1971)]{1971A&A....12..312W} Welin, G.\ 1971, \aap, 12, 312 


\bibitem[Yen et al.(2017)]{2017arXiv170802384Y} Yen, H.-W., Takakuwa, S., Chu, Y.-H., et al.\ 2017, arXiv:1708.02384 

\bibitem[Yoo et al.(2017)]{2017arXiv170904096Y} Yoo, H., Lee, J.-E., Mairs, S., et al.\ 2017, arXiv:1709.04096 

\bibitem[Zurlo et al.(2017)]{2017MNRAS.465..834Z} Zurlo, A., Cieza, L.~A., Williams, J.~P., et al.\ 2017, \mnras, 465, 834 

\bibitem[Zhu et al.(2007)]{2007ApJ...669..483Z} Zhu, Z., Hartmann, L., Calvet, N., et al.\ 2007, \apj, 669, 483 


\bibitem[Zhu et al.(2010)]{2010ApJ...713.1143Z} Zhu, Z., Hartmann, L., \& Gammie, C.\ 2010, \apj, 713, 1143 


\end{thebibliography}
\end{document}